\documentstyle[fleqn,epsfig,run2col,amstex]{article}
\input epsf
\newcommand{\DO}{D\raise1pt\hbox{$\not$}O}
\newcommand{\PT}
  {\hbox{$\textstyle p$\kern-2pt\lower5pt\hbox{$\scriptscriptstyle T$}}}
\newcommand{\QT}
  {\hbox{$\textstyle Q$\kern-2pt\lower4.5pt\hbox{$\scriptscriptstyle T$}}}
\newcommand{\KT}
  {\hbox{$\textstyle k$\kern-0pt\lower2.5pt\hbox{$\scriptscriptstyle T$}}}
\newcommand{\qt}
  {\hbox{$\textstyle q$\kern-0.5pt\lower4.5pt\hbox{$\scriptscriptstyle T$}}}
\newcommand{\avQT}
  {\hbox{$\langle\QT\rangle$}}
\newcommand{\avkt}
  {\hbox{$\langle\KT\rangle$}}
\newcommand{\avqt}
  {\hbox{$\langle\qt\rangle$}}

\def\ppbar{$p\overline{p}~$}             
\def\pbarp{$\overline{p}p~$}             
\def\pt{$p_T~$}                          
\def\met{\mbox{${\hbox{$E$\kern-0.6em\lower-.1ex\hbox{/}}}_T~$}} 
\def\etal{{\it et al.}}                 
\def\pbarp{$\overline{p}p $}            
\def\met{\mbox{${\hbox{$E$\kern-0.6em\lower-.1ex\hbox{/}}}_T$}} 
\def\etal{{\sl et al.}}                   

\newcommand{\Dzero}{D\O\ }

\def\dk{\Delta\kappa}

\def\be{\begin{equation}}
\def\ee{\end{equation}}
\def\bea{\begin{eqnarray}}
\def\eea{\end{eqnarray}}

\newcommand{\figLumUnc}
{
\begin{figure}[thb]
\epsfxsize=8cm 
\centerline{\epsfbox{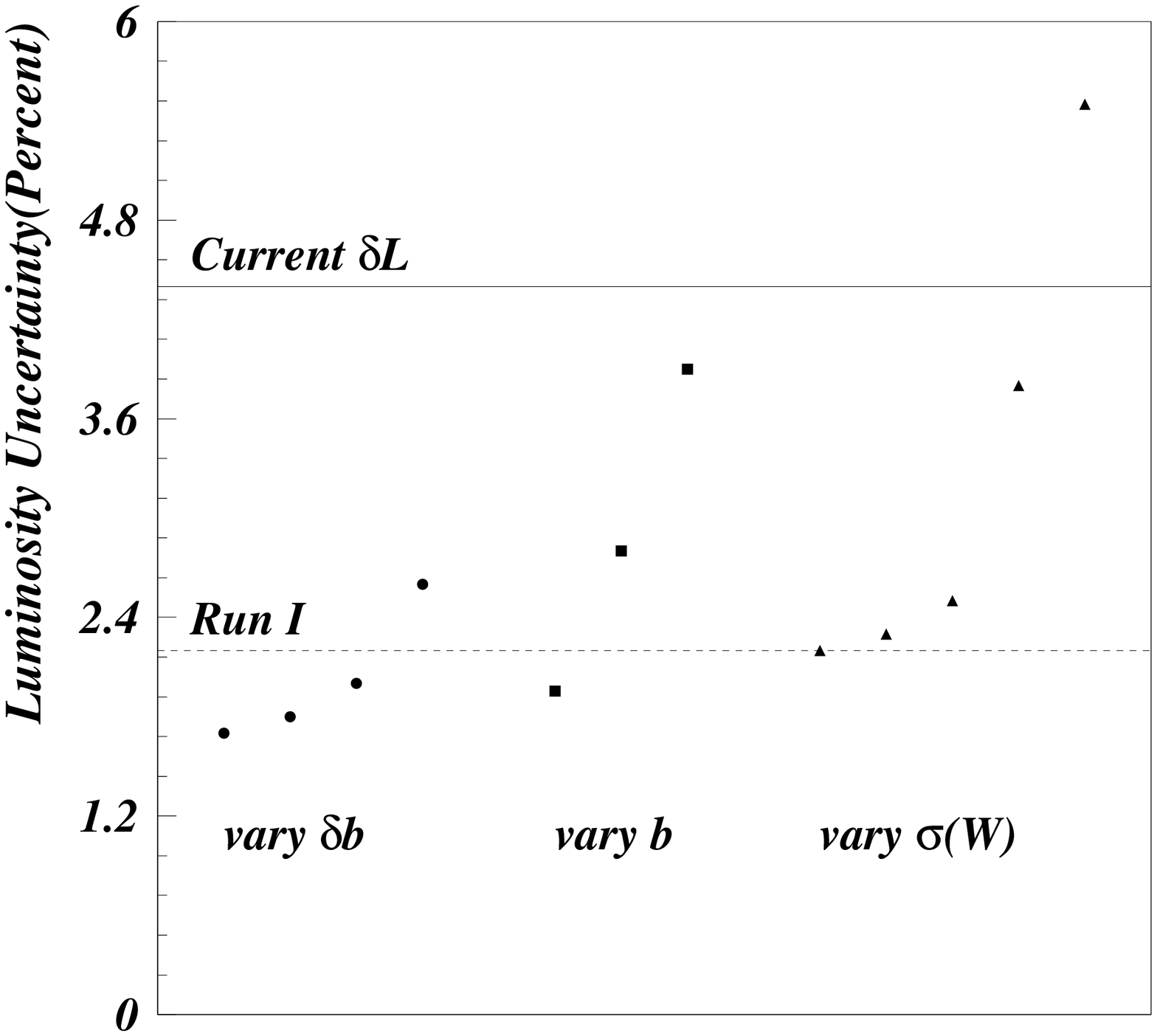}}
\caption{Impact of changing the overall background level ($b$), the
uncertainty on the background level ($\delta b$), and the predicted
$\sigma(W)$ in determining ${\cal{L}}$. Except when being varied, the
experimental values are kept to the Run~I determinations. Except when
varying the uncertainty on the cross section, $\delta\sigma(W)$ is set 
to $0$. The circles denote 
${\delta b\over b} = 0.05,0.10,0.15,0.25.$
The squares denote background fractions, $b=5,10,15\%$. The triangles denote
${\delta\sigma(W)\over\sigma(W)}=0.001,0.005,0.01,0.03,0.05.$
The solid line shows the current values of the luminosity uncertainty
at $D\O$ and the dashed line shows the uncertainty if the rate of $W$
boson production was used and $\sigma(W)$ was known perfectly.
}
\label{fig:lumunc}
\end{figure}
}

\begin{document}
 \title{\bf Report of the Working Group on Photon and Weak Boson Production }
 \author{{\bf Conveners:} U.~Baur (Buffalo), E.L.~Berger (ANL), 
H.T.~Diehl (FNAL), and D.~Errede (UIUC)\\
{\bf Subgroup Conveners:} {\sl Photon Production:} J.~Huston (MSU),
J.~Owens (FSU), and J.~Womersley (FNAL) 
{\sl Weak Boson Production:} D.~Casey (MSU) and T.~Dorigo (Harvard),
{\sl Diboson Production:} U.~Baur (Buffalo) and H.T.~Diehl (FNAL)\\
{\bf Contributing authors:} L. Apanasevich (Rochester), M.~Begel
(Rochester), Y.~Gershtein (Brown), M.~Kelly (Michigan), 
S.~E.~Kuhlmann (ANL), S.~Leone (Pisa), D.~Partos (Brandeis),
D.~Rainwater (FNAL), W.K.~Sakumoto (Rochester), G.~Steinbr\"uck (Columbia),
M.~Zieli\'{n}ski (Rochester), and V.~Zutshi (Rochester) 
}

\begin{abstract}
This report discusses physics issues which can be addressed in photon 
and weak boson production in Run~II at the Tevatron. The current
understanding and the potential of Run~II to expand our knowledge of
direct photon production in hadronic collisions is discussed. We explore 
the prospects for using the $W$-boson cross section to measure the
integrated luminosity, improving the measurement of the $W$ and $Z$
boson transverse momentum distributions, the $Z\to b\bar b$ signal, and
the lepton angular distribution in $W$ decays. Finally, we consider the
prospects for measuring the trilinear gauge boson couplings in Run~II.
\end{abstract}

\maketitle

\section{Introduction}

For the next few years, the Fermilab Tevatron Collider will be the high
energy frontier of particle physics. The luminosity enhancement provided 
by the Main Injector will significantly increase the discovery reach of the
Tevatron experiments over what has been achieved with Run~I data. It
will also move the experimental program into a 
regime of precision hadron collider physics. This will make it possible
to address open questions of high energy physics from several
complementary directions.

Understanding the mechanism of production for photons, $W$, and $Z$-bosons
is important for several reasons. First of all, it
provides an opportunity to directly test the Standard Model
(SM). Second, photon and weak boson production often constitute an
irreducible background in searches for new physics. Finally, a detailed
understanding of the production mechanism for these particles is
necessary to control the 
systematic errors in precision measurements, such as the determination of
the $W$ mass (see Ref.~\cite{wgiii}). 

In this report, we consider several aspects of the production of
photons, $W$, 
and $Z$ bosons which are of interest for Run~II of the Tevatron Collider.
In Sec.~\ref{s:dphot}, direct photon production is discussed. A detailed 
overview of our current experimental and theoretical understanding of 
direct photon production in hadronic collisions is presented. Direct
photon production has long been considered a probe of QCD and a source
for extracting the gluon distribution of the proton. Unfortunately, not
all existing 
(fixed target and Tevatron collider) datasets are consistent, and our
current theoretical understanding of direct photon production,
especially at small photon transverse momenta, is
incomplete. Recent theoretical developments, however, offer optimism
that the long-standing difficulties in direct photon production can
finally be resolved. The
enormous number of photon-jet events expected in Run~II may help to
shed light on these issues. In
particular, photon~-- jet correlations should be helpful in sorting out 
the source of disagreement between theory and experiment. In addition,
the kinematic reach in transverse momentum will be greatly extended in
Run~II. 

In Sec.~\ref{s:topics} several important topics associated with $W$ and
$Z$-boson production are discussed. For many measurements in Run~II, 
knowledge of the
integrated luminosity is essential. The integrated luminosity can be
extracted either from the total inelastic cross section or from the
cross section of a theoretically well understood process with high
statistics, such as inclusive $W$ 
production. Using the total inelastic cross section to determine the
integrated luminosity for Run~I has led to inconsistencies which may
well persist in Run~II. It may thus be advantageous to use the
$W$ production cross section as an alternative. In
Sec.~\ref{s:lum} we present a brief overview of the magnitude of the 
experimental uncertainties in such a measurement. The total uncertainty
in the $W$ cross section is found to be dominated by the uncertainty from
the parton distribution functions, which is considered in more detail 
in the Report of the
Working Group on Parton Distribution Functions~\cite{pdfUnc}. 

In Sec.~\ref{s:topics}, we also consider the transverse momentum
distribution of the $Z$ boson, $Z\to b\bar b$ decays, and the prospects
to measure the lepton angular distribution in $W$ decays. The $Z$ $p_T$
distribution is of interest as a test of QCD, and as a tool for
reducing uncertainties in the transverse momentum distribution of the
$W$. This is important for a precise determination of the mass of the
$W$ boson (see Ref.~\cite{wgiii}). Searching for a light Higgs boson in
the range between 110~GeV and 180~GeV is one of the prime objectives for 
Run~II. $H\to b\bar b$ decays dominate for Higgs boson masses 
$M_H<135$~GeV. The ability to separate the Higgs boson signal in the 
$Wb\bar b$ and $t\bar tb\bar b$ channels from the large QCD background
depends critically on on the 
$b\bar b$ invariant mass resolution, and thus on the measured $b$-quark jet
energies. $Z\to b\bar b$ decays offer a testing ground for algorithms
designed to improve the jet energy measurement for $b$~jets and are also 
useful as a calibration tool. The measurement of the lepton 
angular distributions in $W$ decays serves as a probe of NLO QCD. The
measurement carried out by D\O\ in Run~I is statistics limited. While a
QCD calculation is preferred, large deviations
from QCD are not excluded. In Run~II, this measurement will allow for a much 
improved test of the QCD prediction.

In Sec.~\ref{s:diboson} of this report, we discuss di-boson
production. Vector 
boson pair production provides a sensitive ground for direct tests of
the trilinear gauge boson couplings. A brief overview of the $WWV$
($V=\gamma,\,Z$), $Z\gamma V$ and $ZZV$ couplings is presented and
recent advances in our theoretical understanding of the NLO QCD
corrections to di-boson production are described. After a brief review
of the limits on trilinear couplings obtained in Run~I, the prospects for
strengthening existing bounds in Run~II are discussed. In addition to improving
the measurements of $WWV$ and $Z\gamma V$ couplings, it will be possible 
to determine the $ZZV$ couplings via $ZZ$ production with an accuracy of 
about 15\% in Run~II, and to observe the so-called ``radiation zero'' 
in $W\gamma$ production.

\section{Direct Photon Production\label{s:dphot}}

The use of direct photon production as an electromagnetic probe of hard 
scattering dynamics has a history which covers more than twenty years.
As in other electromagnetic processes such as lepton pair production or
deep inelastic scattering, the point-like coupling of the photon to charged 
particles offers some simplifications over purely hadronic probes. 
Compared to hadronic jet production, direct photon production 
offers the apparent advantages of having fewer subprocesses at lowest
order 
and of avoiding the complications of jet definitions when measuring or 
calculating a cross section. This latter point means that one can extend
the range of transverse momenta to smaller values for direct photons than
for jets. However, in actual practice, these apparent simplifications must
be tempered by having to deal with backgrounds from neutral meson decays, a 
lower event rate compared to jet production, and complications from
photons 
produced during jet fragmentation, to name just a few. Nevertheless, 
direct photon data provide information which complements that obtained 
from other hard scattering processes. Furthermore, photons may be
important signatures of physics beyond the SM. Therefore, it is
necessary to understand the ``conventional'' sources 
of photons before one can fully exploit them in signatures designed to
look for new physics.

In this Section, recent work concerning the phenomenology of initial-state
gluon emission in direct-photon production in hadron collisions is
reported. In Sec.~\ref{s:ph1},
high-mass direct-photon pairs are used to explore the impact of such
radiation in terms of effective parton transverse momenta, $k_T$.  At
fixed-target energies, data on high-\PT\ inclusive $\pi^0$ and
$\pi^0\pi^0$ production are used to further clarify the arguments
presented. We then review
progress towards fully resummed QCD descriptions and present
comparisons of a phenomenological $k_T$ model to recent fixed-target
and collider data. Possibilities for more extensive studies with data 
from Run~II, and the additional information they can
provide for these considerations, are explored in Sec.~\ref{s:ph12}.  
A consistent picture of the observed deviations of NLO perturbative QCD 
(pQCD) calculations from inclusive
direct-photon and $\pi^0$ data is now emerging, and we comment on the
implications of these results for the extraction of the gluon
distribution, $G(x)$, in Sec.~\ref{s:ph13}.

Run~II has the potential to significantly expand our knowledge of direct 
photon production. Issues related to our understanding of the relevant 
production mechanisms in the kinematic range accessible during Run~II 
are reviewed in Sec.~\ref{s:ph2}. In this Section, we also discuss 
observables which may help improve our understanding of  
direct photon production as well as experimental issues 
which can affect the quality of the data. In Sec.~\ref{s:new} some 
predictions for photon-jet correlations are presented. The potential of 
observables other than the usual single photon $p_T$ distribution to
help elucidate the underlying dynamics is discussed. 

Run~I data are available on the associated production of a $\gamma$  
carrying large transverse momentum along with a charm quark $c$ whose 
transverse momentum balances a substantial portion of that of the 
photon~\cite{cdf}.  An intriguing possibility is that the data may be used
to measure the charm quark density in the nucleon as well as to probe
dynamical correlations predicted by QCD.  These possibilities are 
discussed in Ref.~\cite{bbg} where predictions are obtained from 
a full next-to-leading order perturbative QCD calculation of 
$p +\bar{p}\rightarrow \gamma + c + X$ at high energy. The associated
production of a photon and a heavy quark is not discussed further in
this report.

\subsection{Present Status of Direct Photon Production in Hadronic
Collisions$^1$\label{s:ph1}}
\footnotetext{$^1$ Contributed by: L.~Apanasevich,
M.~Begel, Y.~Gershtein, J.~Huston, S.~E.~Kuhlmann, D.~Partos,
J.~Womersley, M.~Zieli\'{n}ski, and V.~Zutshi.}

\subsubsection{Introduction}
 
\begin{figure}[ht]
\centering\leavevmode
\vglue1truept\vspace{-4.25cm}
\hglue1truept\hspace{-1cm}
\epsfxsize 10 cm 
\epsfbox{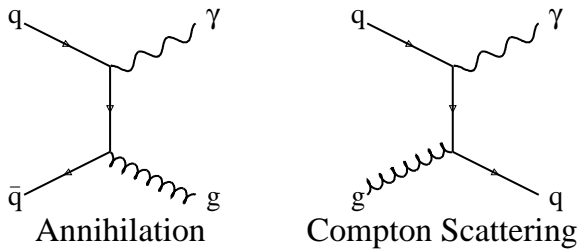}
\vglue1truept\vspace{-6.75cm}
\caption{Leading order diagrams for direct-photon production.}
\label{fig:diagrams}
\vspace{-0.75cm}
\end{figure}

Single and double direct-photon production at high \PT\ have long been
viewed as ideal processes for testing the formalisms of pQCD. NLO
calculations are available for both
processes~\cite{aurenche-nlo,bbf,bqiu,bailey,binoth}. While the importance of
including gluon emission through the resummation formalism was
recognized and available for some time for the di-photon
process~\cite{RESBOS,fergani}, it is only recently that this approach
has been developed for inclusive direct-photon
production~\cite{nason,kidonakisowens,laenen,lilai,li,sterman}. A
complete theoretical description of the direct-photon process is of
special importance as it has long been expected to provide one of the
best measurements of the gluon distribution in the proton.  The
quark--gluon Compton scattering subprocess ($gq\rightarrow\gamma q$)
shown in Fig.~\ref{fig:diagrams} provides a major contribution to
inclusive direct-photon production.  The gluon distribution ($G(x)$)
is relatively well constrained for $x<0.1$ by deep-inelastic
scattering~(DIS) and Drell-Yan~(DY) data, but less so at
larger~$x$~\cite{huston-uncertainty}.  Fixed-target direct-photon data
can constrain $G(x)$ at large~$x$, and consequently has been
incorporated in several modern global parton distribution
analyses~\cite{cteq4,grv92,mrst}.  More recently, however, both the
completeness of the theoretical NLO description of the direct-photon
process, and the consistency of the available data sets have been the
subject of intense
debate~\cite{mrst,baerreno,huston-discrepancy,ktprd,aurenche-dp,aurenche-pi0,kimber}.
Direct-photon measurements in collider data, and especially the data
expected from the forthcoming Run~II at the Tevatron, provide an
important testing ground for novel approaches and improvements in the
understanding of the direct-photon process, and therefore can help
resolve the present arguments.

The understanding of single and double direct photon yields, and of
the more copious high-\PT\ $\pi^0$ production, is of importance for
searches for the Higgs in the $\gamma\gamma$ decay mode at the LHC.
In addition, Higgs production, both at the Tevatron and the LHC, can
be affected by soft-gluon emission from the initial-state partons, and
separation of signal and background can benefit from a reliable
resummation formalism or equivalent parton-shower Monte Carlo
descriptions~\cite{csaba-higgs,csaba-thesis,LHCworkshopdocument}.

\subsubsection{Parton Transverse Momentum}

\begin{figure}[t]
\centering\leavevmode
\vglue1truept\vspace{-1cm}
\hglue1truept\hspace{0.5cm}
\epsfxsize 7 cm 
\epsfbox{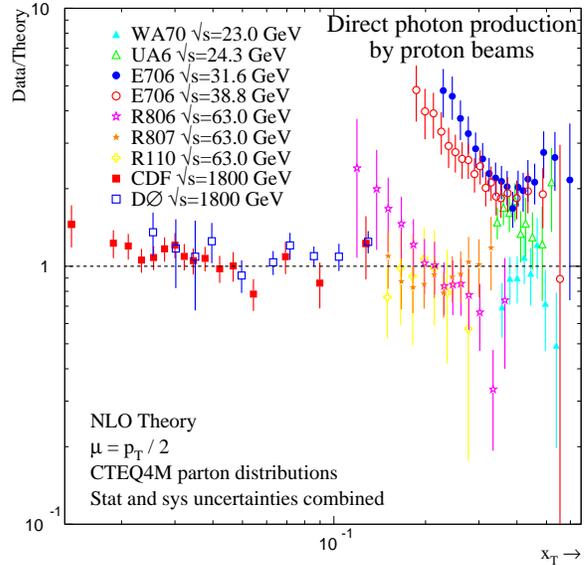}
\vglue1truept\vspace{-1.75cm}
\caption{Comparison between proton-induced direct-photon data and NLO pQCD
calculations for several experiments as a function of photon $x_T$
($=2p_T/\sqrt{s}$).  (The CDF and \DO\ data are from Run~Ia, see
Figs.~\ref{fig:newCDF} and~\ref{fig:newD0} for the Run~Ib update.)}
\label{fig:dpdiscrepancy_p}
\end{figure}

A pattern of deviation has been observed between measured
direct-photon cross sections and NLO calculations
(Fig.~\ref{fig:dpdiscrepancy_p}).
The origin of the disagreement has been ascribed to the effect of
initial-state soft-gluon radiation~\cite{huston-discrepancy,ktprd}.
Correlations between any produced high-\PT\ particles probe aspects of
the hard scatter not easily accessible via studies of single inclusive
particle production.  In particular, studies of high-mass pairs of
particles such as direct photons and $\pi^0$'s can be used to extract
information about the transverse momentum of partons, \KT, prior to
the hard scatter.  Whatever the source, any transverse momentum
between the partons will appear as a net \PT\ imbalance among the
outgoing particles produced in the hard scatter, and is therefore
reflected in the vector sum of the individual \PT\ values of the
outgoing particles (\QT).  If the outgoing particles are pairs of
photons or leptons, then this variable should provide a good measure
of \avkt, with \avkt/parton $\approx\langle\QT\rangle/\sqrt{2}$.  When
the outgoing particles are partons, they will hadronize, but the
reconstructed jets can also yield a measure of \avkt.

\begin{figure}[t]
\centering\leavevmode
\vglue1truept\vspace{-1cm}
\hglue1truept\hspace{0.5cm}
\epsfxsize 7 cm 
\epsfbox{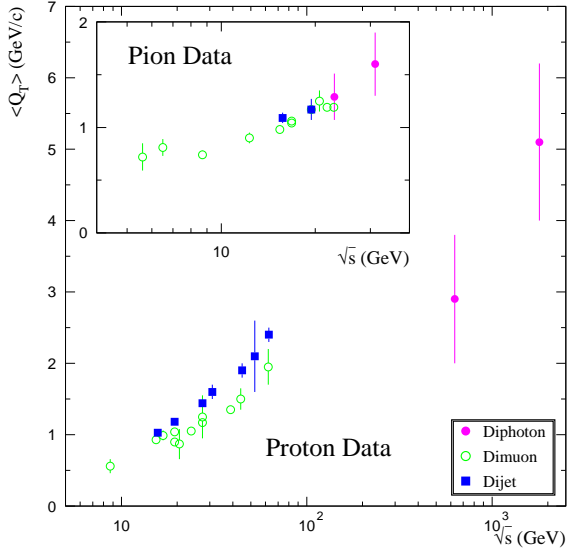}
\vglue1truept\vspace{-1.75cm}
\caption{\avQT\ of pairs of muons, photons, and jets 
produced in hadronic collisions versus $\sqrt{s}$.}
\label{fig:DYkt}
\end{figure}

Evidence of significant \KT\ has been found in several measurements of
dimuon, diphoton, and dijet production; a collection of \avQT\
measurements is displayed in Fig.~\ref{fig:DYkt} for a wide range of
$\sqrt{s}$~\cite{begel,WA70-kt,CDF-ddp,DY-kt,qt}.  The values of
\avQT\ are large, and increase with increasing $\sqrt{s}$.  The dijet
\KT\ measurements (Fig.~\ref{fig:DYkt}) agree qualitatively with the
dimuon and diphoton results, though they have somewhat higher mean
values.  Such a shift is expected since there is also potential for
final-state soft-gluon emission in dijet events.  The values of \avkt\
per parton indicated by these data are too large to be interpreted as
due only to the size of the proton.  From these observations, one can
infer that the \avkt\ per parton is of order 1~GeV/$c$ at fixed-target
energies, increasing to 3~GeV/$c$ to 4~GeV/$c$ at the Tevatron collider,
whereas \avkt\ would be expected to be of the order of 0.3~GeV/$c$
to~0.5~GeV/$c$ based solely on proton size.

\begin{figure}[t]
\centering\leavevmode
\vglue1truept\vspace{-1cm}
\hglue1truept\hspace{0.5cm}
\epsfxsize 7 cm 
\epsfbox{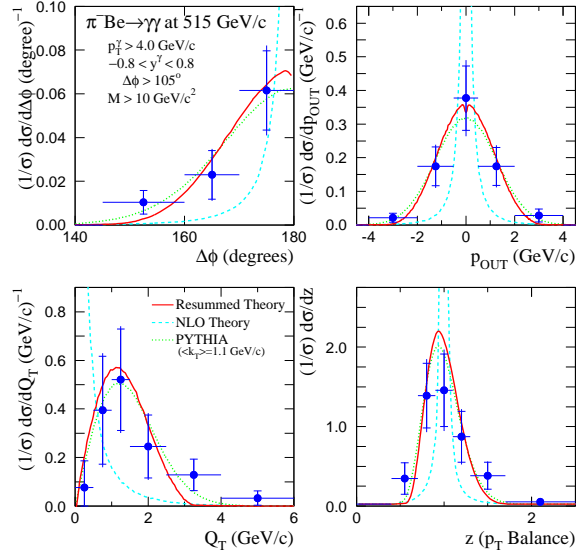}
\vglue1truept\vspace{-2.25cm}
\caption{Diphoton $\Delta\phi$, $p_{OUT}$, \QT, and $z$ distributions 
for E706 $\pi^-$Be data at $\sqrt{s}=31.1$~GeV~\cite{begel}.
Overlayed on the data are the results from NLO~\cite{bailey} (dashed)
and resummed~\cite{RESBOS} (solid) calculations.  {\tt
PYTHIA}~\cite{pythia57} results (dotted) with $\avkt=1.1$~GeV/$c$ are
also shown.}
\label{fig:ddp-kt}
\end{figure}

\begin{figure}[t]
\centering\leavevmode
\vglue1truept\vspace{-1cm}
\hglue1truept\hspace{0.5cm}
\epsfxsize 7 cm 
\epsfbox{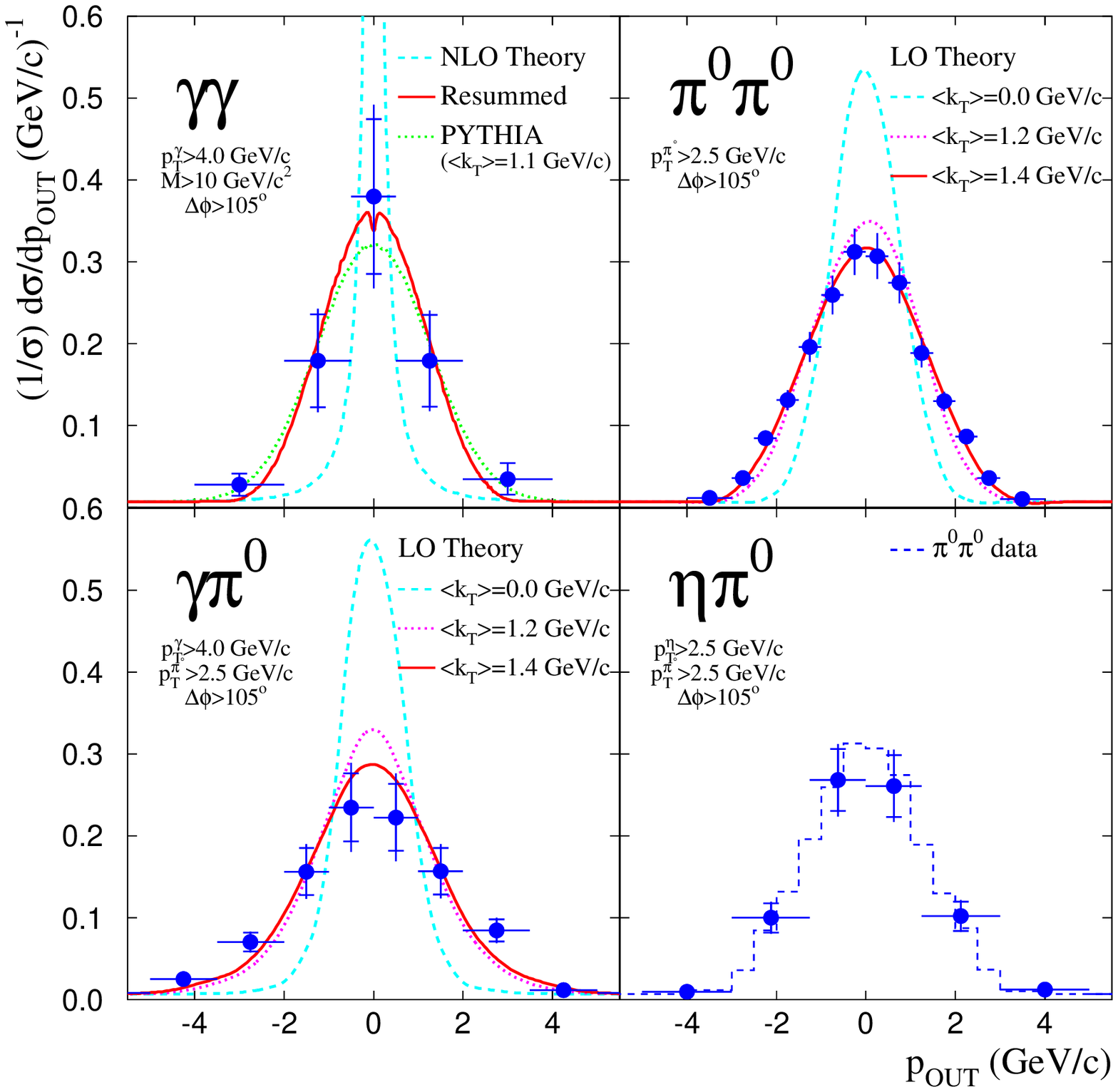}
\vglue1truept\vspace{-2.25cm}
\caption{$p_{OUT}$ distributions for $\gamma\gamma$, $\pi^0\pi^0$,
$\gamma\pi^0$, and $\eta\pi^0$ for E706 $\pi^-$Be data at
$\sqrt{s}=31.1$~GeV~\cite{begel}.  Overlayed on the diphoton
data are the results from NLO~\cite{bailey} and resummed~\cite{RESBOS}
pQCD calculations. {\tt PYTHIA}~\cite{pythia57} results with
$\avkt=1.1$~GeV/$c$ are also shown.  Overlayed on the $\pi^0\pi^0$ and
$\gamma\pi^0$ are the results from LO pQCD calculations~\cite{owens}
for various values of \avkt\ and fixed $\avqt=0.6$~GeV/$c$.  The
$\pi^0\pi^0$ data have been overlayed on the $\eta\pi^0$ data for
comparison.}
\label{fig:pout_all}
\end{figure}

The \PT\ imbalance between the outgoing particles can also be examined
using kinematic variables other than \QT. Given some finite \KT, the
two outgoing particles no longer emerge back-to-back; the azimuthal
angle between the particles, $\Delta\phi$, will differ from
$180^\circ$.  The transverse momentum normal to the
scattering plane, $p_{OUT}$, and the \PT-balance quantity,
\begin{equation}
z= -(\vec{\PT}_1 \cdot \vec{\PT}_2 / {\PT}_2^2) = ({\PT}_2 / {\PT}_1)
\cos\Delta\phi, 
\end{equation}
are two useful variables, each with two possible 
values per pair of objects.

High-mass direct-photon pairs have been measured at the
Tevatron~\cite{CDF-ddp,begel,chen-thesis}.  Distributions as a
function of $\Delta\phi$, $p_{OUT}$, \QT, and~$z$ for such events from
E706~\cite{begel} are shown in Fig.~\ref{fig:ddp-kt}.  Overlayed on
the data are the results from both NLO~\cite{bailey} and
resummed~\cite{RESBOS} pQCD calculations.  There are large differences
in the predicted shapes.  At leading order, each of these
distributions would consist of a $\delta$ function.  While the NLO
prediction has finite width due to the radiation of a single hard
gluon, the resummed theory, which also includes the effects of
multiple soft-gluon emission, is in better agreement with the data.
This is particularly true for \QT, where the NLO calculation tends
towards infinity as $\QT\rightarrow0$, while the resummed ({\tt
RESBOS}~\cite{RESBOS}) calculation follows the shape of the data and
goes to zero.  Also shown in Fig.~\ref{fig:ddp-kt} are the
distributions from {\tt PYTHIA}~\cite{pythia57}, where
\KT\ effects are approximated by a Gaussian smearing technique.
{\tt PYTHIA} provides a reasonable description of the di-photon data
using a value for \avkt\ consistent with the measurements displayed in
Fig.~\ref{fig:ddp-kt}.  Comparisons between CDF and \DO\ data lead to
similar conclusions~\cite{csaba-thesis}.  There is also good agreement
between the WA70 di-photon data~\cite{WA70-kt} and resummed
pQCD~\cite{fergani}.  The increased statistics expected for Run~II
should allow for more detailed comparisons between di-photon data and
theory.

Similar evidence for \KT\ effects is seen in analyses of high-mass
$\pi^0\pi^0$, $\eta\pi^0$, and $\gamma\pi^0$ pairs by
E706~\cite{begel}.  This is illustrated by Fig.~\ref{fig:pout_all}
which shows a comparison of the $p_{OUT}$ distribution from each of
these samples.  The LO pQCD calculation~\cite{owens}, which
incorporates \KT\ effects using a Gaussian smearing technique similar
to that used in {\tt PYTHIA}~\cite{pythia57}, provides a reasonable
characterization of \KT-sensitive variables such as $\Delta\phi$ and
$p_{OUT}$ for \avkt\ similar to that measured for di-photons.  The
\avkt\ values needed to provide good matches to the data for
$\pi^0\pi^0$ and $\gamma\pi^0$ are slightly larger than for
$\gamma\gamma$, but that is expected since $\pi^0$'s emanate from
final-state quarks and gluons that can produce additional gluon
radiation. (We use \avqt=0.6~GeV/$c$~\cite{qt} for the \PT\ due to
fragmentation.)

\subsubsection{\KT\ Phenomenology}

\begin{figure}[th]
\centering\leavevmode
\vglue1truept\vspace{-3cm}
\hglue1truept\hspace{-1cm}
\epsfxsize 10 cm 
\epsfbox{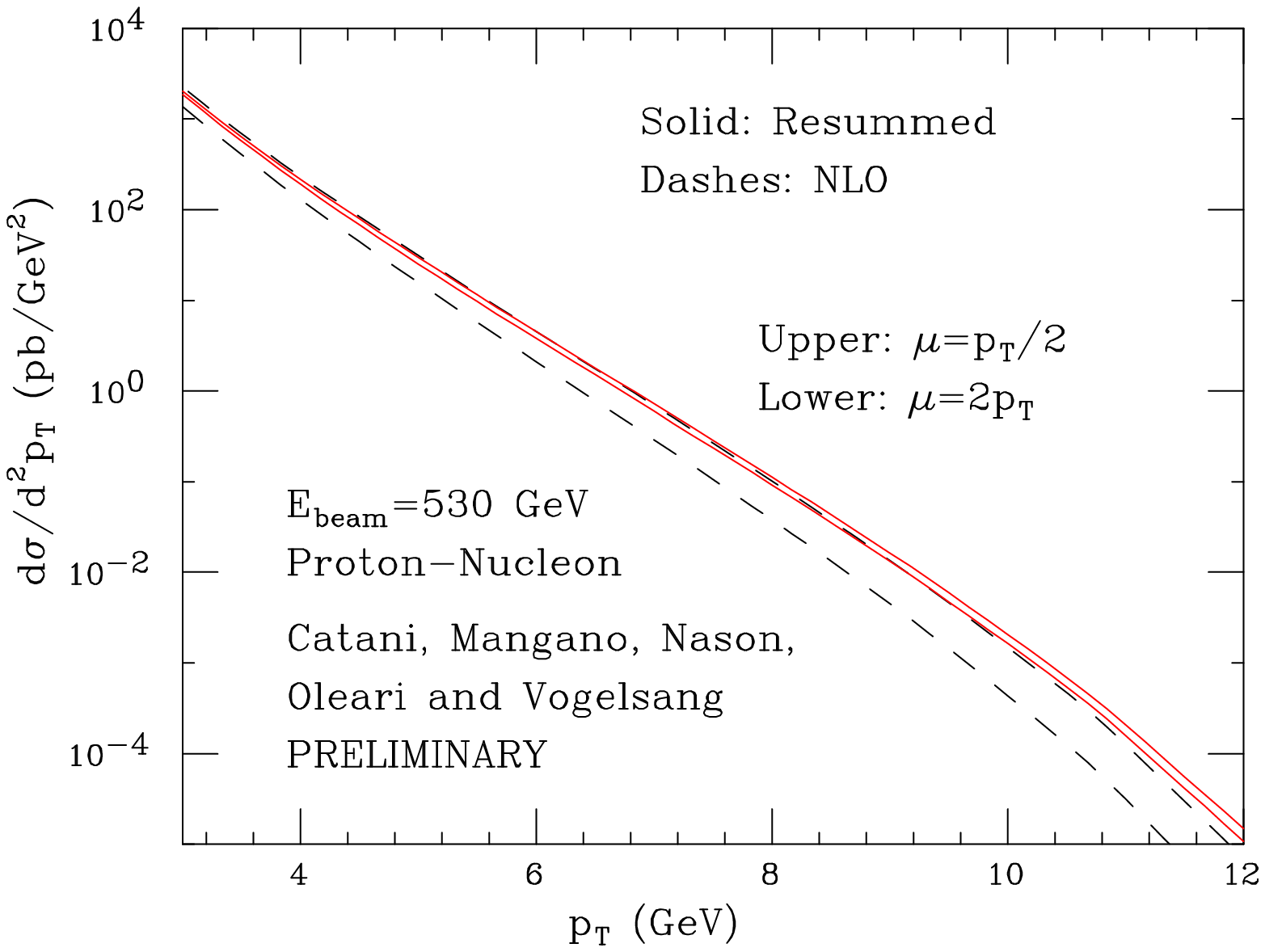}
\vglue1truept\vspace{-4.5cm}
\caption{Comparison between a threshold resummed and a NLO theory calculation
for direct-photon production for two scale choices: $\PT/2$
and~$2\PT$~\cite{nason}.}
\label{fig:nason}
\vspace{0.5cm}
\vglue1truept\vspace{-0.5cm}
\hglue1truept\hspace{0.25cm}
\epsfxsize 7 cm 
\epsfbox{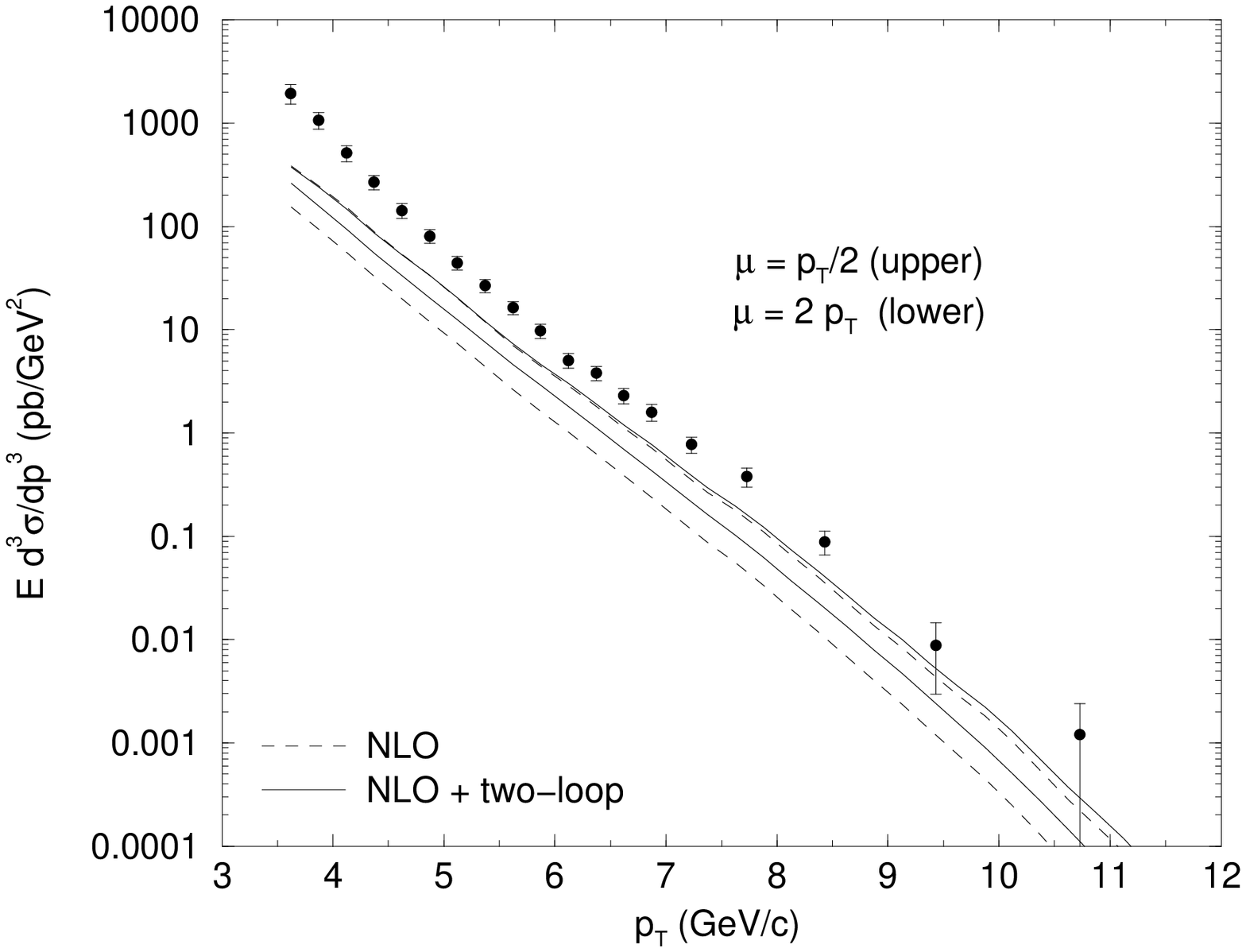}
\vglue1truept\vspace{-.75cm}
\caption{NLO and NNLO results for direct-photon production
in hadronic collisions compared to E706~\cite{E706-kt} $pBe$ data at
$\sqrt{s}=31.6$~GeV~\cite{kidonakisowens}.}
\label{fig:kidonakis}
\end{figure}
\begin{figure}[t]
\centering\leavevmode
\vglue1truept\vspace{-3cm}
\hglue1truept\hspace{12cm}
\epsfxsize 9 cm 
\epsfbox{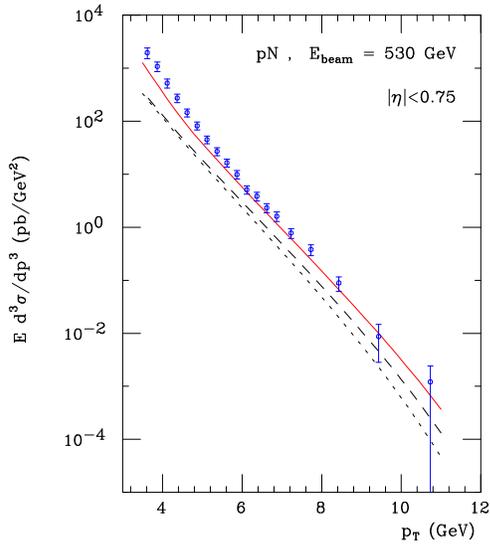}
\vglue1truept\vspace{-1.5cm}
\caption{Direct-photon cross section for the E706 data~\cite{E706-kt}.
The dotted line represents the full NLO calculation, while the dashed
and solid lines respectively incorporate pure threshold
resummation~\cite{nason} and joint threshold and recoil
resummation~\cite{sterman}.}
\label{fig:sterman}
\end{figure}

\begin{figure}[tb]
\centering\leavevmode
\vglue1truept\vspace{-1cm}
\hglue1truept\hspace{0.5cm}
\epsfxsize 7 cm 
\epsfbox{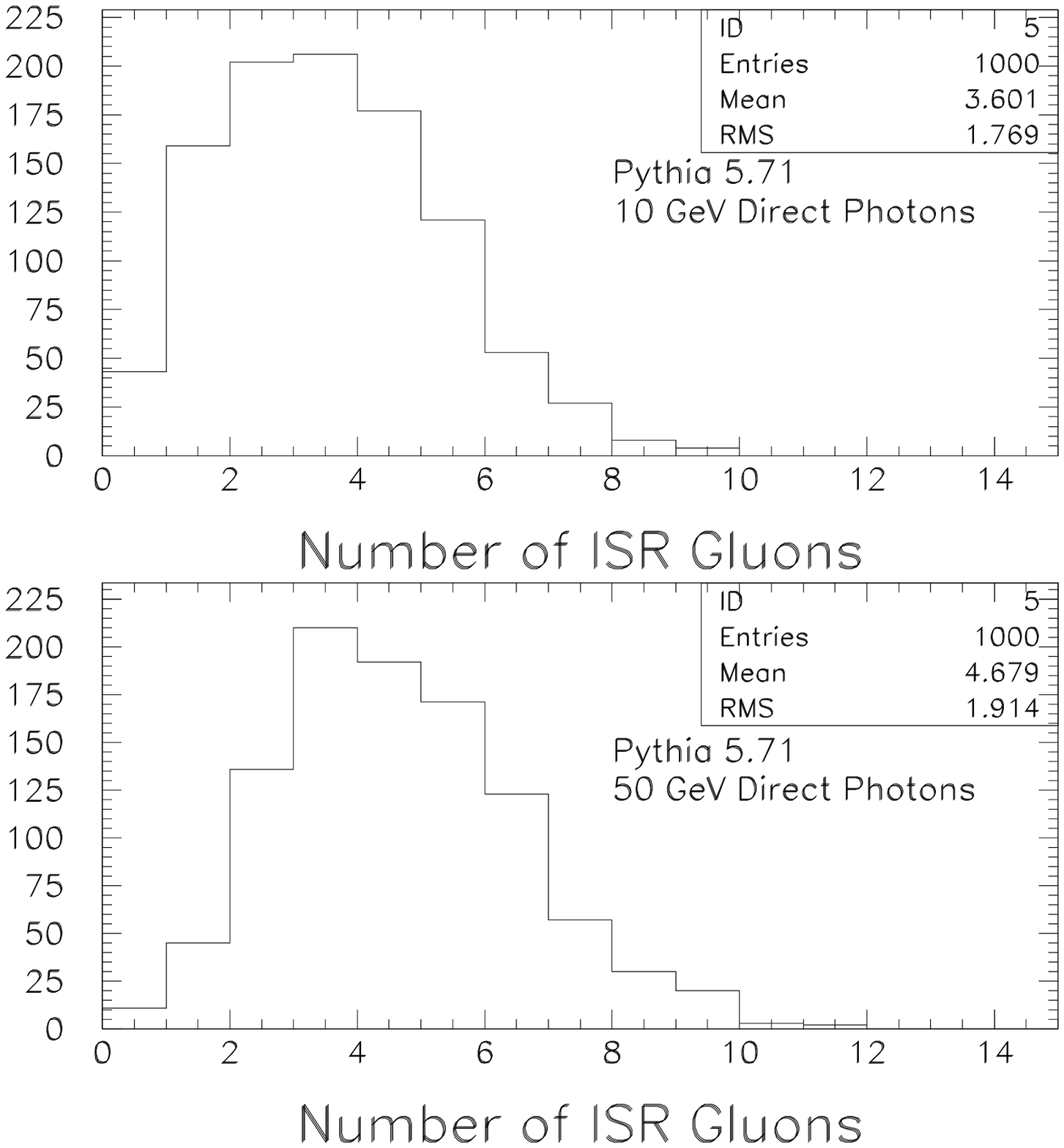}
\vglue1truept\vspace{-0.75cm}
\caption{Number of initial-state gluons in
{\tt PYTHIA} from the direct-photon process at $\sqrt{s}=1.8$~TeV.}
\label{fig:pythia_leading}
\centering\leavevmode
\vglue1truept\vspace{-0.5cm}
\hglue1truept\hspace{0.5cm}
\epsfxsize 7 cm 
\epsfbox{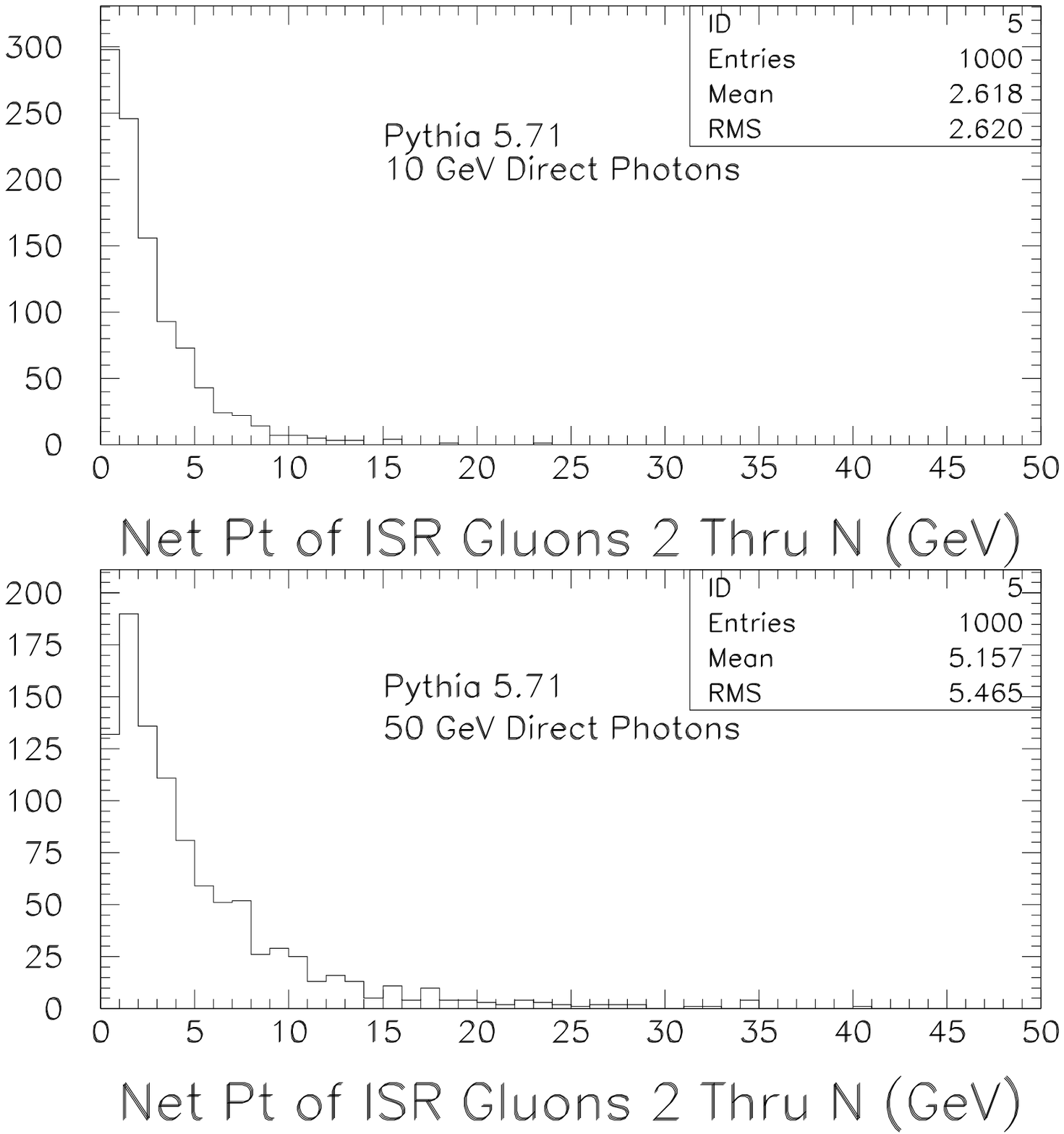}
\vglue1truept\vspace{-0.75cm}
\caption{Net \PT\ due to all but the leading gluon in
{\tt PYTHIA} for direct-photon events.}
\label{fig:pythia_kt}
\end{figure}

Similar soft-gluon contributions are expected to be present in other
hard-scattering processes, such as inclusive production of jets or
direct photons~\cite{contogouris-1981,contogouris-1985,fontannaz,FFF}.
Resummed pQCD calculations for single direct-photon production are
currently under
development~\cite{nason,kidonakisowens,laenen,lilai,li,sterman}.  Two
recent independent threshold-resummed pQCD calculations for direct
photons~\cite{nason,kidonakisowens} do not include \KT\ effects, but
exhibit less dependence on QCD scales than the NLO theory (see
Figs.~\ref{fig:nason} and~\ref{fig:kidonakis}). These
threshold-resummed calculations agree with the NLO prediction for the
scale $\mu\approx\PT/2$ at low \PT, and show an enhancement in cross
section at high~\PT.

A method for simultaneous treatment of recoil and threshold
corrections in inclusive single-photon cross sections has been
developed~\cite{sterman} within the formalism of collinear
factorization.  This approach accounts explicitly for the recoil from
soft radiation in the hard-scattering subprocess, and conserves both
energy and transverse momentum for the resummed radiation. At moderate
\PT, substantial enhancements from higher-order perturbative and
power-law non-perturbative corrections have been found at fixed-target
energies, as illustrated in Fig.~\ref{fig:sterman} in a comparison
with the E706 direct-photon measurement at $\sqrt{s}=31.6$~GeV.
Although the present numerical results are only exploratory estimates
of the size of expected effects, it is already clear that the
phenomenological consequences are significant.

\begin{figure}[t]
\centering\leavevmode
\vglue1truept\vspace{-1.5cm}
\hglue1truept\hspace{0.5cm}
\epsfxsize 7 cm 
\epsfbox{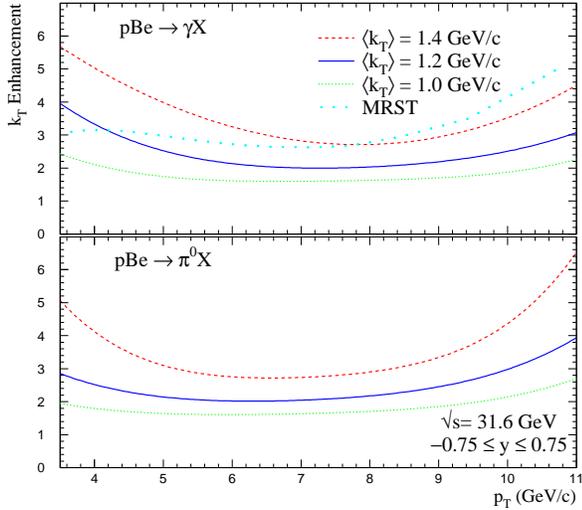}
\vglue1truept\vspace{-2cm}
\caption{The variation of \KT\ enhancements, $K(\PT)$, 
for the E706 $pBe$ data at $\sqrt{s}=31.6$~GeV.}
\label{fig:kfactors}
\end{figure}

Approximate phenomenological approaches to \KT-smearing have been used
in the past to investigate deviations between data and NLO pQCD.  The
underlying hypothesis is that the leading gluon in NLO pQCD
calculations is inadequate for describing the full initial-state
parton shower.  Full parton shower Monte Carlos such as {\tt PYTHIA}
or {\tt HERWIG} have been used to test this idea at collider
energies~\cite{csaba-higgs}.  We used {\tt
PYTHIA}~v5.71~\cite{pythia57} and the direct photon process to extract
the number of initial-state gluons as well as amount of net \KT\
present in initial-state gluons, after subtracting the gluon with the
highest initial state \PT.  The number of initial-state gluons is
shown in Fig.~\ref{fig:pythia_leading}, illustrating that the number
is significantly larger than the NLO pQCD approximation of either~0
or~1.  The net \KT\ present, after subtracting the highest \PT\ gluon,
is shown in Fig.~\ref{fig:pythia_kt} for 10~GeV/$c$ and 50~GeV/$c$
direct photons in $\rm p\bar{\rm p}$ collisions at $\sqrt{s}=1.8$~TeV.
The net \PT\ of such remnant gluons is 2.6~GeV/$c$ for direct photons
with $\PT=10$~GeV/$c$, and 5.2~GeV/$c$ for $\PT=50$~GeV/$c$.

At fixed-target energies, parton-showering models do not provide
sufficient smearing because shower development is constrained by
cut-off parameters that ensure the perturbative nature of the process.
Since traditional NLO calculations do not account for the effects of
multiple soft-gluon emission, a kinematical model was employed to
incorporate \KT\ effects in available pQCD calculations of
direct-photon (and~$\pi^0$) production~\cite{ktprd}.  The relationship
between this phenomenological \KT-smearing and the
Collins--Soper--Sterman (CSS) resummation formalism~\cite{CSS,CSS1} was
considered in some detail in Ref.~\cite{csaba-thesis}.

The same LO pQCD~\cite{owens} program that successfully characterized
high-mass pair production was used to generate K-factors (ratios of LO
calculations for any given \avkt\ to the result for \KT=0) for
inclusive cross sections (Fig.~\ref{fig:kfactors}).  These K-factors
were then applied to the NLO calculations.  This procedure involves a
risk of double-counting since some of the \KT-enhancements may already
be contained in the NLO calculation.  However, the effects of such
double-counting are expected to be small~\cite{ktprd}.

As illustrated in the upper part of Fig.~\ref{fig:kfactors}, the
K-factors for direct-photon production at E706 are large over the full
range of \PT, and have \PT-dependent shapes---a behavior reminiscent
of that obtained from the full resummation formalism~\cite{sterman}.
The lower part of Fig.~\ref{fig:kfactors} displays K-factors for
$\pi^0$ production, based on the same model. The data appear to
require somewhat larger values of \avkt\ in the case of $\pi^0$s.  At
the same values of \avkt\ the K-factors in $\pi^0$ production are
somewhat smaller than for $\gamma$ production.  This is expected
because $\pi^0$'s originate from the fragmentation of the partons.

Figure~\ref{fig:kfactors} also displays the K-factor for photons used
by the MRST group~\cite{mrst} in recent fits to parton distributions.
Their result was obtained through a different technique involving
analytical smearing of the parameterized photon cross section, rather
than an explicit parton-level calculation.  Although the correction is
of similar size, it has a different \PT-dependence.  It should be
noted, however, that despite the similarity in the values of the
K-factors used by the MRST group and the ones presented here, the
\avkt\ values cited by MRST are lower by a factor of~2.  This
difference can be traced to an erroneous relation between parameters
of their analytical smearing function and the transverse momenta of
partons in the hard scatter~\cite{mrst}.  This has already been
pointed out in Refs.~\cite{ktprd,kimber}. The approach used
here~\cite{ktprd} is based on explicit parton kinematics and therefore
does not suffer from this problem.

The treatment of \KT-enhancements proposed in Ref.~\cite{kimber},
based on parton distributions unintegrated over the parton transverse
momenta, suggests possible modifications of the above simple picture.
Reference~\cite{kimber} imposes strong ordering of momentum transfers
of emitted gluons, which prevents transverse momenta of the incoming
partonic system from approaching \PT; \KT\ values are correlated with
the scale at which the parton distributions are sampled.  In their
approach, the K-factors are expected to be smaller than those shown in
Fig.~\ref{fig:kfactors}, and have less \PT-dependence.  Additional
scrutiny of the theoretical ideas should help resolve these
differences.

\subsubsection{High-\PT\ Production}

\paragraph{Fixed Target}

\begin{figure}[t]
\centering\leavevmode
\vglue1truept\vspace{-1cm}
\hglue1truept\hspace{0.5cm}
\epsfxsize 7 cm 
\epsfbox{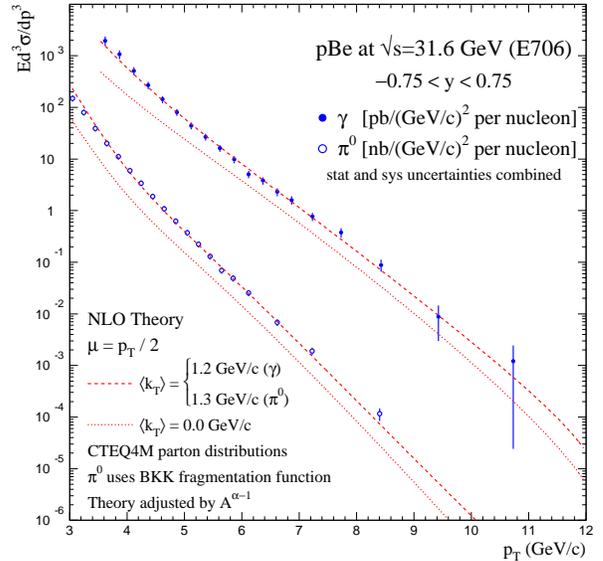}
\vglue1truept\vspace{-1.75cm}
\caption{The photon and $\pi^0$ cross sections for E706 $pBe$ data
at $\sqrt s = 31.6$~GeV~\cite{E706-kt} compared to the \KT-enhanced
NLO calculations.}
\label{fig:530gampi}
\end{figure}

\begin{figure}[p]
\centering\leavevmode
\vglue1truept\vspace{-1cm}
\hglue1truept\hspace{0.5cm}
\epsfxsize 7 cm 
\epsfbox{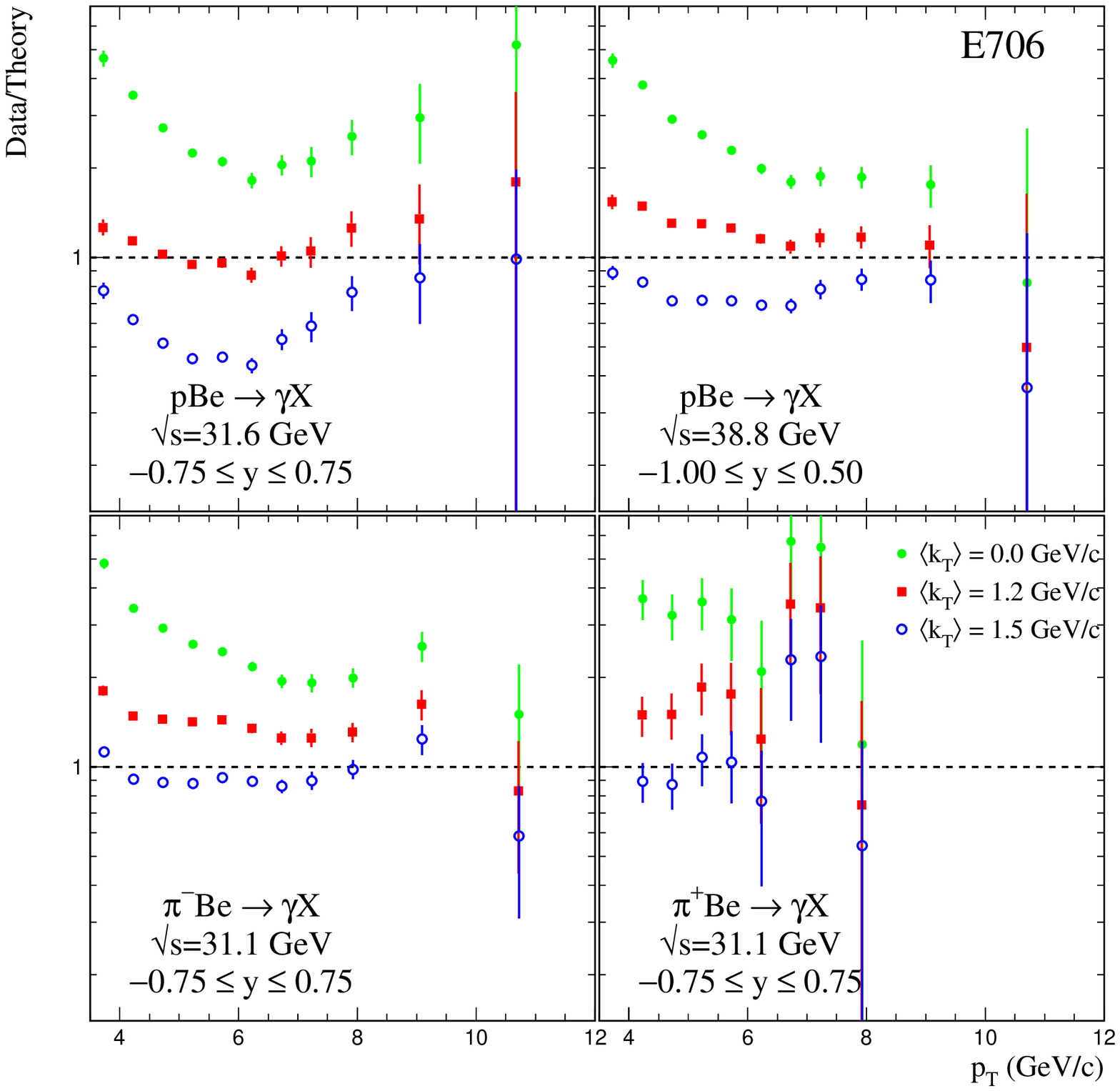}
\vglue1truept\vspace{-1.75cm}
\caption{Comparison between the E706 direct-photon data and 
NLO pQCD calculations with and without
\KT\ enhancements, for several values of \avkt.}
\label{fig:E706dp}
\vspace{1cm}
\hglue1truept\hspace{0.5cm}
\epsfxsize 7 cm 
\epsfbox{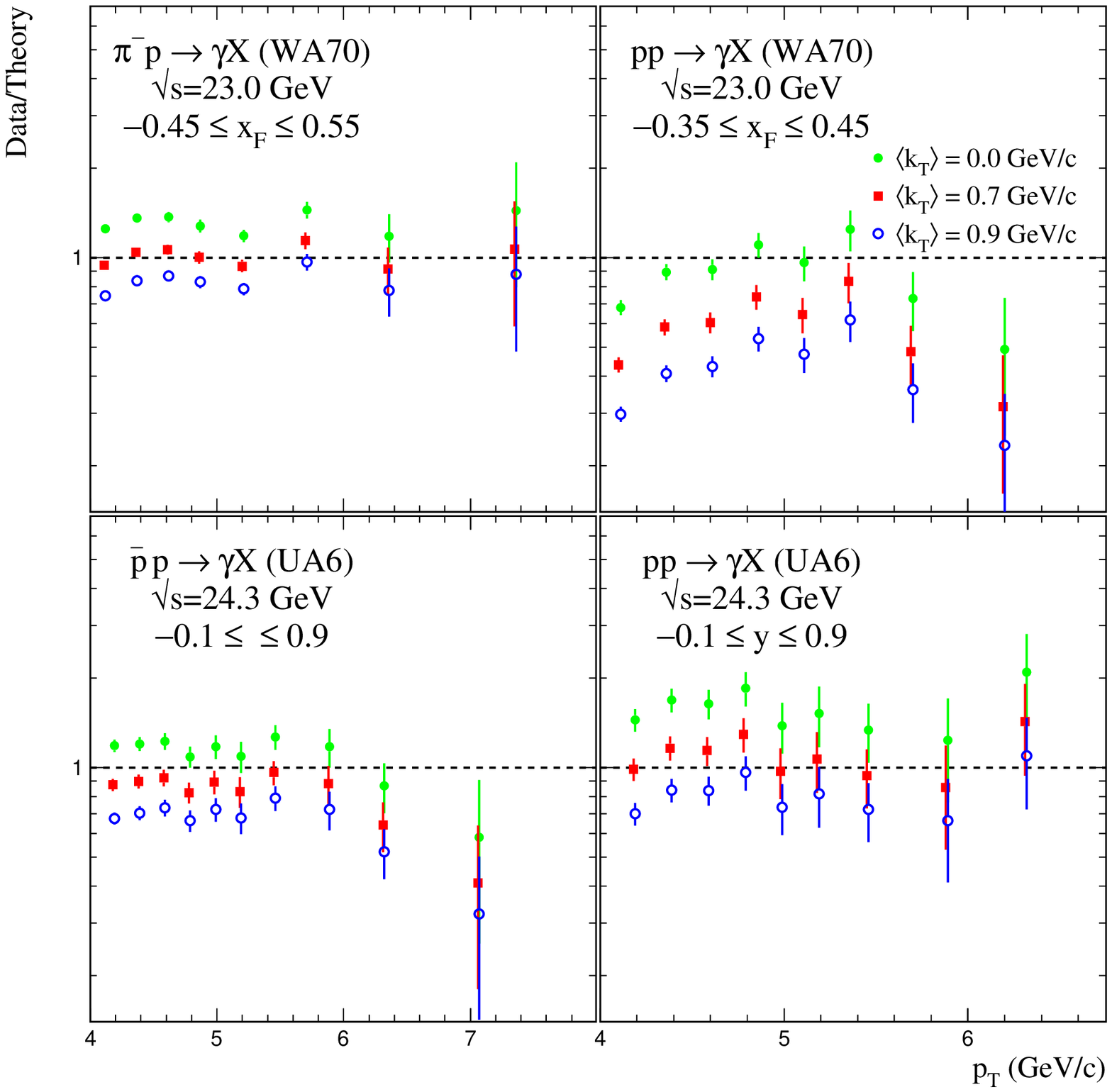}
\vglue1truept\vspace{-1.75cm}
\caption{Comparison between the WA70 and UA6 direct-photon data and 
NLO pQCD calculations with and without
\KT\ enhancements, for several values of \avkt.}
\label{fig:UA6WA70dp}
\vspace{1.5cm}
\end{figure}

Invariant cross sections for inclusive direct-photon and $\pi^0$
production are displayed for the E706 $pBe$ data at $\sqrt{s}=31.6$~GeV
in Fig.~\ref{fig:530gampi}, with overlays from theory~\cite{E706-kt}.
Discrepancies between NLO pQCD theory (dotted curves) and the data are
striking.  The enhancements, generated using \avkt\ values consistent
with the data on high-mass pairs (Fig.~\ref{fig:pout_all}), can
accommodate both the shapes and normalizations of direct-photon and
$\pi^0$ inclusive cross sections.

Comparisons between direct-photon data from E706~\cite{E706-kt},
WA70~\cite{WA70-pi-dp,WA70-p-dp}, and UA6~\cite{UA6-newdp} and
\KT-enhanced NLO pQCD~\cite{ktprd} are shown in Figs.~\ref{fig:E706dp}
and~\ref{fig:UA6WA70dp}.  The values of \avkt\ were based on the data
for high-mass pairs from E706 (Fig.~\ref{fig:pout_all}), and the
di-photon data from WA70~\cite{WA70-ddp,WA70-kt}
($\avkt=0.9\pm0.1\pm0.2$~GeV/$c$).  The center-of-mass energies for
WA70 and UA6 ($\sqrt{s}\approx24$~GeV) are smaller than those for
E706.  Correspondingly, \avkt\ values for these experiments are
expected to be slightly smaller than the values required for E706
(Fig.~\ref{fig:DYkt}).

\begin{figure}[t]
\centering\leavevmode
\vglue1truept\vspace{-1cm}
\hglue1truept\hspace{0.5cm}
\epsfxsize 7 cm 
\epsfbox{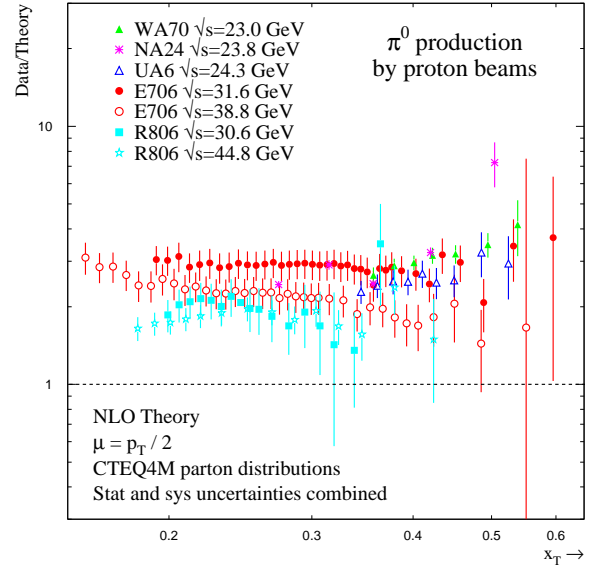}
\vglue1truept\vspace{-1.75cm}
\caption{Comparison between proton-induced $\pi^0$ data and NLO pQCD
calculations for several experiments as a function of $x_T$.}
\label{fig:worldpi0}
\end{figure}

\begin{figure}[p]
\centering\leavevmode
\vglue1truept\vspace{-3cm}
\hglue1truept\hspace{0.5cm}
\epsfxsize 7 cm 
\epsfbox{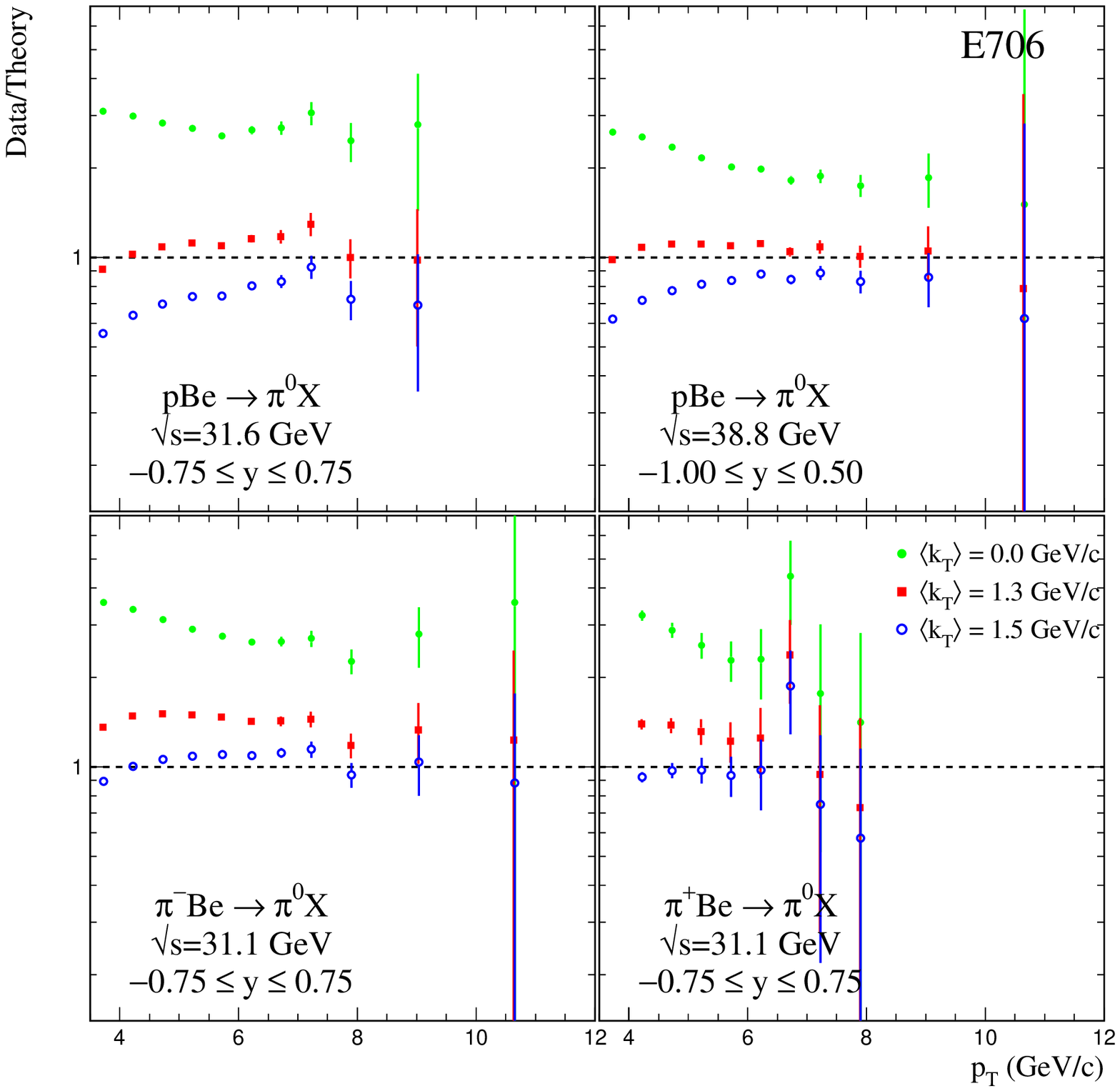}
\vglue1truept\vspace{-1.75cm}
\caption{Comparison between the E706 $\pi^0$ data and 
NLO pQCD calculations with and without
\KT\ enhancements, for several values of \avkt.}
\label{fig:E706pi}
\vspace{1cm}
\hglue1truept\hspace{0.5cm}
\epsfxsize 7 cm 
\epsfbox{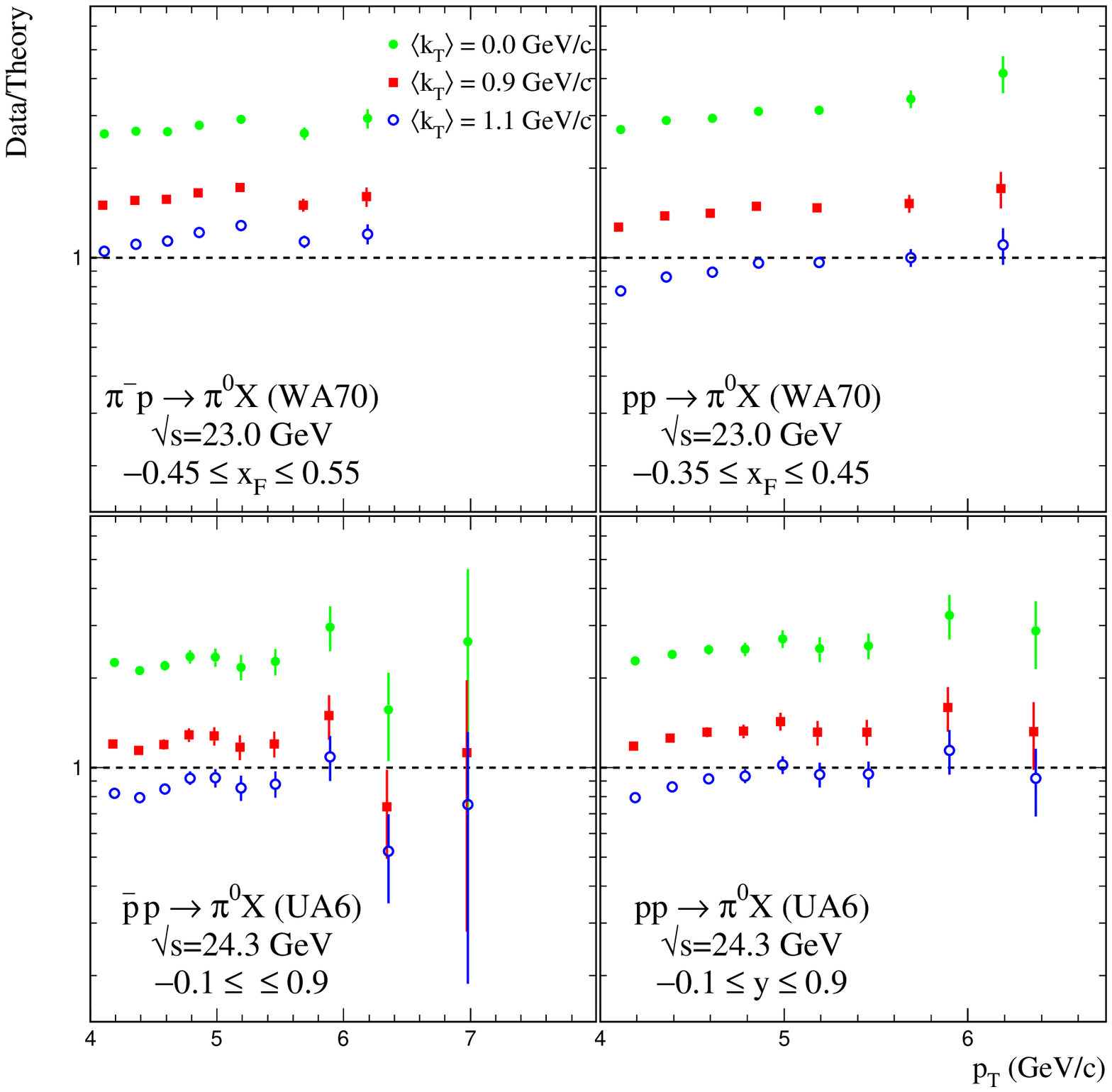}
\vglue1truept\vspace{-1.75cm}
\caption{Comparison between the WA70 and UA6 $\pi^0$ data and 
an NLO pQCD calculation with and without
\KT\ enhancements, for several values of \avkt.}
\label{fig:UA6WA70pi}
\end{figure}

A recent survey of $\pi^0$ production found that current NLO pQCD
calculations significantly undershoot the data~\cite{aurenche-pi0}.  A
comparison between $\pi^0$ data and NLO pQCD is shown in
Fig.~\ref{fig:worldpi0} for several experiments~\cite{apana-thesis}.
The data are consistently a factor of~2 to~3 above theory.  The above
phenomenological model should also be valid for pion production.
Using \avkt\ similar to, but slightly higher than that for direct
photons, good agreement is obtained for $\pi^0$'s measured by
E706~\cite{E706-kt}, WA70~\cite{WA70-pi0}, and UA6~\cite{UA6-pi-eta}
(Figs.~\ref{fig:E706pi} and~\ref{fig:UA6WA70pi}).  The \KT-enhanced
predictions compare well with the $\pi^0$~cross sections, with all
the E706 and UA6 direct-photon data, and with the $\pi^-$~beam
direct-photon cross sections of WA70.

\begin{figure}[t]
\centering\leavevmode
\vglue1truept\vspace{-1cm}
\hglue1truept\hspace{0.5cm}
\epsfxsize 7 cm 
\epsfbox{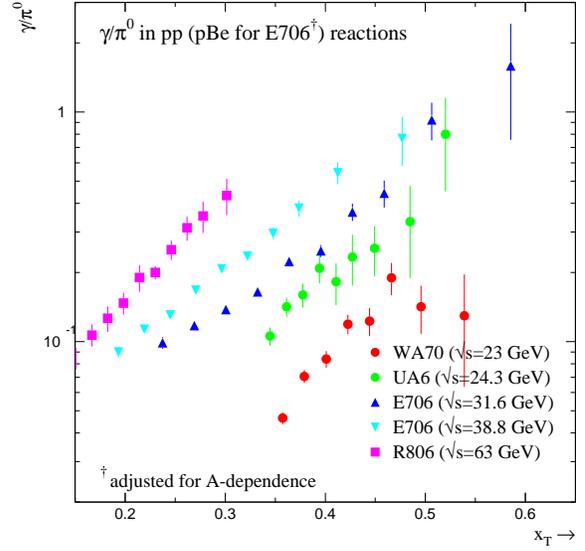}
\vglue1truept\vspace{-1.75cm}
\caption{Ratios of direct-photon cross sections 
to the $\pi^0$ cross sections as a function of \PT\ for various
experiments at several values of $\sqrt{s}$.}
\label{fig:photonTOpi0}
\end{figure}

\begin{figure}[p]
\centering\leavevmode
\vglue1truept\vspace{-3cm}
\hglue1truept\hspace{0.5cm}
\epsfxsize 7 cm 
\epsfbox{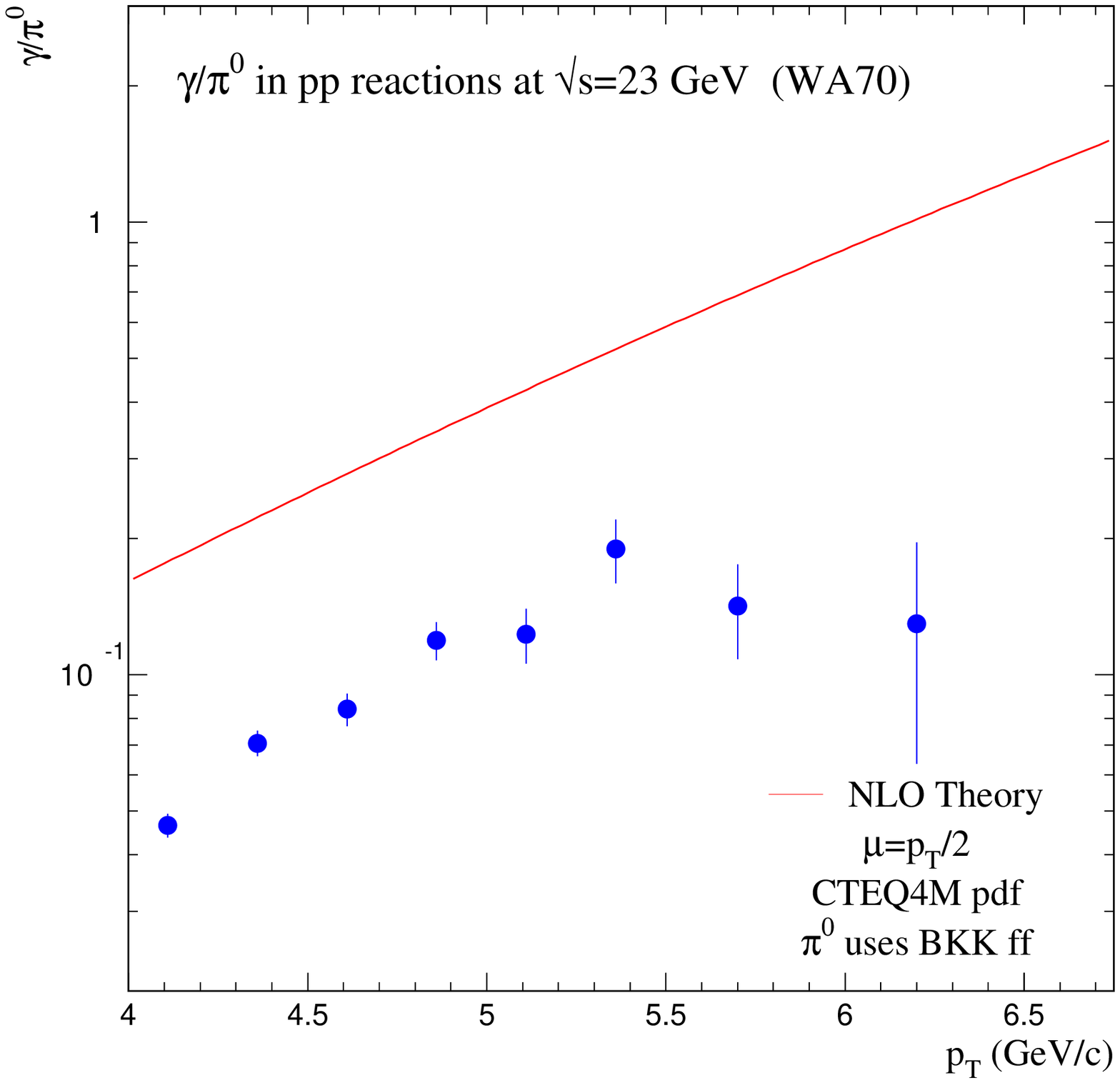}
\vglue1truept\vspace{-1.75cm}
\caption{$\gamma/\pi^0$ comparison for the WA70 $pp$ data 
at $\sqrt{s}=23$~GeV.  Overlayed are the results from the NLO pQCD
calculations.}
\label{fig:WA70photonTOpi0}
\vspace{1cm}
\hglue1truept\hspace{0.5cm}
\epsfxsize 7 cm 
\epsfbox{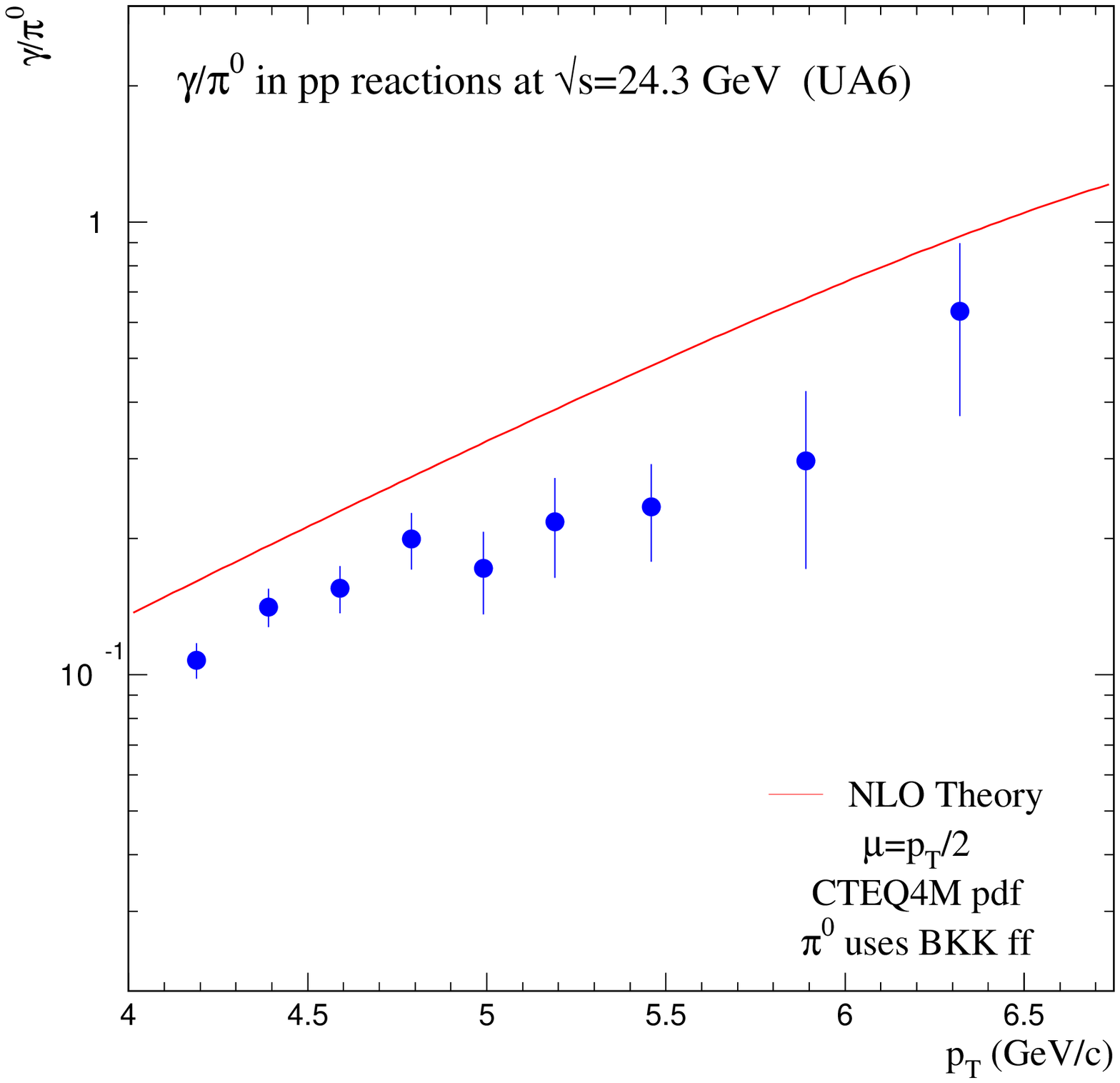}
\vglue1truept\vspace{-1.75cm}
\caption{$\gamma/\pi^0$ comparison for the UA6 $pp$ data 
at $\sqrt{s}=24.3$~GeV.  Overlayed are the results from the NLO pQCD
calculations.}
\label{fig:UA6photonTOpi0}
\end{figure}

\begin{figure}[p]
\centering\leavevmode
\vglue1truept\vspace{-3cm}
\hglue1truept\hspace{0.5cm}
\epsfxsize 7 cm 
\epsfbox{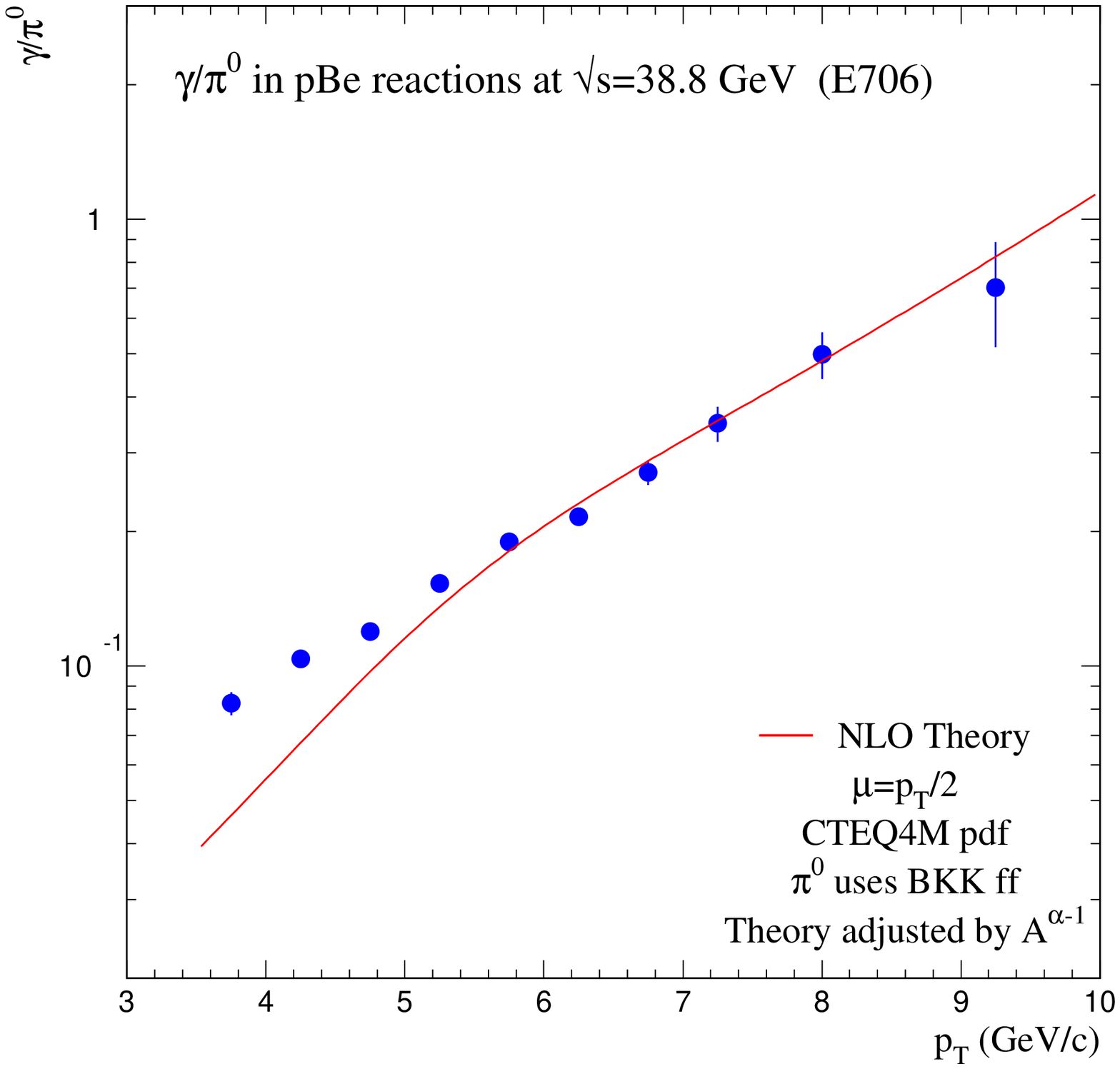}
\vglue1truept\vspace{-1.75cm}
\caption{$\gamma/\pi^0$ comparison for the E706 $pBe$ data 
at $\sqrt{s}=38.8$~GeV.  Overlayed are the results from the NLO pQCD
calculations.}
\label{fig:E706photonTOpi0}
\vspace{1cm}
\hglue1truept\hspace{0.5cm}
\epsfxsize 7 cm 
\epsfbox{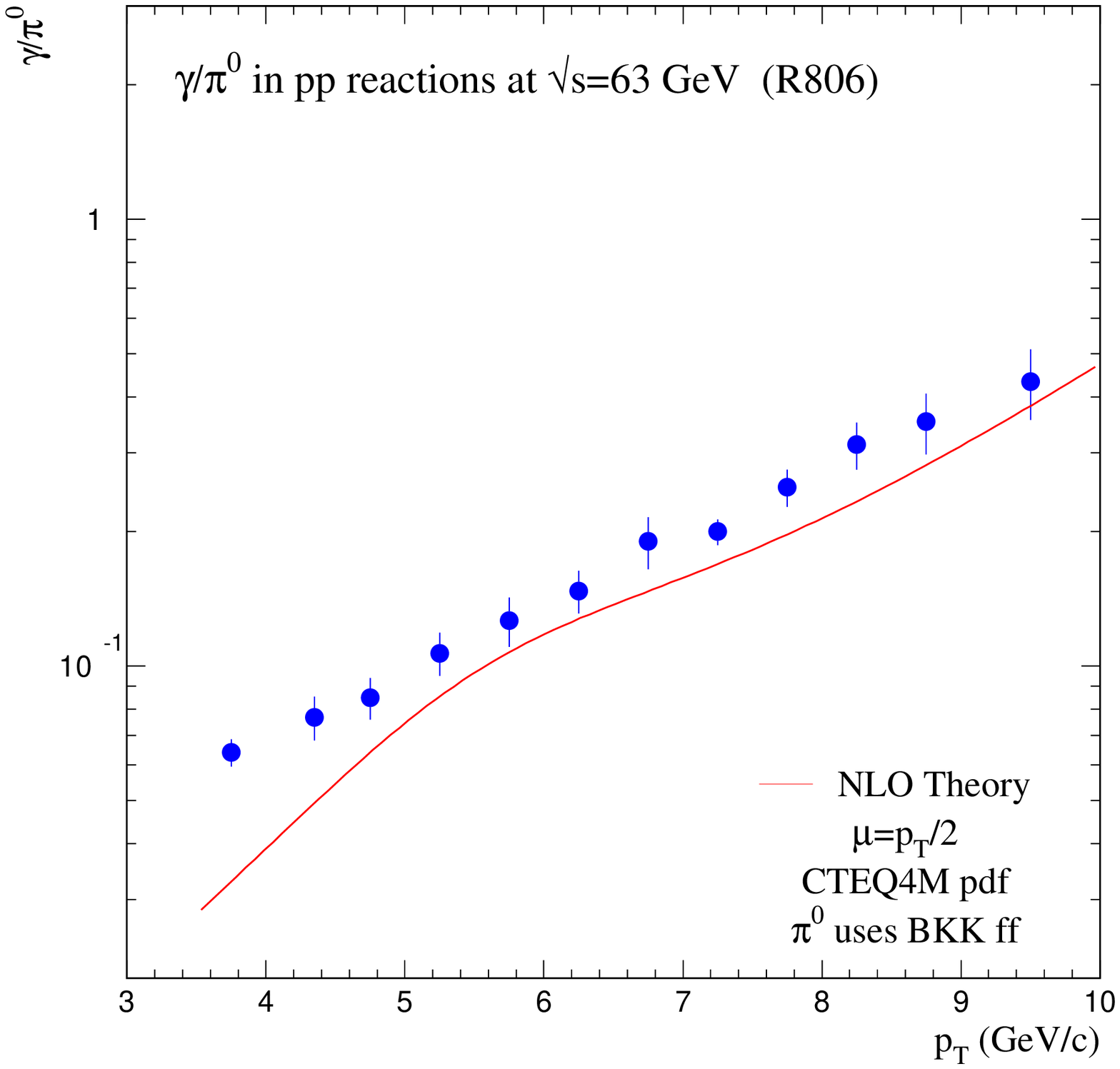}
\vglue1truept\vspace{-1.75cm}
\caption{$\gamma/\pi^0$ comparison for the R806 $pp$ data 
at $\sqrt{s}=63$~GeV.  Overlayed are the results from the NLO pQCD
calculations.}
\label{fig:R806photonTOpi0}
\end{figure}

Since \KT-smearing affects similarly direct-photon and $\pi^0$ data,
the ratio of direct-photon to $\pi^0$ production should be relatively
insensitive to \KT.  Experimental and theoretical uncertainties also
tend to cancel in such ratios.  Figure~\ref{fig:photonTOpi0} shows the
ratios of cross-sections~\cite{apana-thesis} for direct-photon to
$\pi^0$ production for WA70, UA6, E706, and
R806~\cite{R806-total-dp,R806-pi0,R806-ratio}.  The results from
WA70~(Fig.~\ref{fig:WA70photonTOpi0}) and
UA6~(Fig.~\ref{fig:UA6photonTOpi0}), at approximately the same
$\sqrt{s}$, appear to differ significantly.  The ratio for NLO theory
differs from that of WA70 by a factor of three, but only by
$\approx30$\% from the UA6 data.  The WA70 and UA6 $\pi^0$ results
agree (Fig.~\ref{fig:worldpi0}) and most of the difference is
therefore in the direct-photon cross section.  Similar $\gamma/\pi^0$
comparisons are shown for E706 at
$\sqrt{s}=38.8$~GeV~(Fig.~\ref{fig:E706photonTOpi0}) and R806 at
$\sqrt{s}=63$~GeV~(Fig.~\ref{fig:R806photonTOpi0}).  The same 30\%
level of agreement can also be found for the E706 data at
$\sqrt{s}=31.6$~GeV and the R806 data at $\sqrt{s}=31$~GeV and $45$~GeV.

\begin{figure}[p]
\centering\leavevmode
\vglue1truept\vspace{-5cm}
\hglue1truept\hspace{-0.5cm}
\epsfxsize 9 cm 
\epsfbox{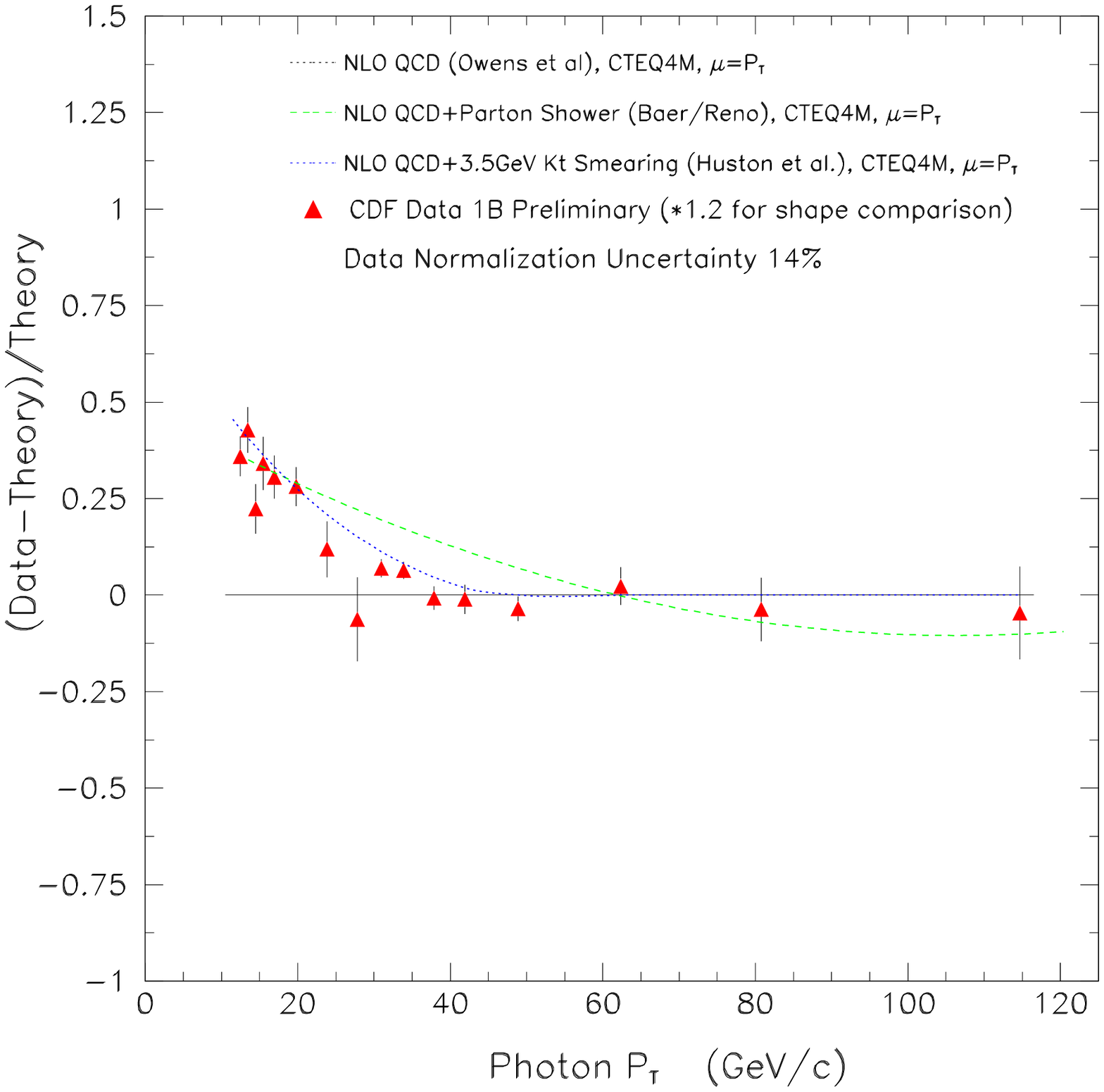}
\vglue1truept\vspace{-0.75cm}
\caption{A comparison of the CDF isolated direct-photon data 
at $\sqrt{s}=1.8$~TeV from Run~Ib with a NLO pQCD prediction and two
implementations of soft gluon corrections to the NLO
prediction~\cite{newCDF}.}
\label{fig:newCDF}
\vspace{1cm}
\vglue1truept\vspace{-1cm}
\hglue1truept\hspace{-0.25cm}
\epsfxsize 8 cm 
\epsfbox{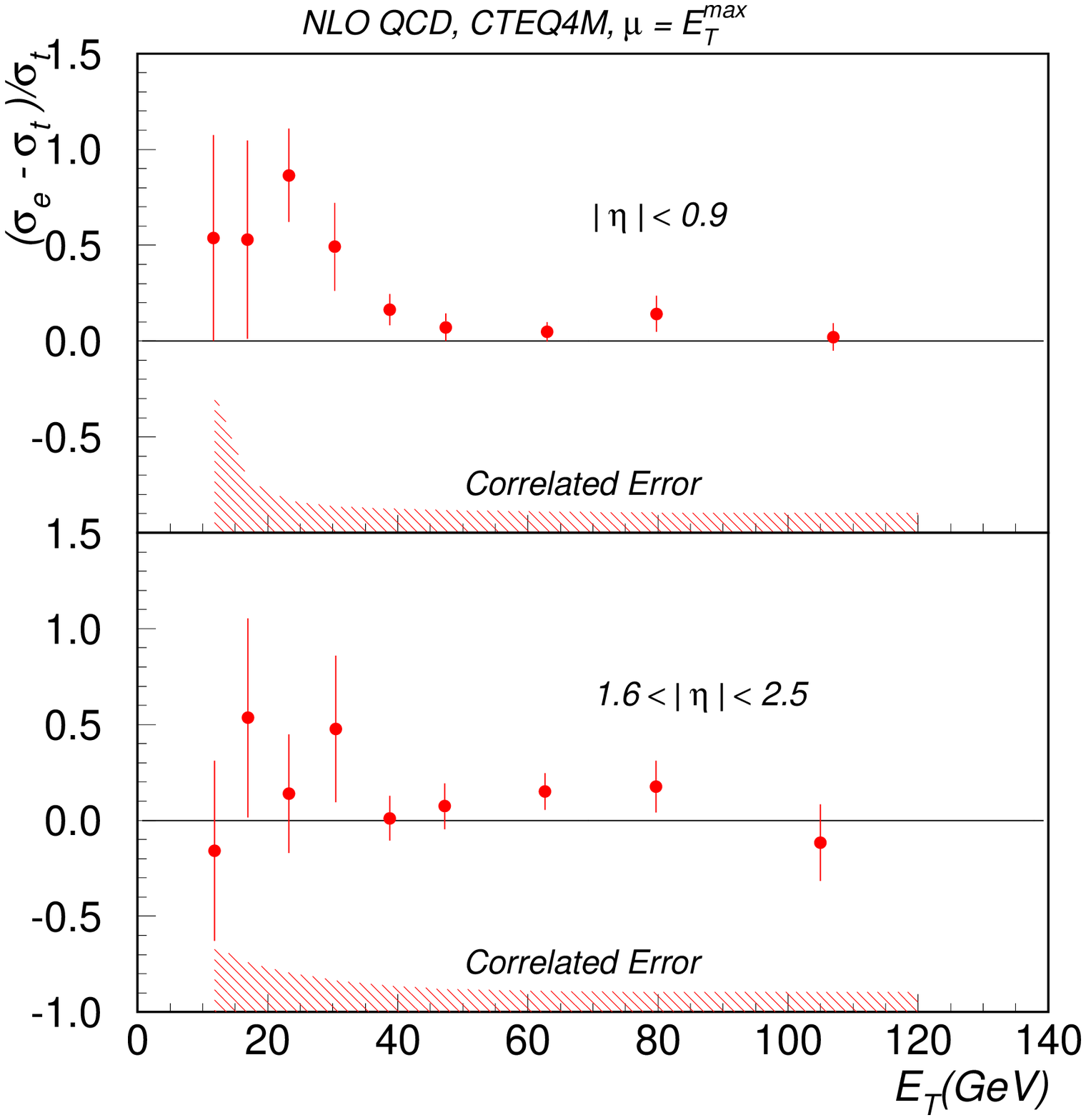}
\vglue1truept\vspace{-2cm}
\caption{A comparison of the \DO\ isolated direct-photon data 
at $\sqrt{s}=1.8$~TeV from Run~Ib with a NLO pQCD calculation in both
the central and forward rapidity regions~\cite{newD0}.}
\label{fig:newD0}
\end{figure}

\begin{figure}[tb]
\centering\leavevmode
\vglue1truept\vspace{-1cm}
\hglue1truept\hspace{0.5cm}
\epsfxsize 7 cm 
\epsfbox{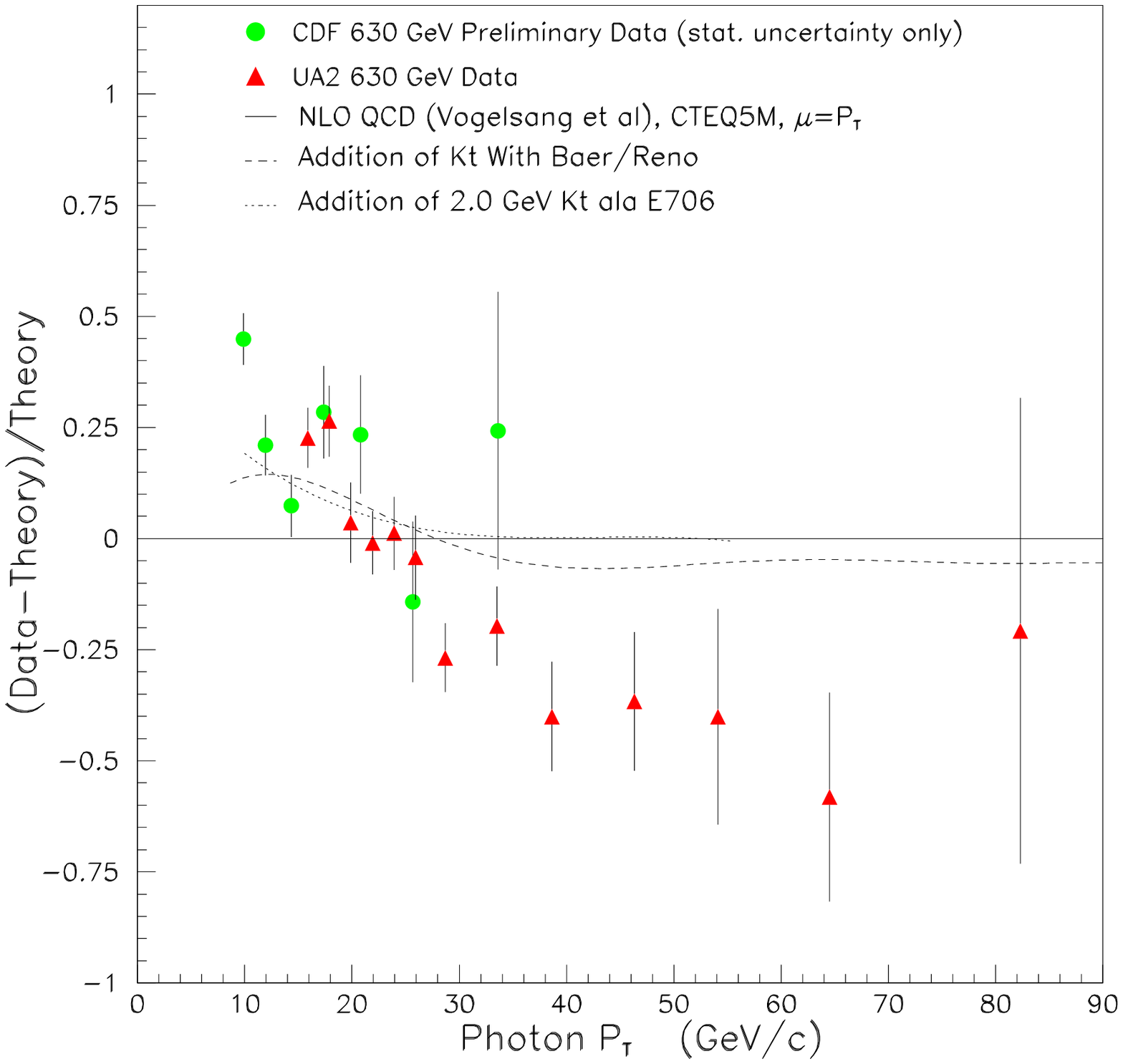}
\vglue1truept\vspace{-1.75cm}
\caption{A comparison between CDF and UA2 data at $\sqrt{s}=630$~GeV.}
\label{fig:CDFUA2630}
\end{figure}

The discussion of direct-photon data from fixed-target and ISR
experiments~\cite{aurenche-dp}, while rejecting the \KT\
interpretation for the observed deviations, pointed to limitations for
the applicability of NLO calculations at lower-\PT values and to
inconsistencies among experiments used in comparisons to NLO
theory. Given the shortcomings of a purely NLO description, a global
comparison of the available data to fully resummed theory may provide
useful insights into these issues.

\paragraph{Run~I Results}

The consequences of \KT\ smearing are expected to depend on
$\sqrt{s}$~(Fig.~\ref{fig:DYkt}).  At the Tevatron collider, where
\PT\ is large compared to \KT, the above model of soft-gluon radiation
leads to a relatively small modification of the NLO cross section.
Only the lowest end of the \PT\ spectrum is modified significantly,
and the K-factor exhibits the expected $\sim 1/\PT^2$ behavior for a
power correction.

A comparison of the results from Run~Ib, shown in
Figs.~\ref{fig:newCDF} and~\ref{fig:newD0}~\cite{newCDF,newD0},
confirms the expected deviation in shape at low \PT.  Using di-photons,
CDF has measured $\avkt=3.6\pm0.8$~GeV/$c$ at
$\sqrt{s}=1.8$~TeV~\cite{CDF-ddp}.  Employing this value, the
phenomenological model adequately describes the shape of the data in
Fig.~\ref{fig:newCDF}.  Also shown in this figure is an implementation
of soft-gluon corrections using an enhanced parton
shower~\cite{baerreno}.  The phenomenological-\KT\ model provides a
better agreement with data than is available in the enhanced parton
shower model.  The agreement between the phenomenological model
implementation of \KT\ smearing and the direct-photon data can also be
seen in preliminary results from CDF at
$\sqrt{s}=630$~GeV~(Fig.~\ref{fig:CDFUA2630}).

The CDF data in Fig.~\ref{fig:newCDF} have been normalized upwards by a
factor of~1.2 for the benefit of a shape comparison.  Without this
normalization, then the CDF data lie below the NLO pQCD calculation at
high~\PT.  Figure~\ref{fig:CDFUA2630} also contains a comparison
between CDF and UA2~\cite{UA2-dp88,UA2-dp-ddp} data at
$\sqrt{s}=630$~GeV, where a similar deficit is observed at high~\PT\
for the UA2 data.  There is currently no explanation for this effect.

\subsubsection{Expectations for Run~II\label{s:ph12}}

\begin{figure}[tb]
\centering\leavevmode
\vglue1truept\vspace{-1cm}
\hglue1truept\hspace{0.5cm}
\epsfxsize 7 cm 
\epsfbox{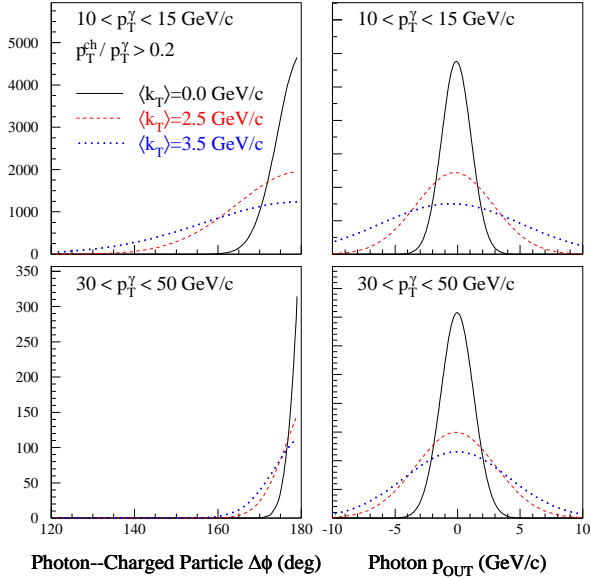}
\vglue1truept\vspace{-1.75cm}
\caption{Left: Distributions of azimuthal angle difference 
$\Delta\phi$ between the direct photon and a recoil charged track for
several values of \KT\ and for two photon \PT\ ranges.  Right: Similar
dependences for the photon $p_{OUT}$.  
}
\label{fig:gammatrack}
\end{figure}

\paragraph{\KT\ Studies at Low-\PT}

To study the level of \KT\ induced by multiple gluon emissions at the
collider, the experiments have employed the relatively low-statistics
di-photon data.  In Run~II, both CDF and \DO\ will have precision
magnetic tracking, which will permit studies of \KT\ effects using
``two-arm'' data on pairs consisting of a direct photon and a recoil
charged track, in a spirit similar to that of the fixed-target
investigations of $\gamma\pi^0$ and $\pi^0\pi^0$ pairs discussed
before.  The advantage of this approach is that it will obviate the
need for jet reconstruction (difficult at low \PT), and minimize
complications from jet energy scale calibration.  Of course,
photon+jet systems are also of interest, but may be harder to study in
the range of interest for checking the effects of \KT\ smearing.

We simulated the expected behavior of photon--track systems using the
same LO Monte Carlo~\cite{owens} employed in previous fixed-target
studies.  The results are illustrated in Fig.~\ref{fig:gammatrack} for
$\Delta\phi$ and $p_{OUT}$ of the photon, for two representative
ranges of photon \PT\ that span the region where the \KT\ effects
appear to be important in the inclusive photon cross sections from
Run~I. (For the measurement of $p_{OUT}$ for photon, the scattering
plane is defined by the colliding beams and the recoil track.) The
sensitivity to the value of \avkt\ in the range of~0 to~3.5~GeV/$c$ is
clearly seen for both variables. The $\Delta\phi$ distributions become
narrower with increasing photon \PT, as expected from simple kinematic
arguments.  For a fixed \avkt, the width of the $p_{OUT}$
distributions is relatively insensitive to \PT, and can therefore be
particularly useful for mapping out the dependence of \KT\ on event
kinematics (especially on the \PT\ of the photon).  To properly
interpret the widths of such distributions in terms of \KT\ induced by
gluon-radiation, it is important to subtract the amount generated in
the fragmentation of partons into the charged particles (as was done
in the fixed-target analyses).

Additional handles on interpreting the data can be obtained through
studies of \PT\ distributions of charged particles from fragmentation
of partons recoiling against direct-photon triggers. In the presence
of significant initial-state \KT, such distributions are expected to
become softer than expected from standard fragmentation functions
(determined {\it e.g.} from $e^+e^-$ data), since the \KT-kick tends to
increase the photon \PT\ while taking it away from the recoil
side. This should be observed most clearly at low photon \PT's, where
the effect from \KT\ is greatest.

\paragraph{Photon Purity}

\begin{figure}[tb]
\centering\leavevmode
\vglue1truept\vspace{-1cm}
\hglue1truept\hspace{0.5cm}
\epsfxsize 7 cm 
\epsfbox{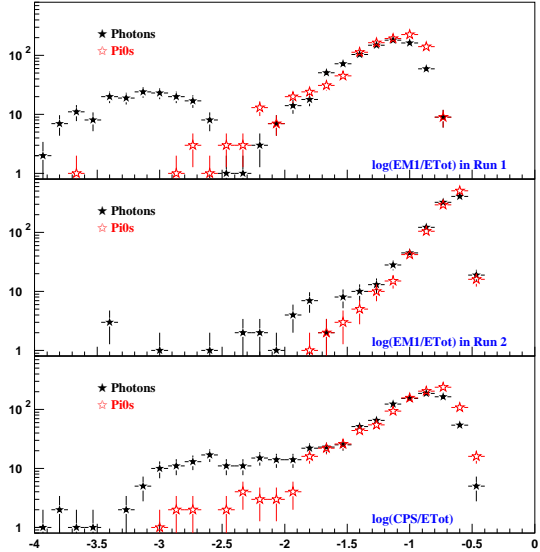}
\vglue1truept\vspace{-1.75cm}
\caption{Comparisons of simulated discriminations between photons 
(solid) and $\pi^0$'s (open) for the \DO\ detector in the Run~I
configuration (top, discriminant variable is $\log(E_{\rm EM1}/E_{\rm
TOT}$); for the same discriminant in the Run~II configuration
(middle); and for the Run~II configuration and $\log(E_{\rm
CPS}/E_{\rm TOT})$ discriminant (bottom).}
\label{fig:proc1}
\end{figure}

\begin{figure}[tb]
\centering\leavevmode
\vglue1truept\vspace{-1cm}
\hglue1truept\hspace{0.5cm}
\epsfxsize 7 cm 
\epsfbox{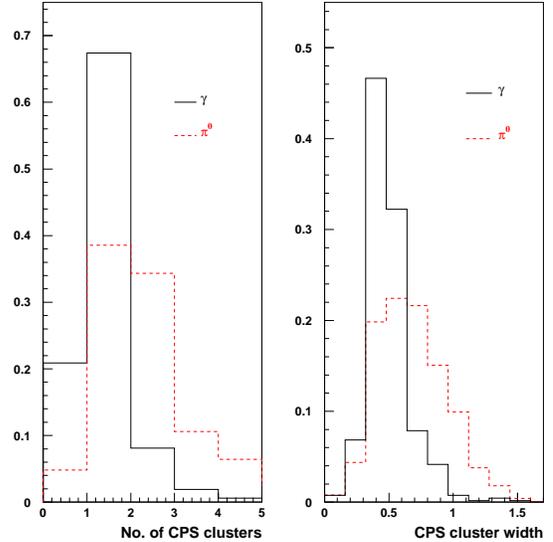}
\vglue1truept\vspace{-1.75cm}
\caption{Additional $\gamma$ {\it vs} $\pi^0$ discrimination 
with the \DO\ detector in Run~II is possible using differences in the
respective distributions for the number of reconstructed preshower
clusters matched with the calorimeter shower (left) or for the width
of the preshower cluster (right).  Solid lines are for $\gamma$'s,
dashed for $\pi^0$'s.}
\label{fig:proc2}
\end{figure}

\begin{figure}[tb]
\centering\leavevmode
\vglue1truept\vspace{-1cm}
\hglue1truept\hspace{0.5cm}
\epsfxsize 7 cm 
\epsfbox{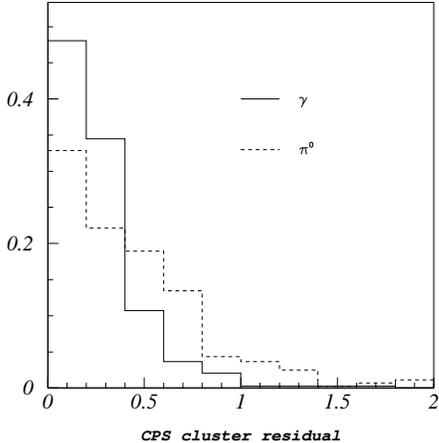}
\vglue1truept\vspace{-1cm}
\caption{The residual of CPS cluster position relative to the photon or
$\pi^0$ in the R-$\phi$ plane in the \DO\ detector. Clusters from
$\pi^0$'s have larger residuals, especially when one of the decay
photons does not convert in the solenoid.}
\label{fig:cpsresidual}
\end{figure}

\begin{figure}[tb]
\centering\leavevmode
\vglue1truept\vspace{-1cm}
\hglue1truept\hspace{0.5cm}
\epsfxsize 7 cm 
\epsfbox{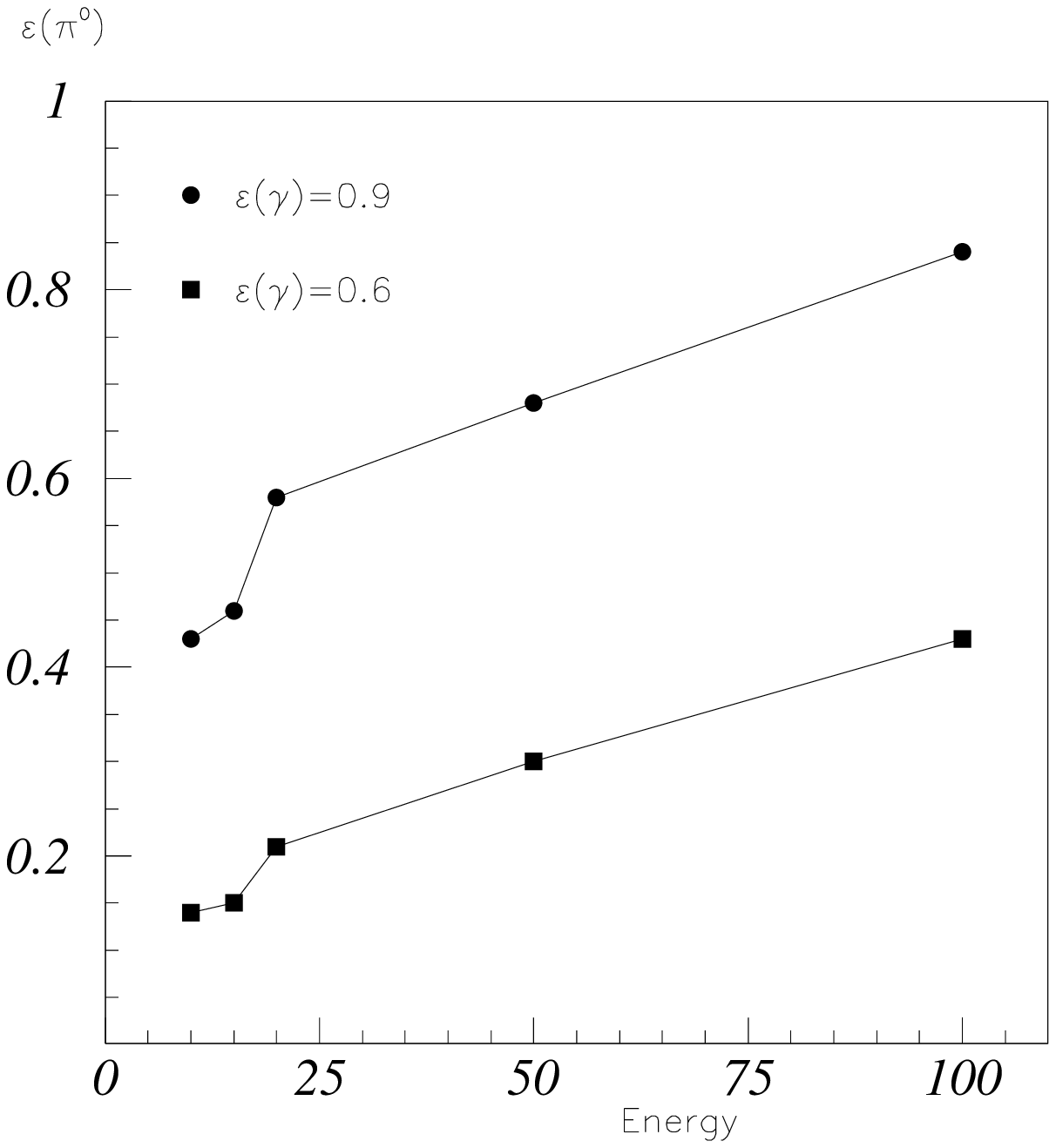}
\vglue1truept\vspace{-1cm}
\caption{Efficiency for a $\pi^0$ to pass covariance matrix $\chi^2$ cuts
corresponding to photon efficiencies of 90\% and 60\%.  (Preliminary
study for the \DO\ detector in Run~II.)}
\label{fig:discrimination}
\end{figure}

The particularly interesting connection between two-arm studies and
measurements of the inclusive photon cross section is mainly at
relatively low \PT\ (10~GeV/$c$ to~35~GeV/$c$), where Run~I results indicate
significant deviation of the cross section from expectations from NLO
pQCD (Figs.~\ref{fig:newCDF} and~\ref{fig:newD0}).  This is also the
region of high statistics, and, consequently, where detailed studies
will be possible. It is therefore important to achieve a high
photon-signal purity in this region, which has proved to be difficult
in Run~I.

In the case of \DO, calorimeter response will be modified in Run~II by
the presence of a central solenoid magnet and preshower
detectors~\cite{D0upgrade}.  The separation of photon signal and
background in Run~I was based on the fraction of electromagnetic
energy detected in the first longitudinal layer of the calorimeter
($E_{\rm EM1}$/$E_{\rm TOT}$). This quantity is particularly sensitive
to differences in the early stages of shower development initiated by
single photons, and by photons from decays of $\pi^0$'s of the same
\PT.  In the Run~II configuration, the signal--background
discrimination based on this variable is expected to deteriorate, but
simulations indicate~\cite{zutshi} that Run~I performance can be
recovered by using instead the fraction of energy deposited in the
preshower detectors ($E_{\rm CPS}$/$E_{\rm TOT}$), as illustrated in
Fig.~\ref{fig:proc1} for photons at central rapidities.  Using the
fine-grained shower-profile information from the preshower strips (ca.
7 mm triangles), an additional factor of two rejection of background
(while maintaining high signal-efficiencies) has been achieved in our
Monte Carlo simulations~\cite{zutshi}.  As illustrated in
Fig.~\ref{fig:proc2}, multiple preshower clusters for sufficiently
large separation between photons from meson decays can be resolved, or
inferred from the widths of clusters when the showers are not fully
separated.  Results are shown for $\PT\approx15$~GeV/$c$ and
$\eta\approx0.95$, where the separation of photon signal from
background is particularly difficult.  Another variable useful for the
discrimination is the distance in the R--$\phi$ plane between the
detected CPS cluster and the photon position (the latter calculated
from the primary vertex and the position of the calorimeter
cluster)~\cite{gershtein}.  Figure~\ref{fig:cpsresidual} shows this
distribution for $\PT\approx15$~GeV/$c$ and $\eta\approx0.1$.

Even higher background rejection can be achieved by exploiting
correlations between the calorimeter and preshower shower profiles in
a covariance-matrix approach.  A study was undertaken using a
simplified version of the Run~I covariance matrix, with added
preshower variables~\cite{gershtein}. Figure~\ref{fig:discrimination}
shows the $E_T$ dependence of the $\pi^0$ efficiency for $\chi^2$
cutoffs corresponding to 90\% and 60\% photon efficiency.

In the central region, CDF tools for $\gamma$--background separation
in Run~I (shower width in the electromagnetic shower-maximum detector,
and conversion in the central-preradiator detector) will remain the
same. These tools provide a clear separation of the photon signal and
the $\pi^0$-dominated background. The addition of a new
scintillator-based endplug calorimeter with a preshower and shower-max
detector will offer an extension of these tools to the forward region
in Run~II.

Thus, we expect a better signal purity at low \PT\ in Run~II than was
achieved in Run~I, which will facilitate more precise measurements of
low-\PT\ direct photons and di-photons.

\paragraph{Photons at High-\PT}

At high values of \PT, separating the direct-photon signal from
background and the expected reach in \PT\ are determined by the
collected luminosity. As shown in Sec.~\ref{s:ph2}, for the initial 
luminosity of
2~fb$^{-1}$, the inclusive direct-photon cross section measurement can
be extended beyond 300~GeV/$c$. While the high-\PT\ data will permit
detailed tests of perturbative QCD, it is not expected to be sensitive
to differences in recent parameterizations of gluon distributions at
large~$x$ ({\it eg.}, CTEQHJ and CTEQM). Thus, the determination of the
large-$x$ gluon distribution will have to continue to rely on
direct-photon data from fixed target experiments and on jet data at
the Tevatron.

\subsubsection{Impact on the Gluon Distribution\label{s:ph13}}

\begin{figure}[t]
\centering\leavevmode
\vglue1truept\vspace{-1cm}
\hglue1truept\hspace{0.5cm}
\epsfxsize 7 cm 
\epsfbox{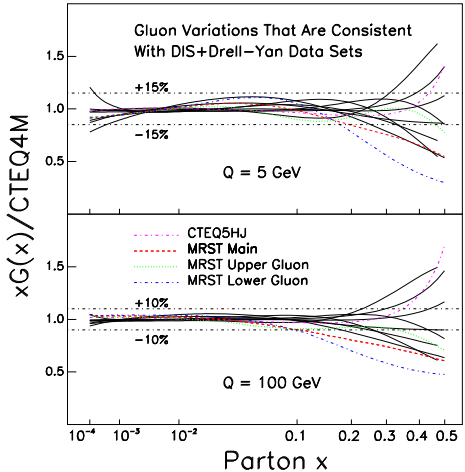}
\caption{The ratio of gluon distributions obtained in the study in
Ref.~\cite{huston-uncertainty} to the CTEQ4M gluon at two different
$Q$ scales. All of these gluon distributions correspond to PDF's which
provide a reasonably good fit to the CTEQ4 DIS/DY data set.}
\label{fig:gluonratio}
\end{figure}

The largest uncertainty in any parton distribution function (PDF) is
that for the gluon distribution.  At low~$x$, the gluon can be
determined indirectly from scaling violations in quark distributions,
but a direct measurement is required at moderate to large $x$.
Direct-photon production has long been regarded as potentially the
most useful source of information on the large-$x$ part of the gluon
distribution. And direct-photon data, especially from CERN fixed
target experiment WA70, have been used in several global
analyses~\cite{mrst,cteq3}.  Another process sensitive to the gluon
distribution, through the gluon--gluon and gluon--quark scattering
subprocesses, is jet production in hadron--hadron collisions.  Precise
data from Run~I are available over a wide range of transverse energy
and, indeed, provide a constraint on the gluon distribution in an $x$
range from about 0.05 to 0.25. However, the low statistical power of
the jet cross section at high~$E_T$, and the dominance of the
$q\bar{q}$ scattering subprocess in that kinematic region, do not
provide for a similar constraint at large~$x$.

Figure~\ref{fig:gluonratio} shows several gluon parton distribution
functions~\cite{huston-uncertainty} that provide a reasonable fit to
the DIS and DY data used in the CTEQ4 fits~\cite{cteq4}. The
excursions shown (normalized to the CTEQ4M gluon) provide an estimate
of the uncertainty in the gluon distribution. The gluon distribution
seems reasonably well-constrained by these data, except at large~$x$.
Also shown in Fig.~\ref{fig:gluonratio} are the gluon distributions
from CTEQ5HJ~\cite{cteq5} (which best fits the Tevatron high-$E_T$ jet
data) and three recent MRST PDF's~\cite{mrst}.  The MRST PDF's
incorporate the WA70 direct-photon data using the \KT-enhancements
described above.  Nevertheless, the theoretical problems associated
with direct-photon production have discouraged the CTEQ collaboration
from using the direct-photon data in their recent fits~\cite{cteq5}.
The recent work on resummation offers hope that this situation will
change.

\begin{figure}[t]
\centering\leavevmode
\vglue1truept\vspace{-1cm}
\hglue1truept\hspace{0.5cm}
\epsfxsize 7 cm 
\epsfbox{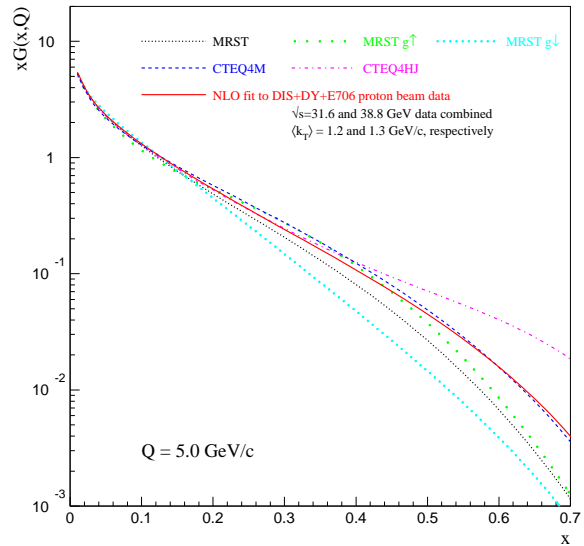}
\vglue1truept\vspace{-1.75cm}
\caption{A comparison of the CTEQ4M, MRST, and CTEQ4HJ gluons, 
and the gluon distribution derived from fits that use E706
data~\cite{ktprd}.  The $g\uparrow$ and $g\downarrow$ gluon densities
correspond to the maximum variation in \avkt\ that MRST allowed in
their fits.}
\label{fig:gluon}
\vspace{-0.5cm}
\end{figure}

The CTEQ4HJ PDF's were determined by increasing the weight for the CDF
jet cross section at high~$E_T$~\cite{cteq4}. In the resultant fit,
the increase was achieved through a significant increase in the gluon
contribution at large~$x$, without inducing serious conflicts with any
of the other experiments used in the CTEQ4 data sets.  This increase
was allowed by the uncertainty in the gluon distribution in this $x$
range, a flexibility not present for any of the quark
distributions. Another demonstration of the uncertainty in the gluon
distribution at large~$x$ can be seen in Fig.~\ref{fig:gluon}, where
the CTEQ4HJ gluon distribution is plotted along with that of CTEQ4M
and three recent MRST PDF's. At $x\approx0.6$, there is over an order
of magnitude spread between the CTEQ4HJ and the MRST $g\downarrow$
gluon distributions.

The CTEQ4HJ gluon distribution (and its successor, CTEQ5HJ) provides the best
description of not only the CDF jet cross section, but that of \DO\ as
well. At $x\approx0.5$, corresponding roughly to $E_T\approx450$~GeV
for central $\eta$, a doubling of the gluon distribution (compared to
CTEQ4M) results in only a~20\% increase in the inclusive jet cross
section.  On the other hand, the fixed-target direct-photon yield of
E706, produced mainly through $gq$ scattering, is proportional to the
gluon distribution.  A fit using the E706 direct-photon data and the
\KT\ K-factors is also shown in Fig.~\ref{fig:gluon}; this result is
very similar to the CTEQ4M gluon.  At the highest reach of the E706
data, the CTEQ4HJ gluon distribution is a factor of~4 to~5 larger than
the CTEQ4M gluon.  With the advent of more complete theoretical
treatments~\cite{sterman} of direct-photon production, the E706 data
should have great impact on the determination of the behavior of the
gluon at large~$x$. This would have implications not only for
fixed-target direct-photon data, but also for collider physics at
highest $Q^2$ scales.

\subsubsection{Conclusions}

Direct-photon physics remains a viable and interesting program for
Run~II of the Tevatron.  The Run~II measurements of single and double
direct photons, and of photon and jet or single track correlations,
will reach larger \PT\ and have improved detection efficiency at low
\PT, compared to Run~I.  Although the data are not expected to improve
directly our knowledge of the gluon distribution at intermediate and
large~$x$, it can do so by providing a testing ground for newly
developed theoretical models and formalisms, and by helping clarify
the currently confused role that multiple gluon emission play in
direct-photon production (and other high-\PT\ processes).  Once this
physics is properly understood, the existing fixed-target data should
provide one of the best constraints on the gluon distribution, as has
been envisioned for a long time.

We have examined the best available experimental information on
production of single and double direct photons (and mesons) at large
\PT\ in both fixed-target and collider energy regimes. Recent
theoretical developments offer optimism that the long-standing
difficulties in the proper description of these processes can finally
be resolved. While there is still no final consensus, the trend of
recent developments has led to an increased appreciation of the
importance of the effect of multiple gluon emission, and to the
emergence of tools for clarifying this issue.

To summarize, measurements of the production of high-mass pairs of
high-\PT\ particles at WA70, E706, CDF, and \DO\ provide consistent
evidence for the presence of large \KT.  NLO pQCD
calculations~\cite{bailey}, which include effects due to the radiation
of a single hard gluon, compare poorly to \KT-sensitive distributions
in di-photon data.  {\tt RESBOS}~\cite{RESBOS}, a NLO pQCD calculation,
which also includes the effects of multiple soft-gluon emission
through the CSS resummation technique, compares well with the shape of
the di-photon data.  LO pQCD calculations~\cite{pythia57,owens} that
incorporate \KT\ effects through Gaussian smearing techniques, provide
reasonable characterizations of distributions for pairs of direct
photons and mesons.

While the apparent inconsistencies between different direct-photon and
$\pi^0$ data sets are not
understood~\cite{ktprd,aurenche-dp,aurenche-pi0}, we found it
instructive to consider results on the $\gamma/\pi^0$ ratio from WA70,
UA6, E706, and R806.  Various experimental and theoretical
uncertainties tend to cancel in such a ratio, which is also relatively
insensitive to \KT-effects.  We find that the ratio from theory agrees
to $\approx30$\% with data from UA6, E706, and R806 over the range
$24~{\rm GeV}<\sqrt{s}<63$~GeV.

LO pQCD has been used to estimate the impact of \KT\ on the inclusive
production of high-\PT\ direct photons and $\pi^0$'s.  This simple
phenomenological model is able to account for differences between NLO
pQCD calculations and inclusive data over a wide range in~$\sqrt{s}$.
While the approximate nature of such models is clear, and has been
discussed in several recent papers, the emerging formalism for the
full (threshold and recoil) resummation of inclusive direct-photon
cross sections appears to vindicate much of the understanding of
effects from multiple gluon emission that has been achieved using
approximate tools. The resummation formalism can be expected to
provide a solid foundation for the treatment of \KT, at which time a
global reexamination of parton distributions, with an emphasis on the
determination of the gluon distribution from the direct-photon data,
should become possible~\cite{sterman-pdf}.

\subsection{Direct Photon Production at the Tevatron$^1$\label{s:ph2}}
\footnotetext{$^1$ Contributed by: J.~Huston and J.F.~Owens}

\subsubsection{Introduction\label{s:intro}}

In this section we summarize features of direct photon production which
are relevant in the kinematic range to be covered in Run~II. In
Sec.~\ref{s:comp}, a  
comparison between Run~I data and the corresponding theoretical
description is presented. Several potential problem areas are
noted. Sec.~\ref{s:exp} contains a 
brief description of the kinematic reach expected for Run~II, based on
an 
integrated luminosity of 2~fb$^{-1}$. Included here is a discussion of
the 
sensitivity to parton distribution functions and to what extent direct photons
at Run~II can help constrain the gluon distribution. In Sec.~\ref{s:new} some 
predictions for photon-jet correlations are presented. The potential of 
observables other than the usual single photon $p_T$ distribution to
help elucidate the underlying dynamics is also discussed. 

\subsubsection{Comparison to Run~I Data\label{s:comp}}

Data for the inclusive cross section for direct photon are available
over 
a wide range in energies from fixed target and collider experiments. By
now it 
is well known that it has not yet been possible to simultaneously
describe 
all of the experimental results with a next-to-leading-order (NLO) QCD 
calculation. A pattern of discrepancies between theory and experiment
exists 
in both the fixed target and the collider data sets. This situation has
been 
reviewed in~\cite{huston-discrepancy,ktprd,aurenche-dp}. An analysis
similar to that for direct 
photon production in Ref.~\cite{aurenche-dp} has been performed for the
case 
of $\pi^0$ production~\cite{aurenche-pi0}. The two processes are closely 
related since $\pi^0$'s decaying to two photons provide much of the
background 
which must be dealt with when extracting the signal for direct photon 
production. Some of these issues are also dealt with in Sec.~\ref{s:ph1} 
where a detailed comparison to fixed 
target data is also presented. While it is clearly of interest to 
understand direct photon production over the entire range of available 
energies, this section will focus on those aspects of the data which can
most 
directly be addressed during Run~II. The first step is to examine the 
theoretical description of the data from Run~I.  

A comparison of NLO QCD predictions to the direct photon data from 
CDF and D{\O} has indicated the presence of a deviation of the data from
theory 
at low values of transverse momentum~\cite{huston-discrepancy,ktprd}. 
This deviation decreases if the effects of soft gluon radiation are
taken 
into account by applying a Gaussian $k_T$ smearing model using a value
of $\langle k_T\rangle$ measured in di-photon production in the two 
experiments.\footnote{The value
of $k_T$ can be directly measured in di-photon events since the photon 
4-vectors can be measured precisely.} Such a $k_T$ treatment
is phenomenologically motivated. 

Recently, there has been progress in more
sophisticated treatments of soft gluon radiation near threshold in the 
parton-parton scattering process ~\cite{laenen,mangano,kidonakisowens}.
At large values of transverse momentum for the photon, the phase space
for the emission of additional gluons in the hard scattering is limited.
This limitation on the emission of real gluons upsets the balance in the 
theoretical expressions between virtual and real emission contributions.
The result is large logarithmic corrections near the threshold for the 
parton-parton scattering subprocesses. These large corrections can be
resummed in a relatively compact formalism. The 
results~\cite{nason,mangano,kidonakisowens} indicate 
that the corrections to existing next-to-leading-order  calculations 
are large as $x_T=2 p_T/\sqrt{s}$ approaches 1. Away from the region at
the edge of phase space it is observed that the corrections to the NLO
results 
coming from the threshold resummation are relatively small over much of
the $x_T$ range covered in the fixed target experiments when the
renormalization and factorization scales are chosen to be $p_T/2$ and 
the resummed results 
show an overall reduction in the sensitivity to the choices of these 
scales. However, the threshold resummation corrections alone are not 
sufficient to explain the discrepancies observed between the theoretical 
predictions and some of the fixed target experimental results. In
addition, 
threshold resummation cannot explain the deviations observed by D{\O} and 
CDF at the low $p_T$ end of the measured distributions.


Another approach to resumming soft gluon effects is that of Ref.~\cite{kimber} 
which uses the DDT~\cite{DDT} or $q_T$-space method. This technique has 
recently been applied to vector boson production~\cite{EV} and compared
with 
the impact parameter method of Ref.~\cite{CSS1}. In Ref.~\cite{kimber} a 
parton-parton luminosity function is defined which depends on the net 
transverse momentum of the pair of colliding partons. The parton distributions
are probed not at the scale of the hard scattering process, but instead
at a scale given by $q_T$. This means that the scale of the parton 
distributions is typically somewhat smaller than, for example, the
transverse 
momentum of the produced photon. This results in an enhancement of the
cross section. In addition, the parton transverse momentum is taken into
account 
when the final photon $p_T$ is calculated. Some enhancement of the fixed 
target predictions is noted using this technique, but it is insufficient
to fully describe all of the fixed target data. For the collider energy
range, 
Ref.~\cite{kimber} quotes only a small effect, which does not appear to be 
sufficient to explain all the observed deviations. 
At present, the resummation is done only with the leading-log 
terms included. More results from this technique are anticipated as 
next-to-leading-log terms are included as well.  

Quite recently a new formalism for simultaneously incorporating both the 
threshold resummation {\it and} the resummation of $k_T$ (or recoil) effects 
has been developed ~\cite{sterman}. This formalism possesses the desirable
property  
of simultaneously conserving both energy and momentum in the resummation 
process. The initial results presented in ~\cite{sterman} indicate that the 
threshold enhancement referred to above is correctly reproduced while, in 
addition, there is a large enhancement due to the newly included recoil 
effects. Detailed applications and studies of the scale dependence and 
the dependence on non-perturbative input parameters are expected soon.
 
The initial indications of discrepancies between theory and experiment
in the collider data came from results taken in Run~Ia.
A comparison of the higher statistics results from CDF in Run~Ib, shown
in 
Fig.~\ref{fig:newCDF},
confirms the shape deviation at low $p_T$ and the agreement with the  
$k_T$ smearing correction obtained using the Gaussian smearing model.
(Also shown is an attempt to implement the soft gluon corrections using 
an enhanced parton shower~\cite{baerreno}.
In this case, the effect seems to drop off more
slowly than the deviation observed in the data.) Note that the $k_T$
smearing correction falls
off roughly as $1/p_T^2$, as expected from a power correction type of
effect. 
This is not true in the case of fixed target experiments, as discussed
in Sec.~\ref{s:ph1}, where the steeply falling parton distribution functions 
enhance the effects of the soft gluon radiation. 

        The D{\O} direct photon cross section for the central region
is also consistent with such a deviation at low $p_T$~\cite{newD0}, 
as shown in 
Fig.~\ref{fig:newD0} 
while no conclusion can
be reached for the D{\O} cross section in the forward region (see the lower 
half of Fig.~\ref{fig:newD0} and Ref.~\cite{newD0}). The possibility 
that the discrepancy between theory and experiment may be dependent on 
rapidity is interesting and is one that can be investigated in more detail 
with the higher statistics expected from Run~II.

\begin{figure}
\centering\leavevmode
\epsfxsize=3 in\epsffile{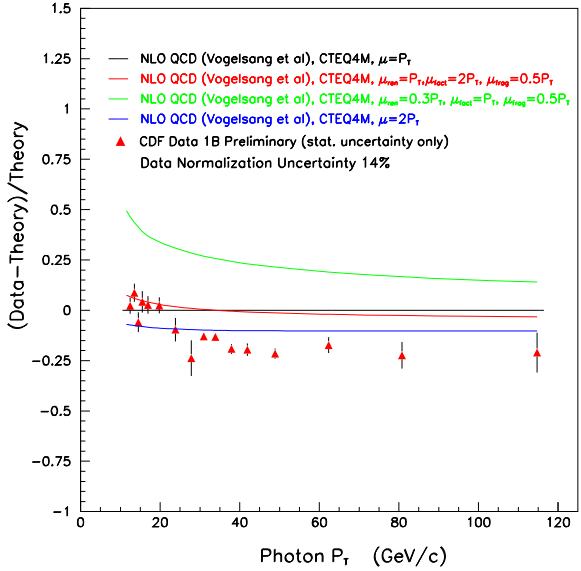}
\caption{Comparison of the CDF Run~Ib data with several NLO curves 
corresponding to different choices of the renormalization and
factorization 
scales.}
\label{cdf_vo}
\end{figure}

In an attempt to achieve a better theoretical description of the data 
it is important to investigate what flexibility exists within the 
conventional QCD hard scattering NLO formalism.
In order to perform perturbative QCD calculations, one must specify the 
renormalization and factorization scales. For the latter, there are two 
scales, corresponding to the factorization of collinear singularities for 
the initial state parton distributions and the final state fragmentation 
functions. For most hard scattering calculations these three scales are 
chosen to be proportional to the characteristic large transverse
momentum - 
that of the photon in this case. Often, all three scales are set equal
to each other. However, this latter step is not necessary and it is
reasonable 
to ask whether or not one can describe the shapes of the CDF and D\O\ data
in 
the region below $p_T = 30$ GeV by a suitable variation of all three of the
scales~\cite{vo}. In Fig.~\ref{cdf_vo} the Run~Ib CDF data are compared to
several 
curves corresponding to different choices of the renormalization and 
factorization scales. One can see that it is possible to get a
steepening of 
the slope at the low-$p_T$ end, but only at the price of an increase in
the overall normalization. Apparently, it is not possible to get both the
shape and the normalization correct by such a strategy.

\begin{figure}
\centering\leavevmode
\epsfxsize=3 in\epsffile{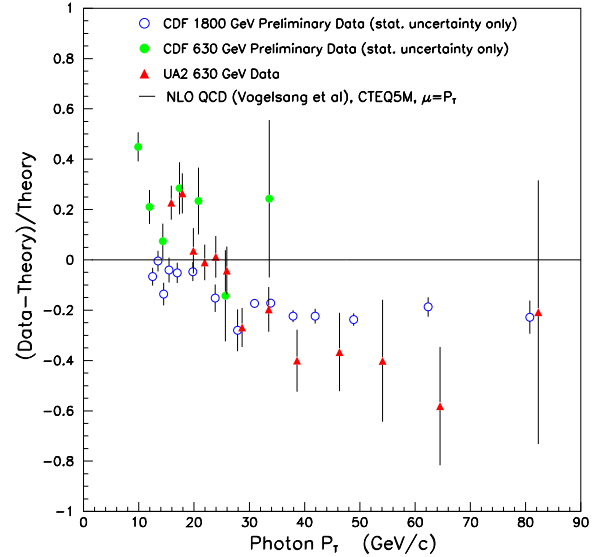}
\caption{A comparison of the CDF photon data from Run~Ib (at both
$\sqrt{s}=630$~GeV and $\sqrt{s}= 1800$~GeV) with a NLO QCD
prediction and the direct photon data from UA2.
}
\label{cdf_ua2}
\end{figure}

In Fig.~\ref{fig:newCDF}, the CDF data has been normalized
upwards by a factor
of 1.2 for an easier shape comparison. If this normalization is taken
out, as
shown in Fig.~\ref{cdf_ua2}, it is evident that the data fall below
the NLO QCD 
prediction at high $p_T$. Also shown is a comparison to the CDF photon
data taken at $\sqrt{s}=630$~GeV and the photon data from UA2~\cite{UA2-dp-ddp}
where a similar deficit is observed at high $p_T$. For most observables, 
typically, the data lie above the NLO QCD  predictions so this is
somewhat of
an unusual situation. It is interesting to note that a measurement of the
photon fraction (of the photon candidate sample) indicates that the photon
fraction seems to be leveling off at approximately 80\%, rather than 
saturating the sample at near 100\% at high $p_T$. The latter outcome would
be predicted from very basic considerations: with a fixed $E_T$
isolation cut,
jets are required to fragment into a $\pi^0$ with a higher $z$ value as the
$p_T$ of the photon candidate increases. Such a fragmentation is
suppressed by
the sharp falloff of the fragmentation function at high $z$. It will be 
interesting and important to understand this behavior. The increased 
statistics of Run~II will allow both the low $p_T$ and high $p_T$
regions to be investigated more thoroughly.  

\subsubsection{Expectations for Run~II\label{s:exp}}

\begin{figure}
\centering\leavevmode
\epsfxsize=3 in\epsffile{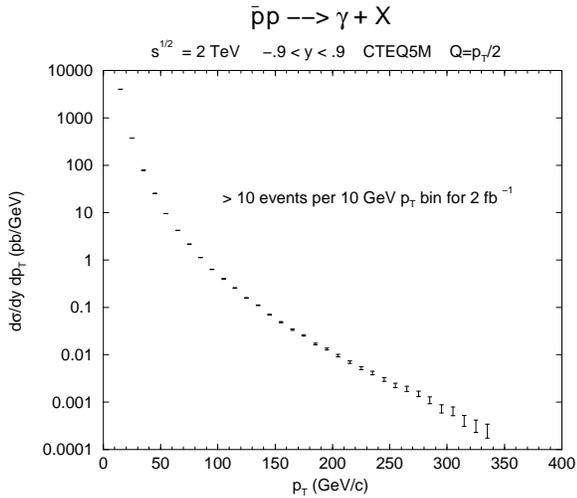}
\caption{Direct photon $p_T$ distribution for Run~II with 
errors based on an integrated luminosity of 2~fb$^{-1}$.}
\label{2fb}
\end{figure}

\begin{figure}
\centering\leavevmode
\epsfxsize=3 in\epsffile{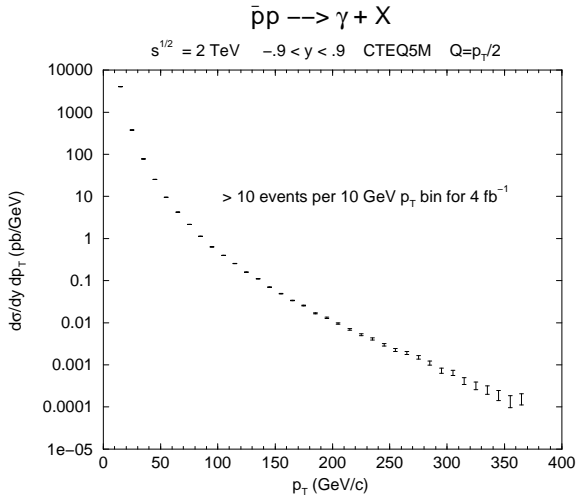}
\caption{Direct photon $p_T$ distribution for Run~II with errors 
based on an integrated luminosity of 4~fb$^{-1}$.}
\label{4fb}
\end{figure}

For the purposes of this section we shall assume an integrated
luminosity 
for Run~II of 2~fb$^{-1}$. As noted in Sec.~\ref{s:comp}, the data on
direct photon production 
from Run~I extend to a transverse momentum of approximately 120~GeV. The 
increased statistics expected from Run~II greatly extend this range as
shown 
in Figs.~\ref{2fb} and~\ref{4fb}. These figures have been generated
using 
the next-to-leading-logarithm program of~\cite{BOO} with the CTEQ5M 
\cite{cteq5} parton distributions and with the renormalization and 
factorization scales set equal to $p_T/2$. The errors shown are
statistical 
only and the results are presented for the transverse momentum range
where 
more than 10 events are expected in a 10~GeV bin of $p_T$ with a total 
integrated luminosity of 2~fb$^{-1}$ and 4~fb$^{-1}$, respectively. 
No efficiency/acceptance corrections, however, have been applied to these
estimates. Typically, these corrections are on the order of 50\%.

\begin{figure}
\centering\leavevmode
\epsfxsize=3 in\epsffile{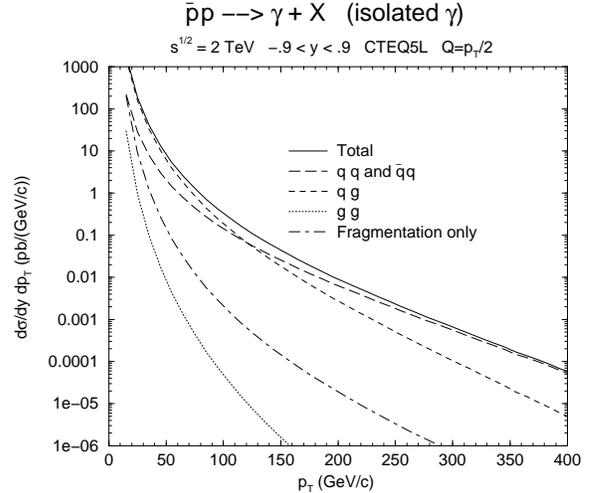}
\caption{Contributions of the various subprocesses for direct photon 
production calculated in the leading-log approximation using the CTEQ5L 
parton distributions.}
\label{subp}
\end{figure}

\begin{figure}
\centering\leavevmode
\epsfxsize=3 in\epsffile{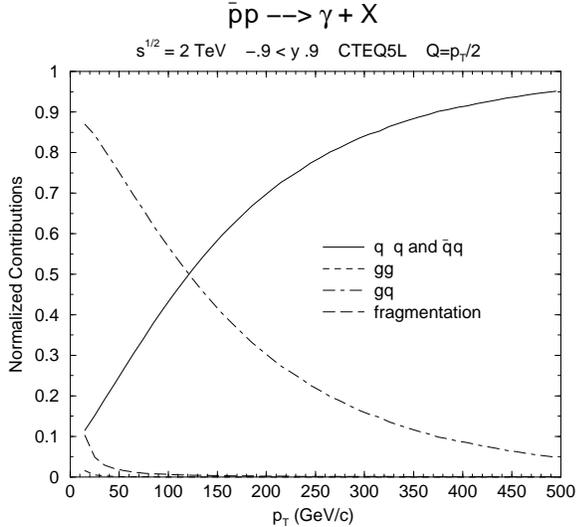}
\caption{The same results as shown in Fig.~\ref{subp} except on a linear 
scale and normalized to the total rate at each value of $p_T$.}
\label{subp_norm}
\end{figure}

\begin{figure}
\centering\leavevmode
\epsfxsize=3 in\epsffile{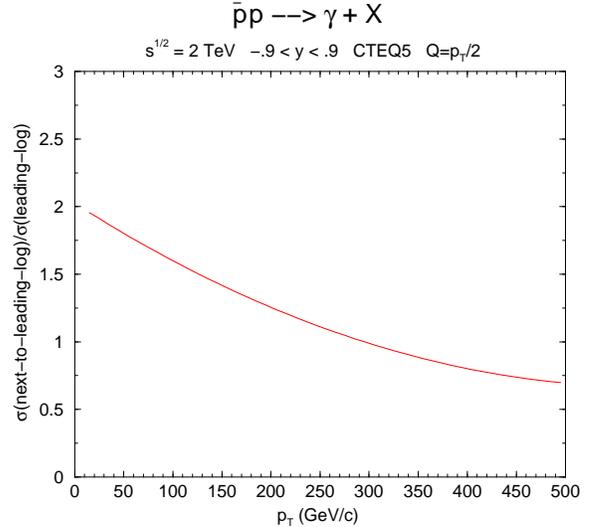}
\caption{NLO/LO ratio for direct photon production.}
\label{ratio}
\end{figure}

One can see that the range of useful statistics extends past 300~GeV in the 
first case and past 350~GeV in the second. This extended coverage in $p_T$ 
corresponds to an increased range of sensitivity for the values of the 
parton momentum fractions of the colliding hadrons. Recall that the inclusive 
cross section for single photon production in the central region is sensitive 
to average values of the parton momentum fraction $x$ approximately equal to 
$x_T=2 p_T/\sqrt {s}$. A range of $x_T$ out to about 0.3 will be covered, 
corresponding to a similar range of $\langle x\rangle$ for the parton
distributions. 
This extended range suggests that the relative ratios of the underlying 
subprocesses should change significantly over the $p_T$ range to be 
covered. To investigate this, the inclusive cross section is displayed in 
Fig.~\ref{subp} along with the contributions from the various parton 
scattering subprocesses. For ease of comparison, the same results are shown 
in Fig.~\ref{subp_norm} on a linear scale relative to the total rate. The 
results in both of these figures were generated using the 
leading-logarithm approximation with the CTEQ5L parton distributions. The 
leading-log approximation was chosen so that the separation between the 
point-like and fragmentation contributions would be unambiguous. Processes 
with two or more 
partons in the final state in addition to the photon can populate
regions of phase 
space where the photon is collinear with one of the partons. These 
topologies 
correspond to the fragmentation process. Thus, the higher order terms mix 
the contributions from the point-like and fragmentation (or bremsstrahlung) 
components. One must define, through the use of appropriate experimental 
cuts, precisely what is meant by the bremsstrahlung component. For the 
purposes of the discussion being presented here the leading-log
predictions are sufficient. Note that the magnitude of the 
next-to-leading-order 
corrections relative to the leading-log predictions for the choice of scale 
used here ($p_T/2$) is slowly varying over the kinematic region being
studied, as shown in Fig.~\ref{ratio}.

The results shown in Figs.~\ref{subp} and~\ref{subp_norm} illustrate
several points worth noting. First, the fragmentation 
component is expected to be a negligible fraction of the total rate.
This is 
due primarily to the imposition of an isolation cut which rejects events
with more than 1 GeV of hadronic energy accompanying the photon in a cone of 
radius 0.4 about the photon direction.\footnote{This cut was used for the 
CDF Run~Ib direct photon measurements. The value of the isolation energy cut 
in Run~II will have to be increased somewhat due to the contributions to the 
isolation energy from the larger number of minimum bias events expected in 
the same crossing.} Such a cut is necessary 
experimentally in order to control the copious background to the photon 
signal from jets fragmenting into $\pi^0$'s.  
This cone isolation energy is almost 
completely saturated by the underlying event energy accompanying the 
hard scatter, leaving little room for energy from the fragmentation of
the jet. 
Note that the precise value 
of the fragmentation contribution relative to the total rate will vary as 
one includes higher order effects, but the overall contribution is still 
expected to be small. Next, one sees the dominance of the $qg \rightarrow 
\gamma q$ subprocess in the region out to about 100~GeV in $p_T$, {\it
i.e.}, 
the range covered by the Run~I data. Beyond this range the $\overline q q 
\rightarrow \gamma g$ subprocess becomes dominant. Gluon-gluon initiated 
processes are not expected to play a 
significant role over the $p_T$ range shown. 

\begin{figure}
\centering\leavevmode
\epsfxsize=3 in\epsffile{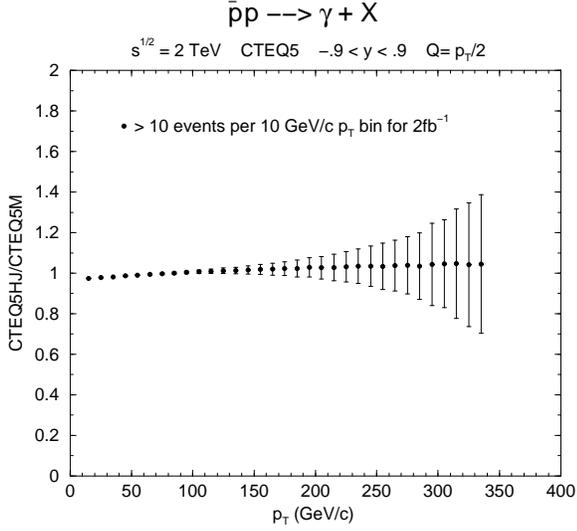}
\caption{Ratio of the direct photon cross sections calculated with 
CTEQ5HJ and CTEQ5M with errors based on an integrated luminosity of 
2~fb$^{-1}$.}
\label{ratio_hj}
\end{figure}

One of the classic applications of direct photon production is to provide 
constraints on the gluon distribution in global fits of parton distributions. 
The gluon distribution is especially uncertain in the region beyond $x
\approx 0.15$~\cite{huston-uncertainty}. Run~I results on high-$p_T$ jet
production from the 
CDF~\cite{CDF_jets} and D{\O}~\cite{D0_jets} collaborations favor a gluon 
distribution which is larger at high-$x$ than was anticipated from
global 
fits which did not emphasize the high-$p_T$ jet data. One such example
is the 
CTEQ5HJ~\cite{cteq5} set of distributions which are favored by both sets
of jet data. One might hope that the direct photon data could shed some
light on 
this issue, but such is not expected to be the case. As shown in 
Fig.~\ref{ratio_hj}, the 
ratio of the CTEQ5HJ and CTEQ5M predictions is consistent with unity
within 
about 5\% over the $p_T$ range under consideration. This can be understood 
by referring back to Fig.~\ref{subp_norm} where it is shown that the $gq$ 
subprocess decreases in importance precisely where one would like to gain 
a constraint on the gluon distribution.

\begin{figure}
\centering\leavevmode
\epsfxsize=3 in\epsffile{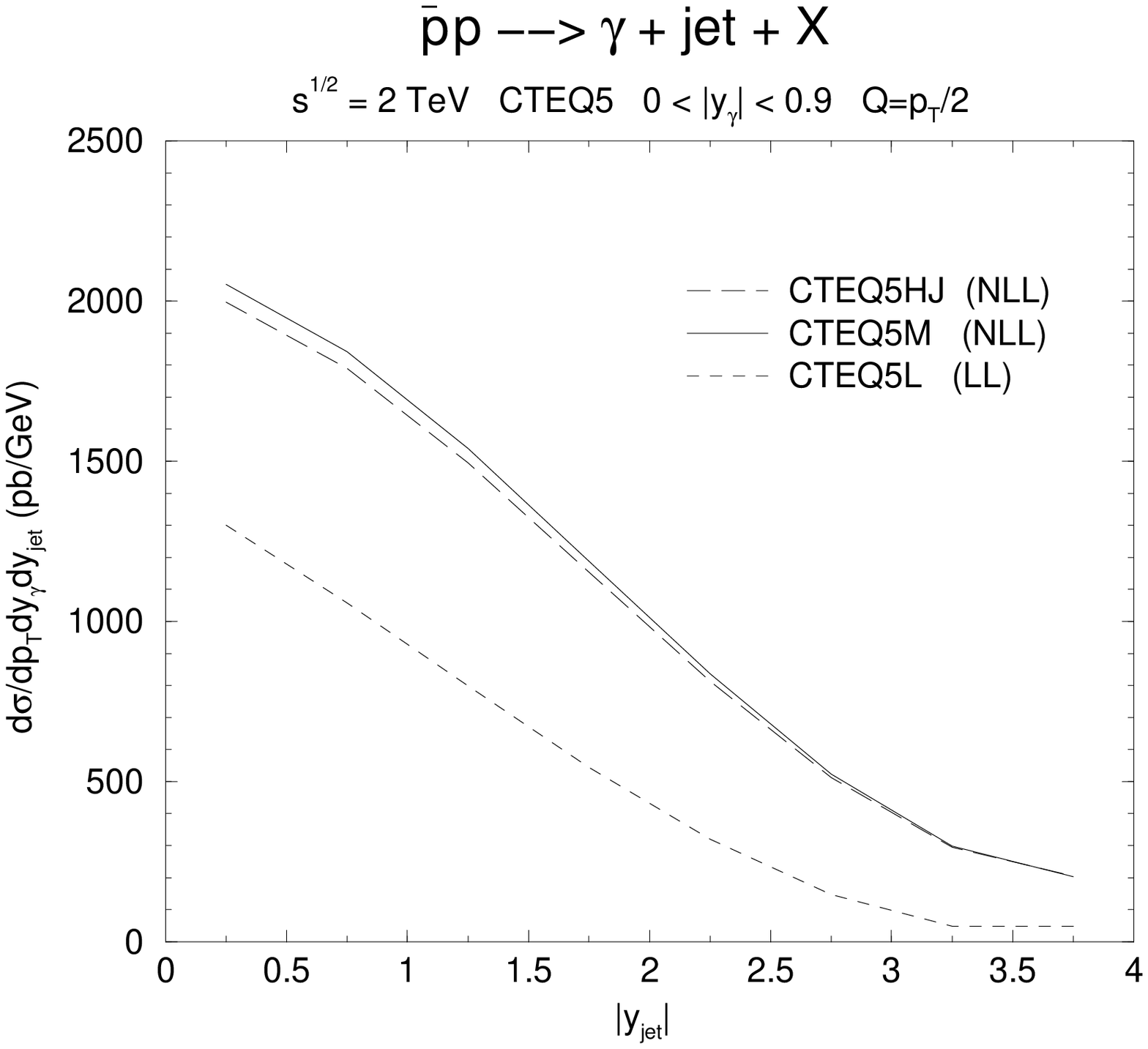}
\caption{Jet rapidity distribution for photons in the central rapidity 
region calculated with the CTEQ5L, CTEQ5M, and CTEQ5HJ parton
distributions with the program of Ref.~\cite{BOO}.}
\label{yj1}
\end{figure} 

\begin{figure}
\centering\leavevmode
\epsfxsize=3 in\epsffile{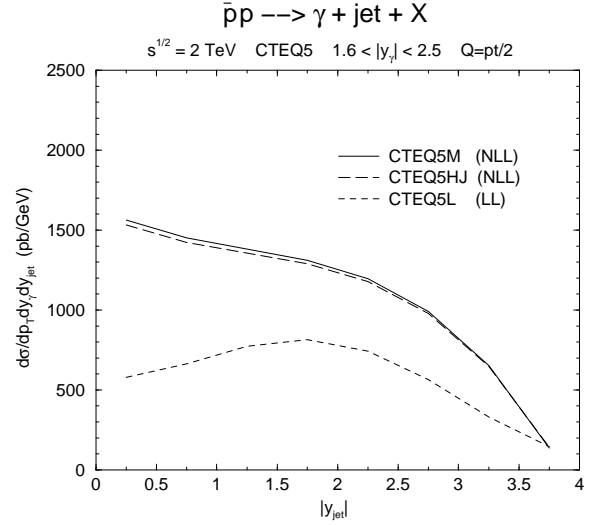}
\caption{Jet rapidity distribution for photon in a forward rapidity
region 
calculated with the CTEQ5L, CTEQ5M, and CTEQ5HJ parton distributions
with the program of Ref.~\cite{BOO}.}
\label{yj2}
\end{figure} 

\subsubsection{New Measurements\label{s:new}}

As noted in the introduction, the process of direct photon production is
of interest for a variety of reasons. In particular, it offers a probe of  
hard scattering dynamics which complements that of jet production. At
this point, the comparison between theory and experiment has not reached the 
quantitative level that one would like, so there is clearly more to be
done. Furthermore, direct photon production has the potential to provide 
constraints on the parton distribution functions, particularly that of
the gluon. Nevertheless, the less than satisfactory description of the
current  data has meant that this role is, as yet, unfulfilled. 
Finally, one must 
understand conventional sources of energetic photons before being able to 
confidently use photons as part of the signal for new physics. With
these points in mind, it is natural to ask whether there are additional
observables 
which could help shed light on some of the theoretical problems.

To date, almost all of the experimental results for direct photon
production 
have been for the transverse momentum distribution of the photon, {\it
i.e.}, 
the single photon inclusive cross section. Clearly, there is additional 
information to be gained by studying the joint rapidity distributions of the 
photon and a jet, {\it e.g.}, $d\sigma/dp_T dy_{\gamma} dy_{jet}$ where
$p_T$ represents the transverse momentum of the photon. This observable has
been measured by the CDF Collaboration~\cite{y_jet} for photons in the the
central region with $16~{\rm GeV} < p_T < 40$~GeV. In this case the theoretical
description of the jet rapidity 
distribution is good. However, no information is yet available for other 
values of photon rapidity. Some leading-log and next-to-leading-log 
predictions for this distribution are 
shown in Fig.~\ref{yj1} for central photons and Fig.~\ref{yj2} for forward 
photons. Notice the characteristic broadening of the jet rapidity distribution 
as the photon is moved forward. Note, too, the change in the ratio between the 
leading-log and next-to-leading-log predictions as the photon rapidity
is varied. Predictions are shown for both the CTEQ5M and CTEQ5HJ parton 
distributions. There is only a slight difference between the two sets,
due primarily to the fact that the curves were generated for $p_T > 
10$~GeV. The resulting low values of $x_T$ result in the parton distributions
being probed in a region where there is little difference between the CTEQ5M
and CTEQ5HJ sets. On the other hand, that means that for these distributions 
there will be relatively less uncertainty due to the parton
distributions and that, therefore, such measurements may be useful in 
helping to understand the 
$p_T$ region where there is a discrepancy between the existing collider
data and the theoretical results. In this regard the D{\O} results for the
inclusive 
photon yield shown in Fig.~\ref{fig:newD0} are interesting as they may
indicate that the theory/data discrepancy has some rapidity dependence. The
lesson here 
is that photon-jet joint observables would be helpful in sorting out the 
source of the disagreement between theory and experiment. Note that if
one wanted to increase the sensitivity to differences between parton
distributions, then a larger minimum $p_T$ cut could be employed. 


As discussed in Sec.~\ref{s:comp}, there is some indication that the photon
yield above a transverse momentum of about 30~GeV may actually be less than 
the theoretical predictions, in contrast to the situation at lower values of 
$p_T$. It was mentioned that this may be related to the observed
behavior of 
the $\gamma/\pi^0$ ratio. This issue is complicated by the necessity of 
placing isolation cuts on the electromagnetic triggers in order to
reduce the 
$\pi^0$ background. NLO Monte Carlo programs can simulate the effect of
these isolation cuts, but the experimental dependence on the parameters
of the  
cuts has not yet been compared to that of the theory. If the theoretical 
treatment of the isolation cuts is wrong, then the comparison of the NLO 
results to collider data must be considered suspect. In recent years the 
question of the theoretical treatment of such isolation cuts has been 
studied by several authors~\cite{BQ,AF,frixione}. What is needed is a
data set showing how the cross section depends on the cone size and the
energy  
threshold utilized in the isolation cuts. Similar comparisons for jet 
production have helped refine the various jet algorithms and have been
very useful in understanding issues related to the theoretical
description of jets. 
 
\subsubsection{Conclusions}

Run~II offers many opportunities to refine our understanding of the production 
of photons in hard scattering processes. The kinematic reach in transverse 
momentum should be greatly extended and the statistical precision of the 
data will also be increased. While there are still problems with the 
theoretical description of the existing data, the Run~II data have the 
potential to shed some light on these issues. In particular, data for 
photon-jet correlations and for the dependence of the cross section on
the parameters of the photon isolation cuts will be helpful. A better 
theoretical understanding of direct photon production will enable this 
process to be 
better used in the study of large transverse momentum processes and the
search for new physics.

\section{Topics in Weak Boson Production$^\dag$\label{s:topics}}
\footnotetext{Contributed by: D.~Casey, T.~Dorigo, M.~Kelly, S.~Leone,
W.K.~Sakumoto, and G.~Steinbr\"uck}

The very large number of $W$ and $Z$ boson events CDF and \Dzero
will collect will yield precision measurements
of the $W$ mass and width~\cite{wgiii}, which are fundamental parameters
of the Standard Model and thus need to be determined with the
highest possible accuracy. The Run~II vector boson datasets will 
of course provide other important advances in the field of 
electroweak physics, and will be the starting point of most new
physics searches; but they will also become standard tools for
the understanding of many sources of systematic uncertainty in otherwise
unrelated physics studies. In this Section we will give an 
overview of some of the additional studies it will be possible
to carry out with vector bosons in Run~II; for rare decays
of $W$ bosons and other topics not covered here see 
Ref.~\cite{tev2000}. 

\subsection { The $\boldmath{W}$ Cross Section as a Luminosity
Monitor\label{s:lum} }

At both D\O\ and CDF, the uncertainty on the total integrated
luminosity is approximately 4.5\%~\cite{DzeroXsec,CDFLum}. Though measuring
the luminosity is difficult to do well, any measurement at the
collider that has an absolute normalization depends on it. In several
important cross section measurements made during Run~I, this
uncertainty contributed greatly to the overall measurement
uncertainty. In particular, for both of the $W$ and $Z$ boson cases,
the uncertainty in the luminosity measurement far outweighed the other
systematic uncertainties in the measurement (see
Table~\ref{tbl:wzxsec}).

There is also a continuing controversy regarding the actual value of
the luminosity at the two detectors at the
Tevatron~\cite{DzeroXsec,CDFLum}. At both D\O\ and CDF, the integrated
luminosity is normalized by the total inelastic cross section in
\pbarp\ collisions. CDF has made this measurement~\cite{CDFpbarp} and
uses it to normalize their luminosity.  D\O\ did not measure the total
inelastic cross section, and chooses to normalize to the world
average, which is $\sim6.2\%$ higher than the CDF
measurement~\cite{DzeroXsec}. This is primarily due to a 2.8 standard
deviation disagreement between the CDF measurement and the E811/E710
measurements~\cite{e811,e710,CDFLumCompare,DzeroLum}.

In the context of Run~II, there are some general concerns regarding
the measurement of the total integrated luminosity, particularly at
high instantaneous luminosity. New luminosity monitors are being
installed at each detectors as part of the modifications for Run~II
and CDF plans to repeat the measurement of the total inelastic cross
section. However, it is not clear that this will result
in a more precise determination of ${\cal{L}}$, and disagreement
regarding the measured \pbarp\ cross section is likely to
persist. Given the high precision of the current $W$ and $Z$ cross
section measurements, increasing precision in the matrix element
calculations (now at NNLO~\cite{vanNeerven}), and the expected
abundance of $W$ and $Z$ bosons produced during Run~II ($\sim10^6$
$W$ bosons and $\sim10^5$ $Z$ bosons within the fiducial volumes of each
detector), we may consider using the rate of $W$ boson production to
normalize the integrated luminosity. In this subsection, we present a
brief overview of the magnitude of the experimental uncertainties in
the integrated luminosity using $W$ boson production, and
discuss the requirements for such a measurement to be competitive with
standard luminosity measurements and in what context it would surpass
the current precision.

\begin{table}
\caption{Measurements of $\sigma(W)$ at the Tevatron. The three quoted 
uncertainties are statistical, systematic, and luminosity.
\label{tbl:wzxsec}}
\begin{tabular}{|l|l|c|}
\hline\hline
Detector & Channel & $\sigma(W)\cdot B$ (nb) \\ \hline
D\O(1A) & $W\rightarrow e\nu$   & $2.28 \pm 0.02 \pm 0.08 \pm 0.10$\\
D\O(1A) & $W\rightarrow\mu\nu$  & $2.02 \pm 0.06 \pm 0.22 \pm 0.09$\\
D\O(1B) & $W\rightarrow e\nu$   & $2.31 \pm 0.01 \pm 0.05 \pm 0.10$\\
D\O(1B) & $W\rightarrow\tau\nu$ & $2.22 \pm 0.09 \pm 0.10 \pm 0.10$ \\
CDF(1A) & $W\rightarrow e\nu$   & $2.49 \pm 0.02 \pm 0.08 \pm 0.09$\\
\hline\hline
\end{tabular}
\end{table}

\subsubsection{$\boldmath{\sigma(W)/{\cal{L}}}$ in Run~I/II at D\O}

As an example, we describe some details of how the Run~I measurement
of the cross section for $W$ boson production at D\O\ would translate
into a determination of the total integrated luminosity.

Reversing the relationship between the integrated luminosity
(${\cal{L}}$) and the production cross section ($\sigma(W)$) in the
calculation done by D\O~\cite{DzeroXsec}, we have the following relation:

\begin{equation}
\label{eq:lum}
{\cal{L}}={N_{cand}(1-f_{QCD})-N_Z\over \epsilon A(1+{A_\tau\over
A})\sigma(W)} 
\end{equation}
where $N_{cand}$ is the number of $W$ boson candidates observed,
$f_{QCD}$ is the fraction of candidates expected from QCD multi-jet
production, $N_Z$ is the number of candidates that are $Z$ bosons in
which one of the electrons was unobserved, $\epsilon$ is the event
identification efficiency, $A$ is the geometric and fiducial
acceptance, $A_\tau$ is the acceptance times branching ratio for
$W\rightarrow\tau\nu$, and $\sigma(W)$ is the predicted cross section
times branching ratio for producing $W$ bosons. 

\begin{table}
\caption{Uncertainties on the components of the measurement of
${\cal{L}}$ if one used the components of the Run~I measurement of the 
$W$ production cross section by D\O.
\label{tbl:runI}}
\begin{center}
\begin{tabular}{|c|c|c|}
\hline\hline
Component & Value & Error on $\cal{L}$ \\ \hline
$f_{QCD}$ & 0.064$\pm$0.014 & 1.5\% \\
$\epsilon$ & 0.671$\pm$0.009 & 1.3\% \\
$A$ & 0.465$\pm$0.004 & 0.9\% \\
$N_{cand}$ & 67078 & 0.4\% \\
$N_Z$ & 621$\pm$155 & 0.3\% \\
$A_\tau/A$ & 0.0211$\pm$0.0021 & 0.2\% \\
$\sigma(W)$ & 22.2$\pm$0.9 nb & 4\% \\
\hline\hline
\end{tabular}
\end{center}
\end{table}

Table~\ref{tbl:runI} show the values of each of the quantities
as measured in Run~I, and the fractional uncertainty each would
contribute to a measurement of ${\cal{L}}$. The total uncertainty from 
measured quantities alone is $2.2\%$. Including a $4\%$ uncertainty in the
prediction of  $\sigma(W)$ increases the total uncertainty to
$4.6\%$. 

We note two things: 1) even the Run~I measurement results in an
uncertainty on ${\cal{L}}$ that is competitive with the directly
measured result\footnote{Of course, we cannot reliably use the theory
prediction without the experimental confirmation from Run~I!} and 2)
by far, the dominant uncertainty is in the prediction of $\sigma(W)$.

In preparation for Run~II, both D\O\ and CDF are undergoing major
upgrades to the detectors. For D\O, this includes the addition of a
solenoid magnet and a complete replacement of the tracking system. The 
total luminosity is expected to increase by a factor of $\sim10$. We now 
consider how each of the factors in Eq.~\ref{eq:lum} will be affected
in Run~II.

The background fraction for events from QCD multi-jet production is
dominated by systematics for the detection of electrons. We do not
expect the uncertainty to decrease dramatically, nor do we expect the
overall background level to change much. It may be possible to reduce
the backgrounds and make them easier to understand if we use the muon
channel; however, no serious study has been made on the subject.

The uncertainty on the lepton identification efficiency is equal parts
$Z$ boson statistics and background subtraction statistics and will
scale by $\sim 1/\sqrt{10}$ (we assume an integrated cross section of
1~fb$^{-1}$ in this section). We expect the tracking efficiency to
increase, but the efficiency of the isolation and \met\ requirements
may decrease due to multiple interactions and decreased resolution due
to the presence of the solenoid in the Run~II detector. Speculating,
these effect may balance each other, leaving an overall reduction in
the uncertainty by $1/3$ to 0.9\%.
 
The geometrical and fiducial acceptance will stay approximately the
same. The uncertainty is dominated by the electromagnetic energy scale
(0.00319 of 0.004) whose uncertainty will be smaller in Run~II due
to an increased number of $Z$ boson events and extra handles provided
by a central magnetic field in the D\O\ detector. Scaling the rest of
the uncertainties 
with the luminosity, we should be able to halve the uncertainty on the
acceptance to 0.5\%.

The uncertainties in the expected number of $Z$ candidates and the
acceptance of electrons and muons from $W\rightarrow\tau\nu$ are
dominated by MC statistics and can be shrunk to a negligible value.

Finally, the number of candidates will increase by a factor of 10 just
from the increased luminosity, by approximately a factor 1.12 from the
increase of the center of mass energy from 1.8~TeV to 2.0~TeV, and by 
another factor of $\sim2$ if one
includes the muon channel, resulting in about $1.4\times 10^6$
candidates, which translates into a statistical uncertainty of about
$0.08\%$.

\begin{table}
\caption{Estimated values for the uncertainties in a measurement of
${\cal{L}}$ in Run~II.
\label{tbl:runII}}
\begin{center}
\begin{tabular}{|c|c|}
\hline\hline
Component & Error on $\cal{L}$ \\ \hline
$f_{QCD}$ &  1.5\% \\
$\epsilon$ & 0.9\% \\
$A$ & 0.5\% \\
$N_{cand}$ & 0.08\% \\
$N_Z$ & 0.0\% \\
$A_\tau/A$ & 0.0\% \\
$\sigma(W)$ &  4\% \\
\hline\hline
\end{tabular}
\end{center}
\end{table}

Table~\ref{tbl:runII} summarizes the expected experimental
uncertainties for measuring ${\cal{L}}$ in Run~II at D\O. A measurement
of the luminosity will be dominated by the background level and
uncertainty and by the uncertainty in the prediction of the cross
section. Figure~\ref{fig:lumunc} shows how changes in the QCD
background level and uncertainty, and uncertainty in the cross section
prediction, affect the overall uncertainty in the resulting luminosity.
In each case, all quantities from experiment were fixed to their Run~I
values except the one being varied. The uncertainty on the cross
section was made negligible when considering the sensitivity to the
background level and uncertainty. We note that decreasing the
fractional uncertainty on the QCD multi-jet background by a factor of
2~--~4 reduces the luminosity uncertainty from 2.2\% to 1.7~-- 1.8\% -- a
factor of $0.3-0.2$. The relatively large uncertainty in the background
level (20\%) translates into 
a significant sensitivity to the background level itself. Essentially,
the trade-off for loosening the selection criteria and allowing more
background into the data sample is that the background must be measured
much more precisely in order to maintain a small uncertainty. In Run~II
(as in Run~I), the best strategy to minimize the effect of the
background uncertainty on the total cross section measurement will be to 
minimize the level of the background altogether -- if the background can 
be made negligible, the effect of the uncertainty of the background on
the cross section (no matter how large) will also be negligible.
The cross section uncertainty continues to dominate the
situation. However, if the cross section uncertainty is kept to
approximately the 
size of the experimental uncertainty, the resulting luminosity
uncertainty is more than a factor of two better than the Run~I value.
\figLumUnc

\subsubsection{Counting on the Prediction for $\boldmath{\sigma(W)}$}
If one determines the integrated luminosity using the rate for $W$
boson production, there are two fundamental issues that must be
resolved. First, one must decide that the calculation for $\sigma(W)$
is reliable in itself; that it agrees with experiment. Second, one
must determine the uncertainty in the calculation, since it will
likely dominate the uncertainty in ${\cal{L}}$. 

The first issue is likely not problematic. If all cross section
measurements were normalized to a specific $\sigma(W)$ calculation,
then the worst impact of a change in that calculation would be to
modify all measurements in the same manner. Additionally, the
advantage would be that all measurements would be easy to compare,
since the controversy over the measurement of the total inelastic
cross section of \pbarp\ collisions would be circumvented.

The problem of determining the uncertainty on the calculation is more
difficult. Various {\sl ad hoc} methods have estimated the uncertainty
on $\sigma(W)$ to be $3-5\%$~\cite{DzeroXsec}.The uncertainty in the
cross section is dominated (almost exclusively) by uncertainties in
the PDF's which go into the
calculation. (The uncertainties due to higher order QCD and electroweak
corrections 
are likely much smaller than these.) Two efforts to understand the
uncertainties in the PDF's quantitatively are described in the report of 
the Working Group on Parton Distribution Functions~\cite{pdfUnc}. 

\subsubsection{Conclusion}
We have described the current expectations for measuring the integrated
luminosity in Run~II using the rate for $W$ boson production. The
experimental uncertainties, totaling $\sim2.2\%$, are dominated by the
uncertainty in the background from QCD multi-jet interactions. However,
the total uncertainty is dominated by an ill-defined uncertainty on
the prediction for the production cross section for $W$ bosons. With
sufficient progress in the continuing effort to quantify this
uncertainty, we may be able to reliably determine the total integrated
luminosity in Run~II using the rate for $W$ boson production to
$\sim3-4\%$.


\subsection {Determination of the Weak Boson $\boldmath{p_T}$
Production Spectrum }

During Run~II at the Tevatron, CDF and D\O\ will obtain the largest
data set of $e^+e^-$ pairs resulting from via
$p\bar{p}\rightarrow\gamma^*/Z$ to date, pushing the analysis of
vector boson production characteristics over the edge from being
limited by statistical uncertainties, to being limited by systematic
uncertainties. The di-electron final state provides two important
experimental handles. Electrons themselves are among the best-measured
objects at either detector, with far better resolution in energy and
position than most final-state high-$p_T$ objects (such as jets or
muons). Additionally, the di-electron final state provides complete
kinematic information about the hard collision; the four-momentum of
the $Z/\gamma^*$ state is known unambiguously. Since the electroweak
character of the decay to di-electrons is generally uncorrelated with
the QCD characteristics of the production of the vector boson state,
di-electron production in $p\bar{p}$ collisions via the Drell-Yan
production is a sensitive probe for investigating many aspects of
QCD. The rapidity ($y$) distribution is the Drell-Yan analog to deep
inelastic scattering structure function, providing additional
constraints on PDF's. The transverse momentum ($p_{\rm T}$)
distribution is sensitive to predictions from standard perturbative
QCD at high-$p_{\rm T}$ ($\sim Q^2$) and to predictions from
soft-gluon resummation calculations at low-$p_{\rm
T}$~\cite{VBP,ResBos}. Additionally, the low-$p_{\rm T}$ region is
sensitive to non-perturbative effects not calculable in pQCD. These
effects are included via a universal form factor, not unlike PDF's,
whose parameters must tuned to data.

In addition to the intrinsic benefits of precision measurements of
QCD, there are practical benefits for other measurements at the
Tevatron. In the low-$p_{\rm T}$ region, where the cross section is
highest, uncertainties in the phenomenology of vector boson production
have contributed to the uncertainty in the measurement of the mass of
the $W$ boson ($M_{\rm W}$)~\cite{d0W,cdfW}. Diboson, top quark, and
Higgs boson production all have single and di-electron backgrounds from
$W$ and $Z$ boson production that will be more constrained through a
precise measurement of $Z/\gamma^*$ production properties. Also, the
universality of the resummation approach requires further experimental
testing, with implications ranging from the impact on the precise
determination of $M_{\rm W}$, to the production of Higgs bosons and
di-photons~\cite{higgs,RESBOS}.

High mass Drell-Yan $e^+e^-$ pairs are experimentally distinctive: the
electrons typically have large $E_{\rm T}$'s, are separated from each
other in $\eta$ and $\phi$, and tend to be separated from jets and
other activity in an event. In Run~I, CDF and D\O\ collected such
events with electrons in the central ($|\eta_{\rm det}|<1.1$) and forward
($1.1<|\eta_{\rm det}|<2.4/2.5$) regions, providing a coverage in the
$e^+e^-$-pair rapidity of up to $|y|\sim 3$. Since the online and off-line
electron identification efficiencies are comparable (for both
experiments), additional off-line requirements primarily enhance the
rejection of background from QCD multijet events that were
mis-measured as electrons. At CDF, the most powerful discrimination is
provided by tracking and precise electron shower centroid
measurements. To cope with the large jet backgrounds in the forward
detector region, CDF developed the SVX-Plug tracker~\cite{SVXPlug} and
improved the matching of tracks in the vertex tracker with calorimeter
shower positions during Run~I.  Applying such forward tracking
techniques to both $e$'s of a pair reduced the backgrounds from about
10\% to a percent or less. At D\O, the most background rejection was
obtained by additional isolation and shower-shape requirements,
reducing the backgrounds from $\sim 10-15\%$ to $4-7\%$, depending
upon fiducial region. With the addition of central magnetic field and
enhanced tracking in the Run~II detector, the background is expected
to be reduced even further.

To allow direct comparisons of experimental results with QCD
predictions, the experimental results are fully corrected for detector
acceptance, experimental efficiencies, and detector resolution
effects.  For the Run~I data, CDF and D\O\ used Monte Carlo
simulations to determine the corrections for the detector
efficiency, acceptance, and resolution as a function of $p_T$ and
$y$. CDF generated the di-electron signal using {\small
PYTHIA}~\cite{Pythia6104}, processing them with the CDF detector
simulation and reconstruction programs. Additionally, they used
{\small PHOTOS}~\cite{Photos20} to simulate final state QED radiation
from the $\gamma^*/Z \rightarrow e^+e^-$ vertex. {\small PYTHIA} and
the detector simulation were tuned to obtain satisfactory agreement
with data~\cite{CDFZptRunI}. D\O\ generated $\gamma^*/Z$ events using the
resummed prediction from {\small LEGACY}~\cite{legacy}, smearing the
decay electrons with a parameterized detector simulation which
included final-state QED radiation corrections. The parameters and
resolutions in the detector simulation were tuned to obtain agreement
with data~\cite{D0ZptRunI}.

In general, the predictions for vector boson production do not yet
include QED effects, therefore the experimental corrections must
attempt to account for them. At a minimum, final state QED radiation
must be included because the effects are large~\cite{QEDRadCor}, and
CDF and D\O\ included such corrections in their Run~I measurements. As
there is yet no numerical implementation of the QED corrections analogous
to the QCD 
soft gluon resummation formalism, initial state QED effects have not
been considered. Initial state QED radiation effects in $p_{\rm T}$
are expected to be similar to those in QCD, but since the coupling is
much smaller, the $p_{\rm T}$ distribution due to QED effects should be much
softer. It is expected that resummation of the initial state
soft-photon emission can be implemented similarly to the soft gluon
case~\cite{QEDRadCor}.

Using 110~pb$^{-1}$ $e^+e^-$ data from Run~I, both CDF and D\O\ have
measured the Drell-Yan cross section $d\sigma/d p_{\rm
T}$~\cite{CDFZptRunI,D0ZptRunI} and CDF has measured
$d\sigma/dy$~\cite{CDFyMuRunI,CDFyElRunI}. For both CDF and D\O, the
measurement error at the peak of the $p_{\rm T}$ distribution ($\sim
3$~GeV/$c$) was $\sim 6$\%. The measurement error at $|y|=0$ was $\sim
5$\%. The combined efficiency and acceptance for these measurements was
$\sim 33$\%. 

The dominant systematic uncertainties in these measurements are the
efficiency and background corrections. Generally, the uncertainty in
shape is more problematic than the uncertainty in overall
normalization. As mentioned, the level of background is expected to be
reduced in Run~II with the enhanced tracking available in both
detectors. This will also reduce the effect of the
normalization-uncertainty on the final measurement. The uncertainty in
the shape of the background is dominated by statistics from events
that satisfy multijet and direct-$\gamma$ triggers, which also satisfy
the kinematic and fiducial requirements necessary for the $Z/\gamma^*$
analysis. Again, the enhanced statistics of Run~II should allow this
uncertainty to by reduced. The overall electron identification
efficiency is well-known in both experiments--$\delta\epsilon\sim
0.5$\%. The shape as a function of $p_{\rm T}$ is dominated by the
isolation requirement being spoiled by jet activity nearby the
electron cluster. Understanding this effect requires either excellent
{\small GEANT}-type Monte Carlo or a great number of di-electron events
from single-electron triggers, so one can investigate the effects of
hadronic activity on electron isolation in an unbiased manner. We
expect to see improvements in both areas in Run~II, hopefully reducing
the uncertainty in the shape of the efficiency as a function of $p_{\rm
T}$ from $\sim 3-5$\%\ to $\sim 1$\%.

\begin{figure}
\begin{center}
\mbox{\epsfxsize=8cm \epsffile{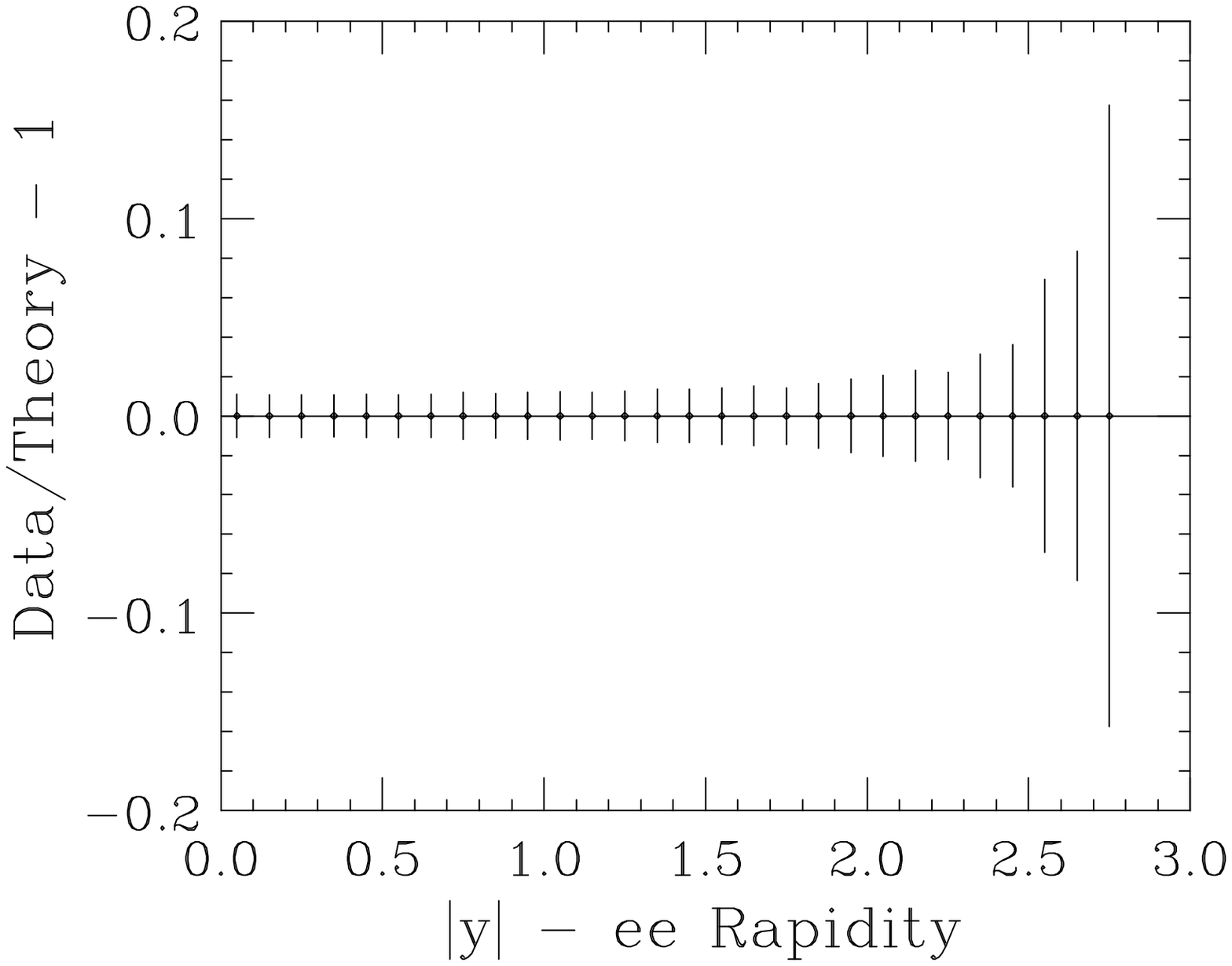}}
\end{center}
\caption{The expected CDF Run~II measurement error on $d\sigma/dy$ of $e^+e^-$
    pairs in the mass range 66~-- 116~GeV/$c^2$. ``Data/Theory'' has been
    arbitrarily set to unity. The error is for 2~fb$^{-1}$.}
\label{dsdyerr}
\end{figure}

\begin{figure}
\begin{center}
\mbox{\epsfxsize=8cm \epsffile{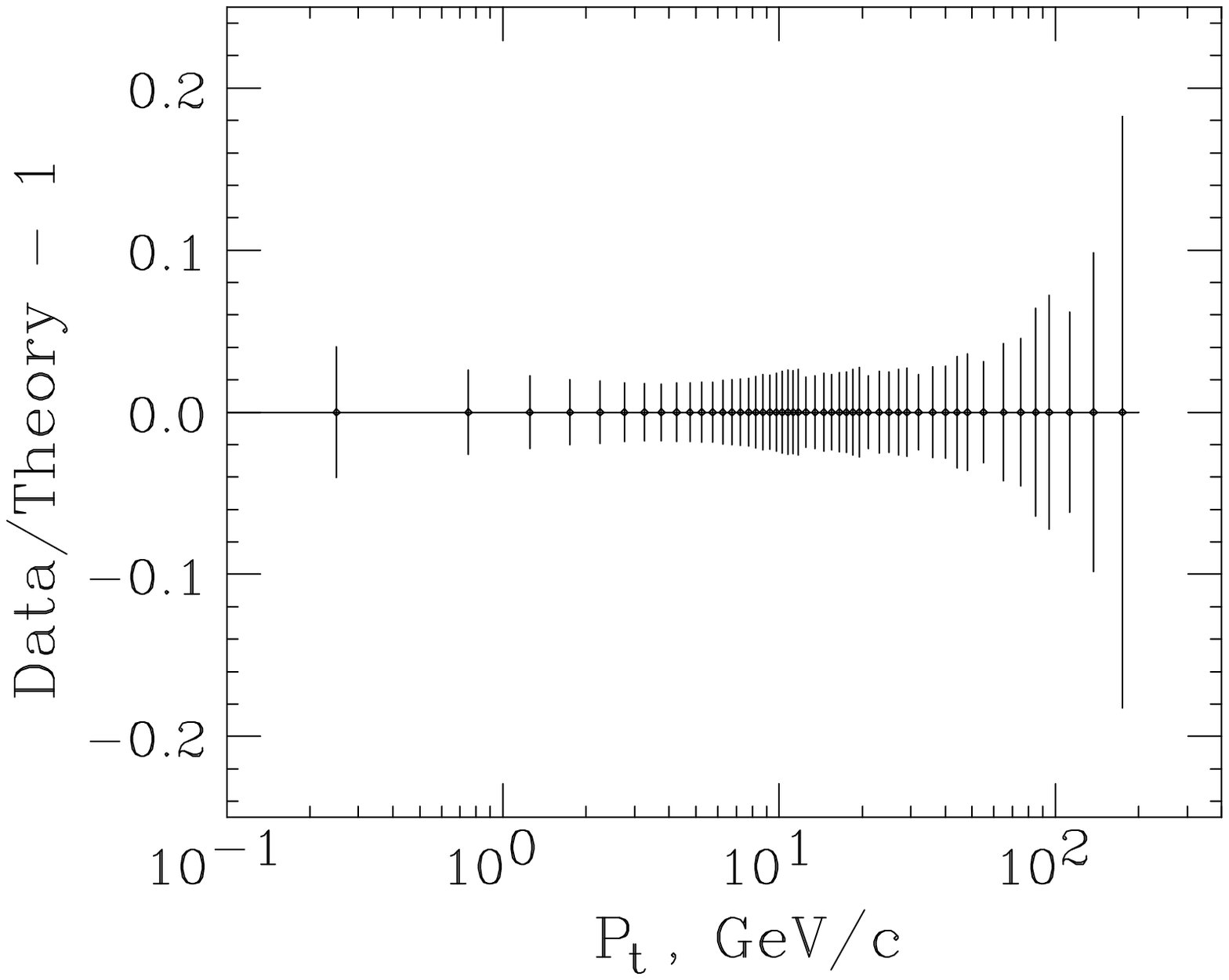}}
\end{center}
\caption{The expected CDF Run~II measurement error on $d\sigma/d p_{\rm T}$
    of $e^+e^-$ pairs in the mass range 66~-- 116~GeV/$c^2$. ``Data/Theory''
    has been arbitrarily set to unity. The error is for 2~fb$^{-1}$.}
\label{dsdPterr}
\end{figure}

The expected Run~II measurement errors for the $d\sigma/dy$ and
$d\sigma/d P_{\rm T}$ measurement can be estimated from the Run~I
measurement errors by assuming that the statistical (``Stat'') and
systematic (``Syst'') errors scale as follows:
\begin{eqnarray*}
  {\rm Stat} \rightarrow 1/\sqrt[2]{N_{\rm ev}} \\
  {\rm Syst} \rightarrow 1/\sqrt[4]{N_{\rm ev}}
\end{eqnarray*}
where $N_{\rm ev}$ is the number of events in a bin. The scaling of
the systematic uncertainties should be considered a rough
parameterization. Scaling the Run~I uncertainties from CDF to an
integrated luminosity of 2~fb$^{-1}$, we obtain predictions for the
total measurement uncertainty in $d\sigma/dy$ (Fig.~\ref{dsdyerr}) and
$d\sigma/d p_{\rm T}$ (Fig.~\ref{dsdPterr}).

With an expected precision of a few percent, the Run~II data can
provide even more stringent constraints on all aspects of our
understanding of $Z/\gamma^*$ productions, with important implications 
on our understanding of resummation in pQCD, and  the precision
determination of $M_{\rm W}$. The precision of the
Run~II $d\sigma/dy$ and $d\sigma/d p_{\rm T}$ measurements will most
likely be limited by systematic uncertainties from corrections to the
detector acceptance/efficiency and resolution. These uncertainties
will be constrained through the improved accuracy of Monte Carlo detector
simulations, and the additional data available in general during
Run~II.

\subsection{Offstream Searches for Vector Bosons }

\subsubsection{Introduction }

Since their discovery in 1983~\cite{rubbia1,rubbia2,wdiscua2,zdiscua2}, 
$W$ and $Z$ bosons have been studied
at hadronic colliders only via their leptonic decays. As a matter of fact
the hadronic decays of these particles are generally so difficult 
to separate from the huge QCD backgrounds that, after 
the extraction of a nice mass bump in the jet-jet mass distribution 
by the UA2 collaboration in 1987~\cite{UA2WZ1,UA2WZ2}, they have laid dormant 
for quite a while as an electroweak physics topic.

\begin{figure}[h!]
\centerline{\epsfig{file=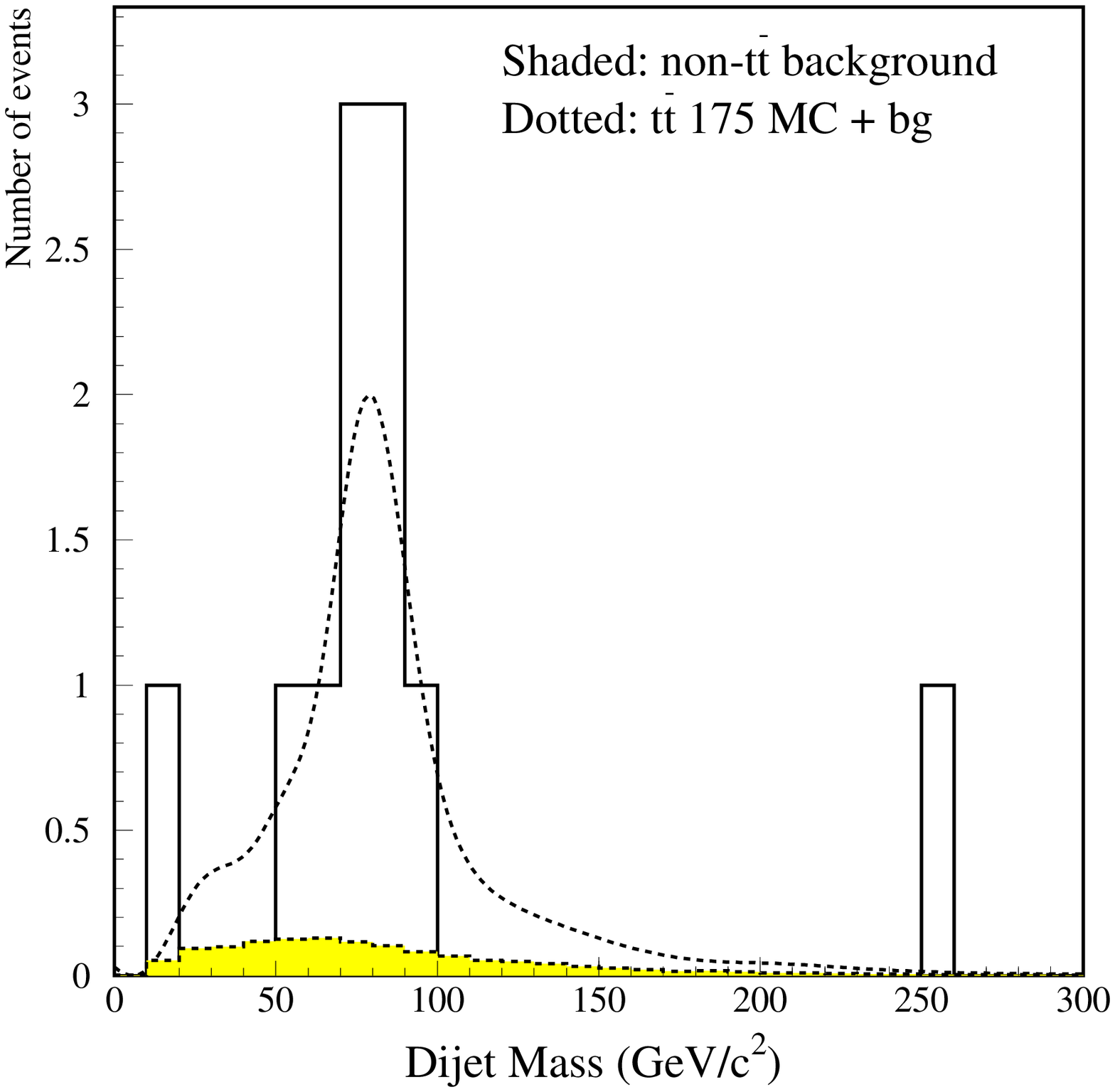, bb= 20 130 540 660,width=8cm,
clip=}}
\caption {The jet-jet mass distribution of untagged jets in
              the very high purity single lepton $t \bar{t}$ 
              candidates sample collected by
              CDF requiring two additional $b$-tagged jets. }
\label{f:wjj}
\end{figure}

Things have started to change with the increase of collider 
luminosity and dataset size. 
During Tevatron's Run~I, hadronic $W$ decays have been successfully
used by CDF and D0 in the discovery and measurement of the top quark
properties both in the single lepton and fully hadronic final states, and 
a handful of jet-jet masses peaking at 80~GeV have been extracted from
a subset of high-purity $t \bar{t}$ events (see Fig.~\ref{f:wjj}).
More recently, a signal of $Z$ decays to $b$-quark pairs has emerged
in the CDF data (Sec.~\ref{s:zbbrun1}).

With Run~II sample sizes it will be possible to search for more such
hadronic signals, and some of them are expected to start becoming 
useful tools for other physics advances. In fact, 
their potential as calibration tools for the jet energy measurement is high, 
provided that they can be collected by unbiasing triggers. 
This seems particularly likely in the
case of $Z \to b \bar{b}$ decays, where the background has been shown 
to be reducible to a manageable size and the hardware tools for 
collecting them with good efficiency and small bandwidth concessions 
are now available; moreover, for $b$-quark jets the absolute energy
scale cannot be fixed by photon-jet balancing techniques, due to the rarity
of events with a photon recoiling against a single $b$-quark jet: a $Z$ peak
may then really be our best chance for that purpose.
Hadronic $W$ decays will also be an ideal calibration tool in $t \bar{t}$
events, but efforts need to be spent on finding their signal in independent
data samples. These may be provided by diboson production processes,
where triggering and background issues are less problematic.
 
\subsubsection { Hadronic Decays of $\boldmath{W}$ Bosons }

Searches for a $W$ mass signal in inclusive jet triggers have been
fruitless in Run~I data. With respect to the lower energy $S\bar{p}pS$
collider, the Tevatron's higher center-of-mass energy is a disadvantage
for once, because in the face of a four-fold increase in signal cross
section the background from QCD processes increases by
an order of magnitude, due to its steep behavior with respect to parton $x$. 
Moreover, no dedicated low-$E_T$ jet triggers were devised either at CDF
or \Dzero during Run~I, given the experiments' focus on the high energy
frontier; 
at the very end of the run, however, a sector of CDF's central tracking chamber
became inoperative due to a broken wire, which allowed 1.9~pb$^{-1}$ of data 
to be collected by a high-bandwidth 12~GeV dijet trigger. The data thus
gathered did not allow the extraction of a $W$ peak either, but lends itself
to fruitful extrapolations to Run~II.

It seems reasonable to investigate the collection of events with very
low jet $E_T$ by special low-luminosity runs, which might become an
attractive option in the event of temporary inoperativeness of tracking 
detectors. Here we examine a scenario where CDF~II or \Dzero gathers 
some 100~pb$^{-1}$ of unprescaled data collected with a trigger
requiring two calorimeter towers above 4~GeV at Level~1 and two jets 
with $E_T>10$~GeV at Level~2, back-to-back in phi --about the smallest 
thresholds that do not saturate the bandwidth. 
About two million $W/Z \to jj$ decays could be then collected,
which would be reduced to 200,000 after optimized kinematic cuts requiring
two central back-to-back jets and little extra-jet activity. 
Toy Monte Carlo studies suggest that in such a scenario the $W$ mass 
could be fit with a $\sim 0.5$~GeV uncertainty, provided the availability 
of a prescaled sample with looser requirements at Level~2: this would yield 
an understanding of the absolute jet energy scale of the detectors to 
better than $1\%$.
Such a dataset could then clearly be used also for excellent tests of 
optimization of jet algorithms, and thus offer benefits to any
search for hadronically decaying massive objects.

\subsubsection{$\boldmath{Z}$ Decays to $\boldmath{b}$-Quark Pairs in
Run~I \label{s:zbbrun1}} 

Thanks to the several million $Z$ decays to $b$-quark pairs collected by the
LEP I and SLD experiments since 1992, the physics of these decays
is extremely well studied and understood. At a proton-antiproton collider 
that particular process had not been identified before, though; 
therefore the extraction of a signal in Run~I
data was interesting in its own right. Moreover, the knowledge of how to
extract a $Z$ peak enables a careful design of a dedicated trigger for Run~II, 
which may allow us to collect a large sample of these events, from which
the mass distribution can be fit and thus insight can be obtained on 
the absolute energy scale for $b$-quark jets, substantially
reducing one of the critical sources of systematic uncertainty in the 
top quark mass measurement.
\begin{figure}[h!]
\centerline{\epsfig{file=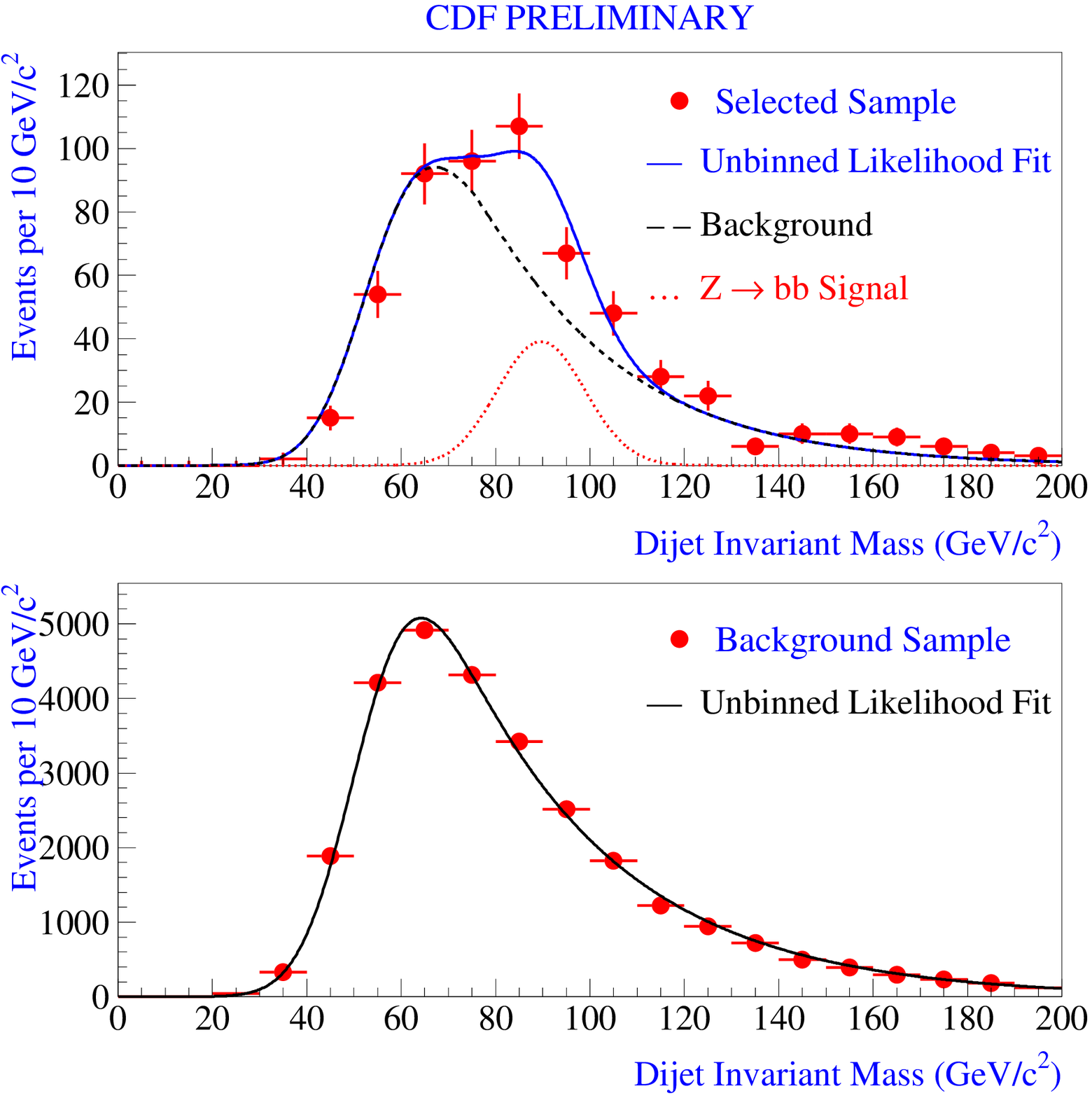, bb= 0 0 540 540,width=8cm,
clip=}}
\caption { Top: the $Z \to b \bar{b}$ peak in the signal sample; 
              bottom: a signal-depleted sample is used to extract 
              the background shape.}
\label{f:zbb_peak}
\end{figure}

A $Z \to b \bar{b}$ signal was extracted in Run~I CDF data using kinematic
tools and $b$-quark vertex tagging. 
Starting from a dataset of about five million events enriched of 
$b$-quark decays collected by a single muon trigger, a very tight 
kinematic selection was devised, which increased the signal to background 
ratio by three orders of magnitude.

The main background to the $Z$ decay to $b$ quarks
is due to direct QCD production of a $b\bar{b}$ pair, while 
non-$b$ backgrounds yielding muons can be completely eliminated by requiring 
the presence of a secondary vertex in each of the two jets; this cut 
also reduces the flavor excitation and gluon splitting contributions
to $b$ quarks in the sample quite effectively. 
Most of the direct $b\bar{b}$ pairs are produced at the Tevatron
by a gluon fusion process, whose high color charge in the initial state
and color flow topology are distinctive characteristics. To exploit the smaller
probability of QCD radiation in the signal, the two leading jets were
required to be back-to-back in $\phi$, and the sum of all other 
calorimeter clusters in the event was required to be smaller than 10~GeV. 
These cuts selected 588 events, whose jet-jet mass distribution was
fit to the sum of QCD background and $Z$ decay to yield a signal of
$91 \pm 30$ events, with a S/N ratio at the $Z$ peak equal to 1/3;
the fit resulted in a $Z$ mass of $90.0 \pm 2.4$~GeV 
(see Fig.~\ref{f:zbb_peak}).
If the same analysis should be replicated with 20 times more
statistics and no detector improvements, this would yield
a relative error in the $b$-jet energy scale smaller than $1\%$. The 
picture could be even rosier as far as statistics goes, due to the extended
lepton coverage and improved silicon tracking the Run~II detectors will
be endowed with, but the strong bias due to the triggering lepton will 
make these findings of difficult use for generic $b$-quark jets. 

Besides its possible use as a calibration tool, however, a 
$Z \to b \bar{b}$ peak provides a fine testing ground for 
algorithms designed to improve the jet energy measurement for $b$ jets, 
which is one of the critical 
points for the discovery of an intermediate mass Higgs boson in Run~II. 
The $b\bar{b}$ final state is the dominant one in Higgs boson decay 
if $M_H < 135~{\rm GeV}/c^2$~\cite{higgswg}. Our ability to extract this
particle from the large QCD background in Run~II (for instance in the 
$W b \bar{b}$ final state, when associated $WH$ production is sought)  
will therefore depend critically on the resolution we can attain 
on the Higgs boson mass as reconstructed from
the measured $b$-quark jet energies: both the possibility to see a bump
in the mass spectrum of jet pairs associated to a leptonically decaying 
$W$ bosons, and the alternative option of applying a mass window cut as 
a selection tool for these events, will strictly depend on the actual 
mass resolution.

The expected resolution for a generic jet-jet resonance at CDF and D0 
was roughly $\sigma_{M_{jj}} = 0.1 M_{jj}$ in Run~I. 
A relative improvement of this number by $30\%$ would significantly 
extend~\cite{higgswg} our discovery reach for the Higgs boson in Run~II. 
In order to achieve that improvement we must study in detail the 
characteristics of $b$-quark
jets emitted in the Higgs decay, and use to their utmost the large amount
of available information provided by the various detector components 
\Dzero and CDF~II are made up of. For example,
three-dimensional tracking in the new SVX II detector may allow 
CDF~II to infer the momentum of the escaping neutrinos in semileptonic 
$b$-quark decays,
greatly improving the energy measurement of the resulting jets; this
plan will work well in Run~II, given the larger acceptance for charged
leptons from semileptonic decays provided by the new detectors.
Furthermore, the possibility of measuring track momenta to higher rapidity 
will allow a fruitful use of tracking information to improve the 
calorimetric measurement of jets.

A detailed study of the observable
quantities of $b$-quark jets produced in $Z \to b\bar{b}$ decays 
followed by one semileptonic $b \to \mu X$ decay 
have been shown to allow a sizable reduction of the width of the 
reconstructed $b \bar{b}$ peak. 
The quantities found useful for this purpose in the CDF analysis
were the muon momentum, the projection of missing transverse energy 
along the jet axes, and the charged fraction of the jets. 

The muon momentum is needed in the correction of jets containing a 
semileptonic decay of $b$ quarks, because the 
minimum ionizing muons do not contribute linearly to the 
energy measured in the calorimeter. The missing $E_T$, projected along
the jet directions in the transverse plane, provides useful information
on the amount of momentum taken away by the neutrino in the muon jet 
and on possible
fluctuations of the energy measurement in both the muon and the away jet. 
The charged fraction of the
jets, defined as the ratio between the total momentum of charged tracks
belonging to a jet and the energy measured in the calorimeter, also helps
reducing the uncertainty in the energy measurement.

\begin{figure}[h]
\centerline{\epsfig{file=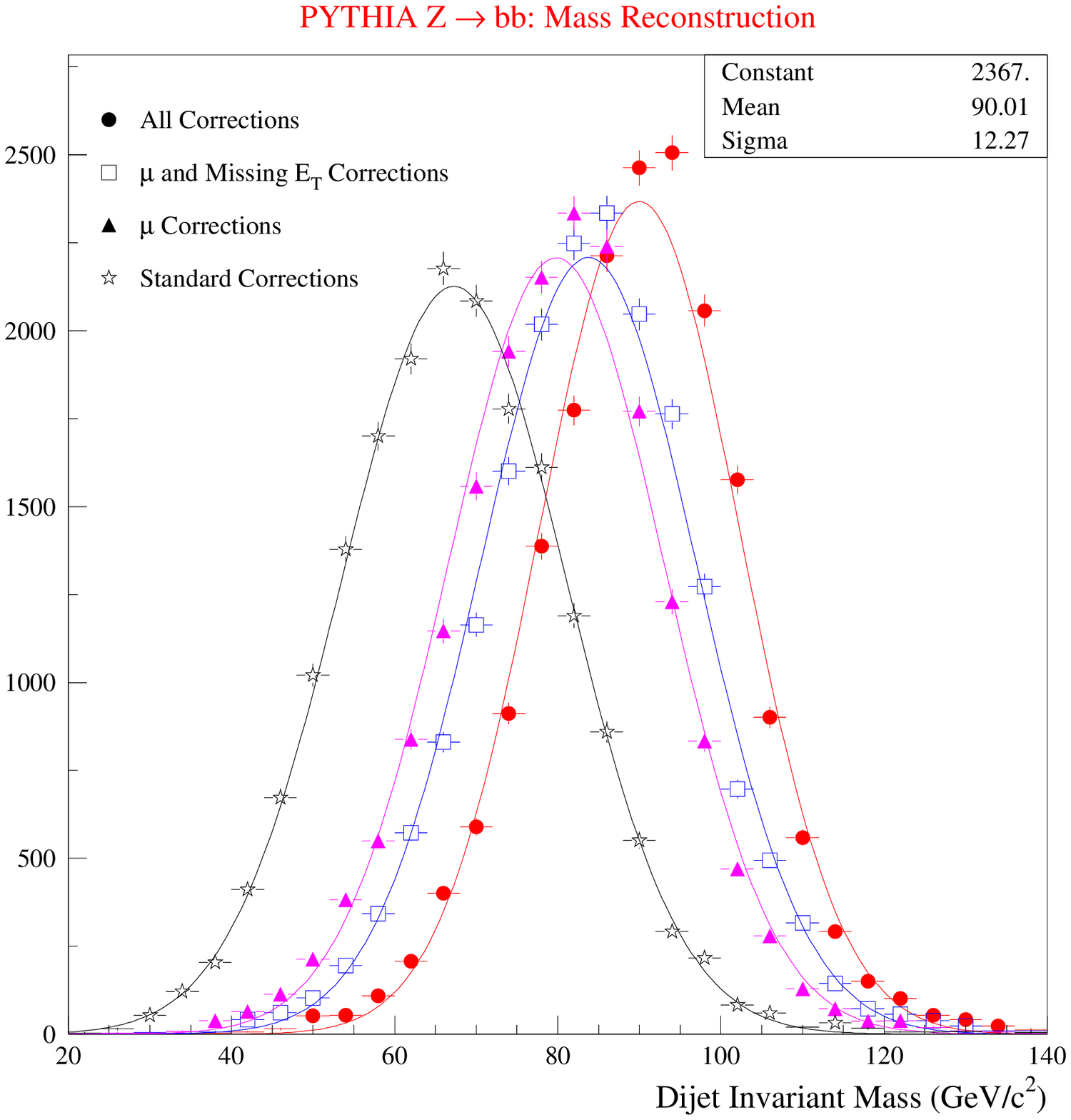, bb= 0 0 540 540,
                    width=8cm, clip=}}
\caption { The four Gaussian fits show the improvement of the mass 
reconstruction for simulated $Z \to b\bar{b}$ events (Pythia V5.7, CDF
detector simulation) when the 
observable characteristics of the $b$-quark decays are properly 
taken into account in the mass reconstruction. \label{f:bb_res}
} 
\end{figure}

By properly accounting for the value of these observables, it was possible
to reduce the relative uncertainty in the dijet mass measurement, 
$\sigma_M / M_{jj}$, by nearly $50\%$ (see Fig.~\ref{f:bb_res}).
 
If the alternative plan (described in the following section) based on 
collecting $Z \to b \bar{b}$ decays by triggering directly on secondary 
vertices in jets at Level~2 should fail, CDF~II will anyways be able 
to observe a peak of several thousand events in the inclusive lepton datasets
by simply replicating the Run~I analysis. These peaks will not allow
a precise calibration tool for inclusive $b$-quark jets, given the
biasing semileptonic decay of one of the two jets, but they will anyways
be extremely useful as a testing ground for the algorithms now under
development which aim at reducing the jet energy resolution.

\subsubsection{$\boldmath{Z}$ Decays to $\boldmath{b}$-Quark Pairs in Run~II}

A strategy to collect $Z \to b\bar{b}$ events in Run~II has been 
studied having in mind the CDF trigger configuration.
We tried to keep the trigger requirements as simple as possible and optimize 
them in order to have 
acceptable trigger rates (less than $2\%$ of the total bandwidth at each level)
and maintain a good signal efficiency.

Most $Z \to b\bar{b}$ events should contain two reconstructable secondary 
vertices in the 
final state.  In Run~II it will be possible to trigger on secondary vertex
information. Exploiting this feature, it should be possible to collect a 
sample of $Z \to b\bar{b}$ events without particular requirements
on the $b$ decay. One purpose of this dataset is to provide a sample to 
calibrate the calorimeter energy scale for jets containing $b$ quarks, which 
may limit our ability to measure the top mass. Therefore the trigger path 
should bias the energy scale measurement as little as possible. 

At Level~1 the XFT (eXtremely Fast Tracker) information will be 
available: transverse momentum $P_{T}$, azimuthal angle $\phi$ and 
charge sign of particles crossing all of the Central Outer Tracker 
(COT) layers.
We require the presence of two central high $P_{T}$ tracks in opposite 
hemispheres. Cuts are chosen on the track $P_{T}$ and $\Delta \phi$  
to maximize the statistical significance ($S/\sqrt{B}$).
The track $P_{T}$ cuts are 6~GeV and 4~GeV, with  $\Delta \phi$ $>$
150$^o$.

At Level~2 the rate is reduced by requiring the tracks to have a finite 
impact parameter. The SVT (Silicon Vertex Tracker) processor will
provide this information. The best significance is found by requiring two SVT
tracks with $120\ {\mu}m < |d| < 1000\ {\mu}m $. 

The computing power of the Level~3 processors should allow the reconstruction
of secondary vertices online. 
We require two jets, each with uncorrected  $E_T>10$~GeV in a cone of
radius R = 0.7, and check that the jets contain two displaced vertices.

 A summary of signal efficiency, trigger cross section and trigger rate
at the three trigger levels is reported in Table~\ref{t:zbbrun2} for a typical 
instantaneous luminosity ${\cal L } = 1.4 \cdot 10^{32}{\rm cm^{-2}s^{-1}}$. 

\begin{table}[hbt] 
 \caption{Summary of the $Z \to b\bar{b}$ trigger efficiency,
              cross section, and rate at each level.\label{t:zbbrun2}}
\footnotesize
\begin{center}
  \begin{tabular}{||l|c|c|c||}  \hline\hline
 Trigger & $\epsilon_{S}$ ($\%$) & $\sigma_{T}$
& $R_T$ (Hz) \\ \hline \hline
Level~1        & $17.0 \pm 0.2$ & 
($5.7 \pm 0.3$) $\mu b$ & $800 \pm 42$ \\ \hline            
Level~2        & $3.10 \pm 0.03$ & 
($32 \pm 3$) $nb$ & $4.4 \pm 0.4$ \\ \hline
Level~3        & $2.4 \pm 0.1$ & 
($3.2 \pm 0.9$) $nb$ &  $ 0.5 \pm 0.1$ \\ \hline \hline
  \end{tabular}  \end{center}
\end{table}   

Since $\sigma (p\bar{p} \to Z \to b\bar{b}X) \simeq 1$~nb, about 
$2\cdot 10^6$ $Z \to b\bar{b}$ events 
will be produced in Run~II, assuming 2~fb$^{-1}$ of data collected
in two years. We expect to collect 48,000 signal events in a total 
of about $6.4 \cdot 10^6$ with this trigger,  corresponding
to a S/B of 0.0075 and a significance $S/\sqrt{B} \simeq 19$.

\begin{figure}[hbt]
\begin{center}
\centerline{ \epsfig{file=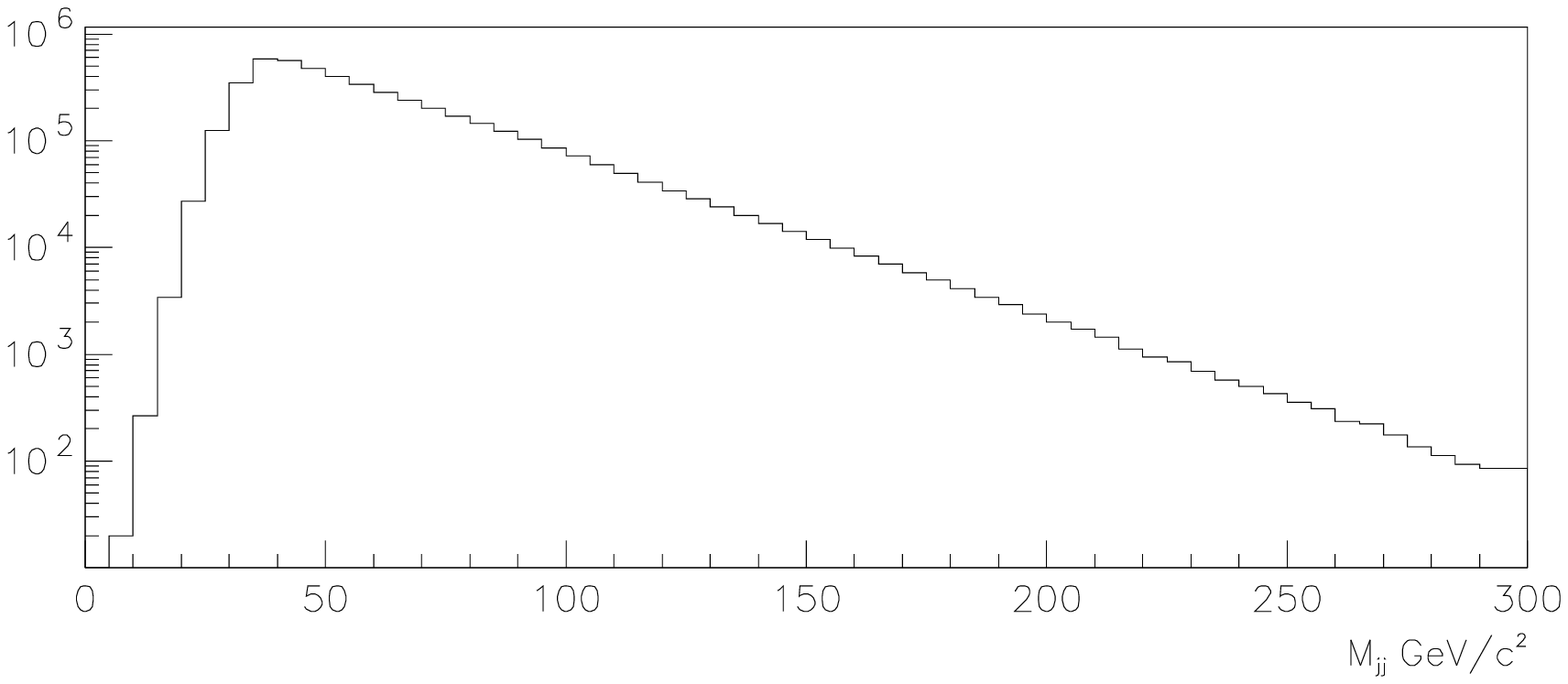,bb=0 0 540 235, width=9cm, 
                      height=3.5cm, clip=}}
\caption{$M_{jj}$ distribution expected from Run~II: signal+background.}
\label{f:dijetspectrum}
\end{center}
\end{figure}
It is also advantageous to require the presence of no additional
jets with $E_T > 10$~GeV in the event, either at trigger level or 
at a pre--analysis level.
This would increase the signal fraction to 0.013 and the significance
to $S/\sqrt{B}= 24$.
We simulated the signal extraction procedure under these assumptions.
We searched for the $Z \to b\bar{b}$ signal in the jet--jet
invariant mass distribution $M_{jj}$. We expect to see an enhancement 
corresponding to the $Z$ mass, since the statistical significance of 
the signal is high. Outside of the signal region the $M_{jj}$ spectrum should 
be approximately described by a decreasing exponential.

The PYTHIA Monte Carlo is used to model the $Z \to b\bar{b}$
invariant mass spectrum. The background $M_{jj}$ distribution expected
for the background is inferred from generic Run~I dijet data by 
assuming that the request of secondary vertices does not modify its shape. 
The $M_{jj}$ distributions for signal and background normalized 
to 2~fb$^{-1}$ have been added (see Fig.~\ref{f:dijetspectrum}). 

\begin{figure}[hbt]
\begin{center}
\centerline {\epsfig{file=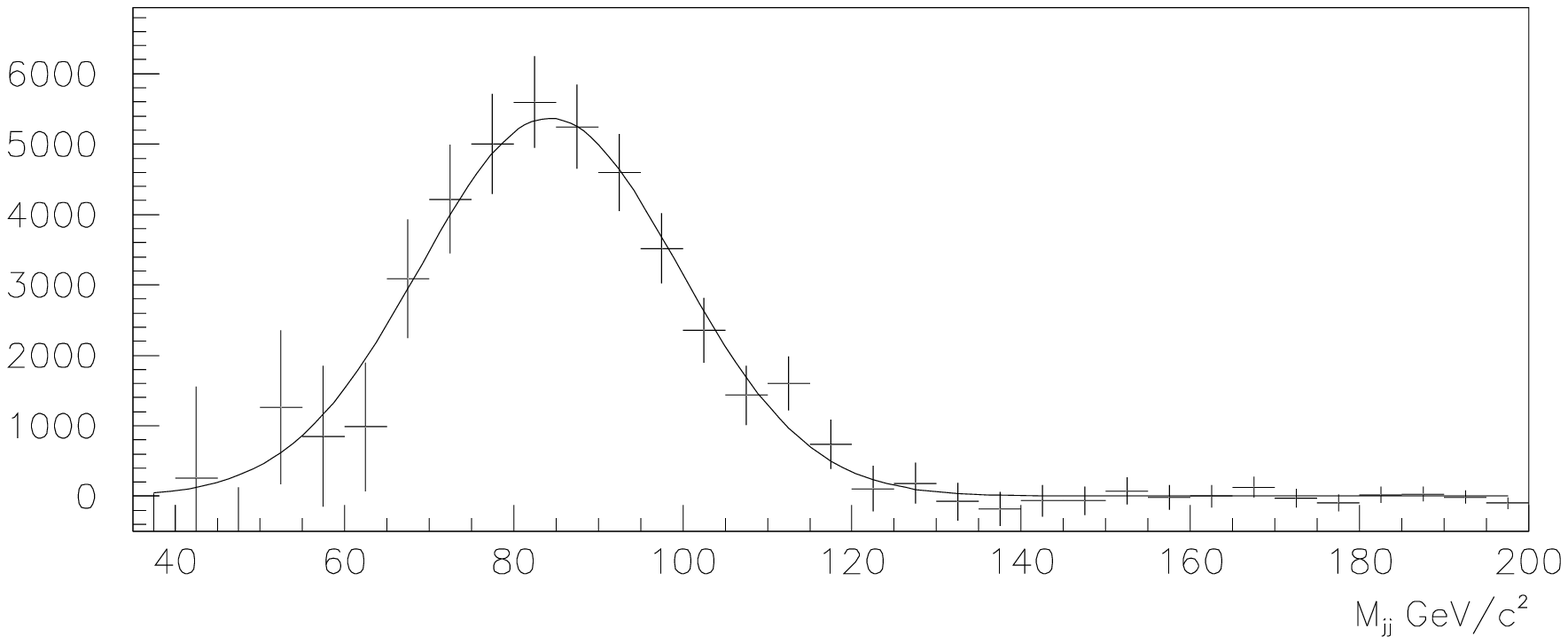,bb=0 0 540 235, width=9cm,height=3.5cm, 
                     clip=}}
\caption{Background-subtracted $M_{jj}$ distribution for $Z$ events 
               obtained from a pseudo-experiment. }
\label{f:pseudo}
\end{center}
\end{figure}

Because of the very low $S/B$ ratio, we cannot distinguish 
by eye the signal peak in Fig.~\ref{f:dijetspectrum}. 
We conducted 1000 pseudo-experiments, varying every time the bin content 
according  to a poissonian distribution with mean equal to the bin content 
itself. 
For each pseudo-experiment the region outside the $Z$ peak was fitted
with a decreasing  
exponential. The value of this function was subtracted, 
bin by bin, from the total spectrum. The results obtained in one of these
pseudo-experiment subtractions is shown in Fig.~\ref{f:pseudo}, together
with a gaussian fit to the excess found. 
With this method we typically obtain for the jet-jet invariant mass peak
position $M_{meas}$ = 84.0~GeV/c$^2$, with a 
statistical uncertainty of 1.5~GeV/c$^2$ on $M_{meas}$.\footnote{The 
jet energy corrections applied here are not optimized for $b$~jets,
whose higher mass and decay properties are sensibly different from those
of generic light-quark and gluon jets: 
the invariant mass peak is thus about 7~GeV lower than the true $Z$
mass.}

The interval of $\pm 2 \sigma$ around the $Z$ peak (55~GeV/c$^2 < M_{jj}
< 115$~GeV/c$^2$) contains about $93\%$ of the signal events and $25\%$ 
of the background events. Therefore in this region $S/B = 0.05$ and the 
statistical significance is $S/\sqrt{B}= 44$.

The error on the $Z$ mass can be directly translated into an error on the
$b$-jet energy scale. Measuring the $Z$ mass with an error 
of the order of 1.5~GeV/c$^2$ will allow to determine the $b$--jet absolute 
energy scale with an uncertainty of $1.7 \%$.

\subsubsection{Other Hadronic Signals\label{s:other} }

Besides its interest for the study of trilinear gauge boson couplings,
discussed in Sec.~\ref{s:diboson}, associated production of two vector
bosons yielding a leptonic and a hadronic decay may provide 
additional handles for the physics of hadronic resonances, both because 
of the ease of collecting these events with good efficiency in 
high-$P_T$ lepton triggers and because of the
larger signal to background ratio with respect to single
boson production processes, due mainly to the reduction of background 
processes with gluons in the initial state. 

$WW$ production is the best example: in a sample of 5~fb$^{-1}$, 
for instance, about 3000 $p\bar{p} \to WW \to l \nu jj$ events can 
be collected by applying standard cuts on the leptonic decay products
and requiring two jets with uncorrected $E_T>15$~GeV; the signal to
noise ratio  
is then close to 1/40 before any optimized selection. Thence an 
observation of the $W$ peak in the dijet mass distribution will be 
relatively easy to obtain. Systematic effects in the mass fits 
due to the low S/N ratio may make this sample of little impact as a 
source of knowledge of jet energy scale when compared
to the high-purity $W \to jj$ samples that single lepton $t \bar{t}$ decays
may provide; but the signal may still be extremely useful for the study 
of jet resolutions.

Another process that will be likely observed in Run~II is associated 
$WZ$ production with a subsequent leptonic decay of the $W$ boson 
and a decay of the $Z$ to $b$-quark jets. In 5~fb$^{-1}$ about 500 
such events can be collected by the lepton triggers, from where 
secondary vertex tagging can considerably increase the signal purity.
Despite its small size, this signal may be of fundamental importance
in checking systematic uncertainties in the standard model Higgs boson
search, which will mainly focus on the very same dataset with very
similar analysis cuts.

\begin{figure}[h]
\centerline{\epsfig{file=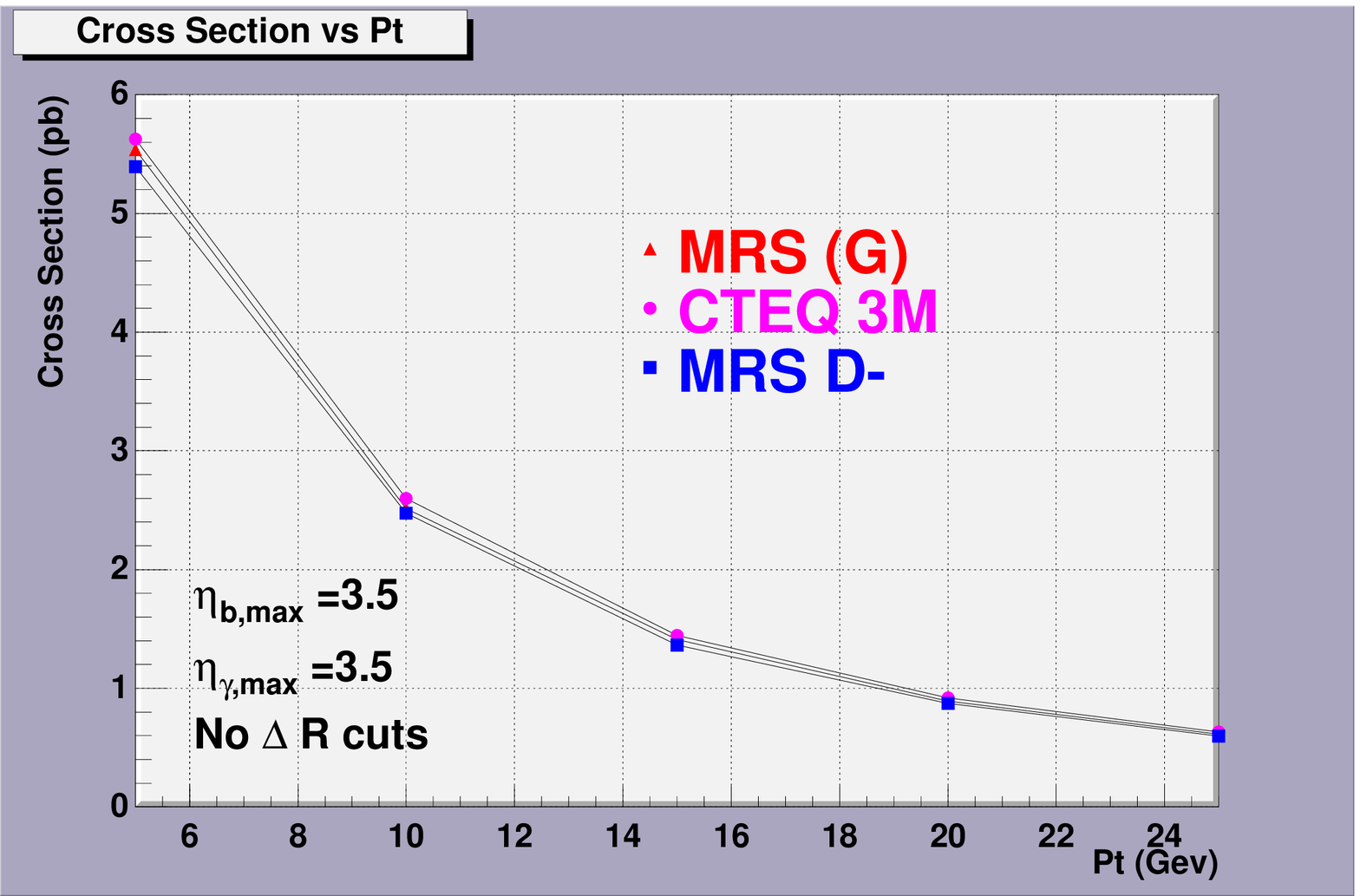, bb= 0 0 588 387,
                    width=8cm, clip=}}
\caption { The cross section times branching ratio for $Z \gamma$
production with subsequent $Z \to b \bar{b}$ decay, for different
choices of PDF sets, 
computed with the Baur-Berger-Zeppenfeld
generator~\cite{baurmc}. \label{f:zgxs} 
} 
\end{figure}
Finally, the possibility of collecting $Z \to b \bar{b}$ in 
photon triggers has been investigated. $Z \gamma$ production
has a small cross section --about 2.5~pb for a photon 
with $E_T>10$~GeV and two $b$-quark jets (see Fig.~\ref{f:zgxs})-- 
but the heavy flavor decay provides a quite distinctive signature, 
and the process could be easily put in evidence
if a sufficiently low trigger threshold were set on the photon transverse 
energy, or alternatively if a $\gamma b$ trigger could be devised. The 
advantage of the process producing an additional photon over inclusive $Z$ 
production lays in the suppression of the irreducible background from
direct QCD $b \bar{b}$ production: since initial
state photons can only be present in quark-antiquark annihilation diagrams,
at Tevatron energies the $b \bar{b} \gamma$ background is suppressed by
roughly an order of magnitude more than the $Z$ signal with respect to the  
searches in inclusive $b \bar{b}$ 
samples. It is however difficult to devise an unprescaled 
trigger capable of a sufficient efficiency for the signal while maintaining 
the total rate at an acceptable level. In particular, a $\gamma b$ 
trigger would have to require at Level~1 both a photon candidate and one
or two  
charged tracks; the tracks and the photon would then need to be 
separated in azimuthal angle, to reduce the rate of fake photon signals 
from QCD events. There is 
currently no plan to devise such a trigger in CDF~II or \Dzero, while the
lowest unprescaled photon triggers will collect events with 
${E_T}^{\gamma}>25$~GeV,
where the $\gamma b \bar{b}$ cross section is only 0.6~pb. With
5~fb$^{-1}$ it 
will be relatively easy to isolate a signal of one or two hundred events
over a similarly sized background, but, given the small size, its exploitation 
appears dubious.

\subsection { Lepton Angular Distributions in $\boldmath{W}$ Boson Decay }
 
Next-to-leading order perturbative QCD predicts that in $W \to l \nu$
decays an angular
distribution of ($1 \pm \alpha_1 \, \mathrm{cos} \theta^* +
\alpha_2\,  \mathrm{cos}^2 \theta^*$)~\cite{mi} should be observed, 
where $\theta^*$ is the polar angle of the decay lepton in the
Collins-Soper frame~\cite{cs}.  In the presence of QCD corrections, the
parameters $\alpha_1$ and $\alpha_2$ are functions of $p_T^W$, the
$W$ boson transverse momentum.  

The measurement of
$\alpha_2$ serves as a probe of  NLO QCD, using the well understood 
$W$-fermion coupling. By probing the spin structure  of $W$ production, this
measurement provides another method that is independent of purely
QCD analyses, while adding 
to the list of measurements using vector bosons
to study NLO QCD. Moreover, the measurement  of the angular distribution of the
decay leptons is also of importance for the $W$ mass measurements,
because the next-to-leading order QCD
corrections to the angular distribution are a non-negligible 
contribution to the $W$ mass.

\begin{figure}[!htbp]
\centerline{\epsfig{figure=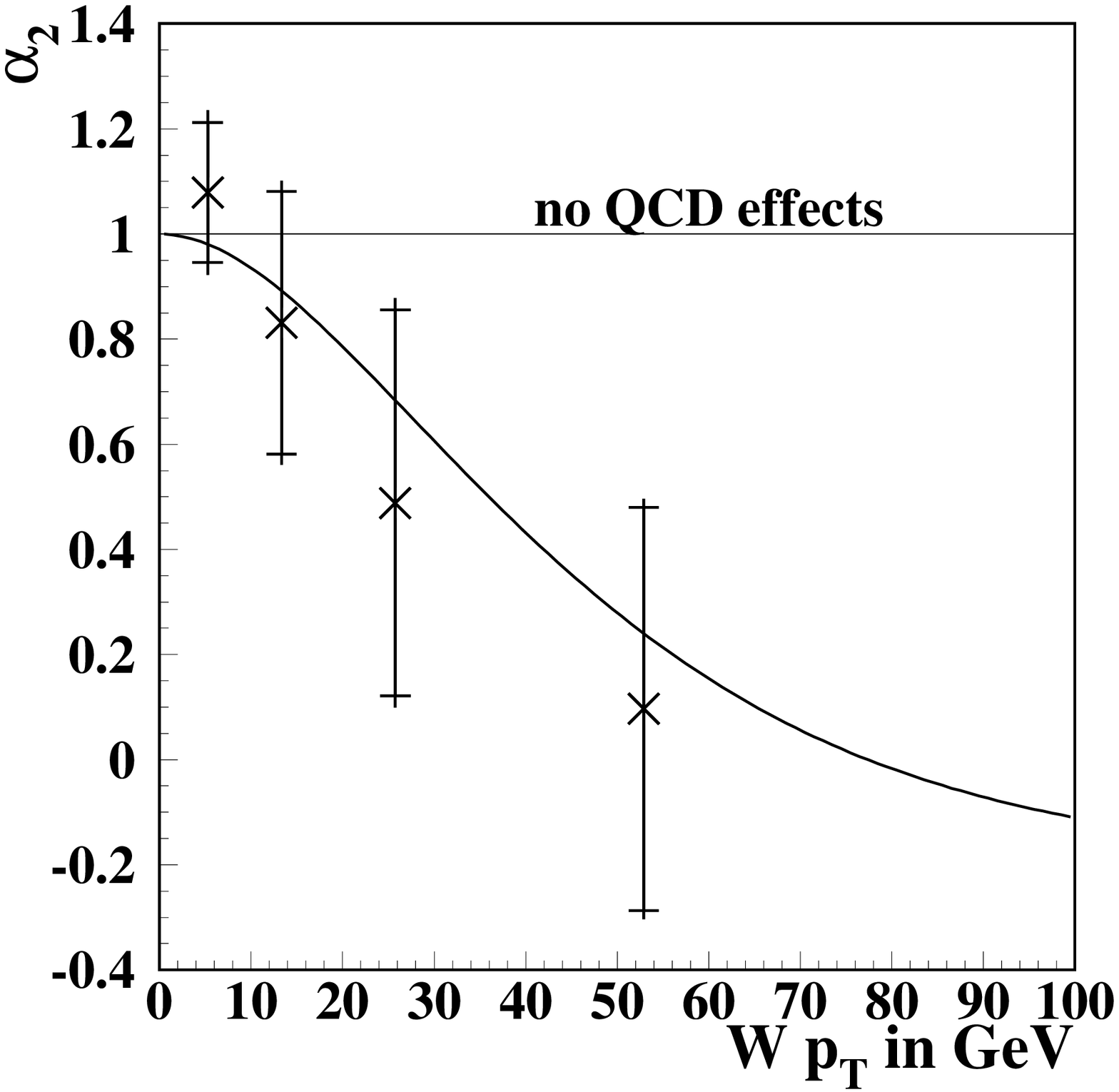,height=8cm,width=8cm}}
\caption{{\Dzero Run~I result: the measured  $\alpha_2$ as a function of \pt 
         with combined statistical and systematic errors compared to the 
         NLO QCD calculation by Mirkes (curve) and calculation in the
        absence of QCD effects (horizontal line). The statistical errors 
        alone are shown as horizontal ticks. \label{fig:RunI}}}
\end{figure}

The measurement of the angular distribution of electrons from $W$ bosons
obtained with Run~I data collected by \Dzero~\cite{ma,thesis} is statistically 
limited (see Fig.~\ref{fig:RunI}). While a calculation that includes
QCD effects is preferred over one that does not, this preference is not strong 
enough to exclude a \pt\ independent angular parameter $\alpha_2$.
With the next collider run starting in the near future, 
it is worthwhile looking at the 
sensitivity of this measurement in Run~II. In the following discussion
we will estimate the size of statistical and systematic errors to this
measurement with 2~fb$^{-1}$ in Run~II.

The expected statistical errors in the determination of $\alpha_2$
should simply scale like the inverse of the square root 
of the number of events. We consequently need to calculate the 
expected number of $W$ boson events under Run~II conditions.
We get a factor of 57 in $W$ boson statistics which breaks down as
follows (see~\cite{tev2000}): 
\begin{equation}
\begin{split}
\frac{N_{W,RunII}}{N_{W,RunI}}= & f_{lum} \times\ f_{2.0\,{\mathrm TeV}}
\times \frac{N_{e}+N_\mu}{N_e} \\
 & \times\ \frac{\epsilon_{tracking}}{\epsilon_{no \: tracking}}
\end{split} 
\end{equation}
where
$f_{lum} = 20$ indicates the increase in luminosity, $f_{2.0TeV}= 1.2$
indicates the increase in $W$ cross section due to the increase of the 
center-of-mass energy from $\sqrt{s}=1.8$~TeV to 2~TeV,
$(N_{e}+N_\mu)/{N_e}=2$ 
is the additional statistics gained by including the muon channel, and
\begin{equation} 
\frac{\epsilon_{tracking}}{\epsilon_{no \: tracking}}=\frac{0.95}{0.8}
\end{equation}
is the increase in efficiency due to tracking capabilities of the upgraded
\Dzero detector.

\begin{figure}[!htbp]
\vbox{
\vspace{-0.2in}
\centerline{\epsfig{figure=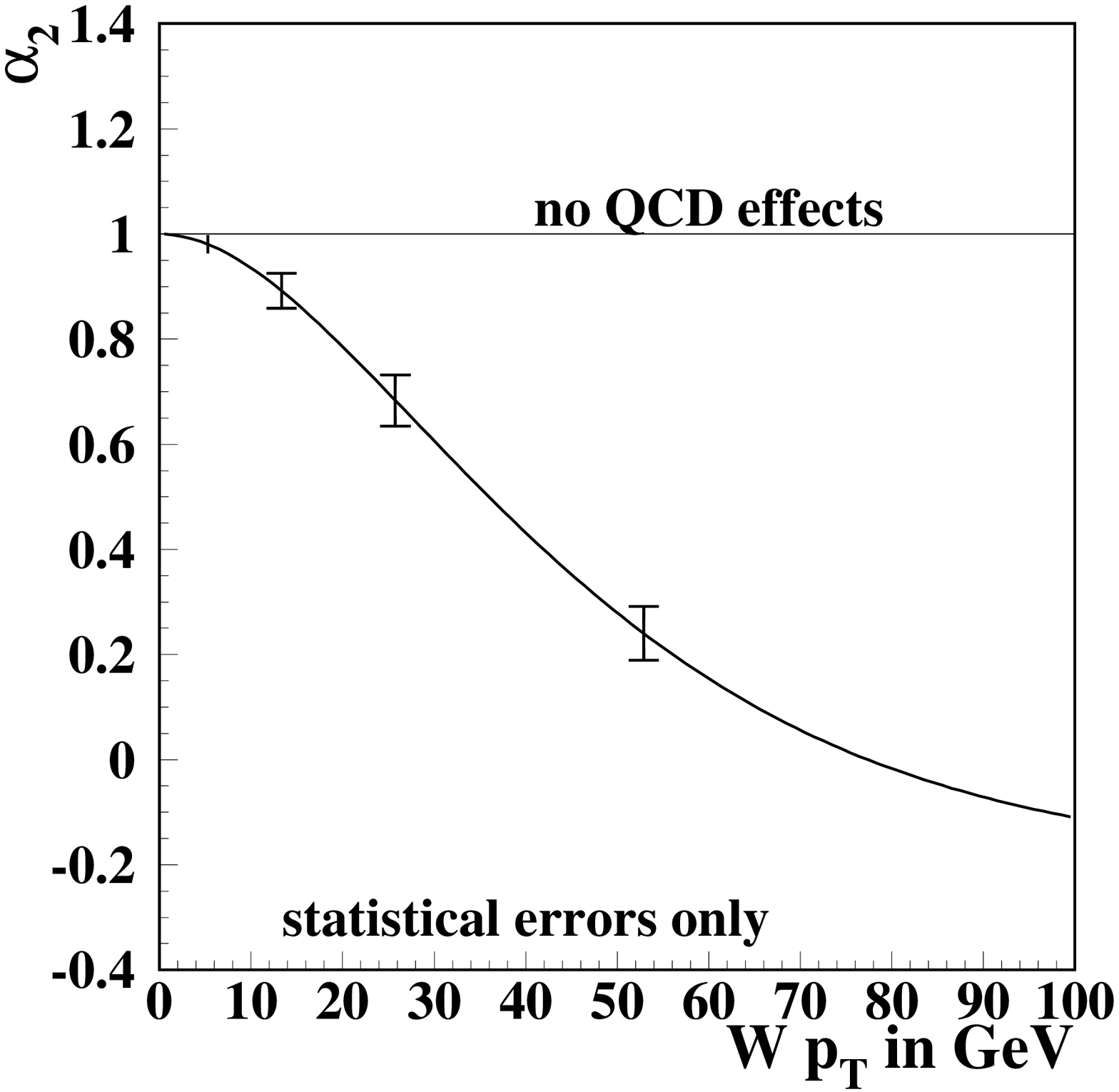,height=8cm,width=8cm}}
\vspace{-0.cm}
\caption{ {Estimated sensitivity of the $\alpha_2$ measurement obtained by 
          scaling statistical errors to Run~II conditions. 
          \label{fig:RunIIstats}}}}
\end{figure}

The statistical errors of the Run~I measurement are therefore scaled by 
$1/\sqrt{57}$, as shown in Fig.~\ref{fig:RunIIstats}. 
The statistical uncertainties to this measurement become quite small,
and a further look at systematic errors is therefore necessary.
Table~\ref{tab:resultsII} shows a summary of statistical and systematic errors
for Run~I and Run~II. Since the modeling of the hadronic  recoil is done 
from $Z$ data, the error due to the hadronic resolution will  scale 
with $Z$ statistics up to a point. 
The estimate of this error is done by scaling the number of
$Z$ events by a factor of $\sqrt{28.5}$ (the same as for $W$ events but
excluding muons). The error on the hadronic response  will also improve
with increased $Z$ statistics. Consequently, the largest errors left are
the ones 
due to the QCD multi-jet background and the electromagnetic scale. 
It will be of crucial importance for this
and other electroweak measurements to better estimate the QCD background. This
goal can partially be reached by taking more events with a QCD monitor
trigger~\cite{thesis}. Of course,
it would be best if the QCD background fraction could be reduced even further.
At this point it is not obvious if this is feasible since 
the current
set of electron identification cuts is already very efficient in reducing this
background.
A Monte Carlo program that correctly models QCD multi-jet events including a 
realistic detector model could also help in better estimating the shape of this
background.
For the low \pt\ region, the dominant error will be the electromagnetic scale.
There is however currently no good estimate by how much this error will 
be reduced.

\begin{table}[!htbp]
\caption{{ Central values and statistical errors for $\alpha_{2}$ and
systematic errors due to 
backgrounds, the hadronic energy scale and resolution, and the choice of
$\alpha_1$.  All systematic errors are for Run~I if not otherwise noted.
\label{tab:resultsII}}}
\footnotesize
\begin{tabular}{||l|l|l|l|l||}	\hline \hline
\pt [GeV]    &   $0-10$ &   $10-20$ &   $20-35$  &   $35-200$ \\ \hline
mean \pt [GeV] &  5.3    &13.3  &25.7  & 52.9         \\ \hline \hline
$\alpha_2$, Run~I& $1.07$  & $ 0.82 $  & $ 0.49 $  & $ 0.10 $    \\ 
& $\pm0.13$  & $\pm0.25 $  & $ \pm0.37 $  & $ \pm0.37 $    \\ \hline
$\alpha_2$, Run~II&    $0.98$   & $ 0.89$  & $ 0.68$  & $ 0.24 $    \\ \hline
Stat. errors, e& $\pm0.024$  & $ \pm0.047 $  & $\pm0.069 $  & $\pm0.069 $    \\
\hskip1.3cm e+$\mu$ & $\pm0.017$  & $ \pm0.033 $  & $\pm0.049 $ &
$\pm0.049 $    \\ \hline 
total syst error &  $\pm0.08$  & $ \pm0.09 $  & $\pm0.12 $  & $\pm0.12 $
\\ \hline 
QCD & $\pm0.04$ & $\pm0.05$ & $\pm0.09$ &  $\pm0.07$        \\ \hline
$Z$&    $\pm0.01$ & $\pm0.02$ & $\pm0.02$&  $\pm0.04$        \\ \hline
top&     $\pm0.00$ & $\pm0.00$ & $\pm0.00$&  $\pm0.02$        \\ \hline
EM scale  & $\pm0.06$ & $\pm0.05$ & $\pm0.03$&  $\pm0.04$        \\ \hline
had scale  & $\pm0.03$ & $\pm0.01$ & $\pm0.04$&  $\pm0.04$        \\ \hline
had. resolution &  $\pm0.02$ & $\pm0.02$ & $\pm0.05$ &  $\pm0.06$   \\
had. res. Run~II  &  $\pm0.004$ & $\pm0.004$ & $\pm0.009$&  $\pm0.011$
\\ \hline 
$\alpha_1 $  & $\pm 0.01$ & $\pm0.05$ & $\pm0.03$&  $\pm0.03$        \\
\hline \hline 
\end{tabular}
\end{table}

In the current measurement we had to fix  $\alpha_{1}$ to the value predicted 
by the QCD calculation since even after summing over both $W$ signs we are
slightly sensitive to  $\alpha_{1}$ due to acceptance effects.
The error due to this choice for $\alpha_1$ was estimated by setting 
$\alpha_1 = 2.0$, the value predicted by the $V-A$ theory in the absence
of QCD  
effects (see Table~\ref{tab:resultsII}). This error is non-negligible.
Since the
central magnet in Run~II will allow for sign identification of electrons,
$\alpha_{1}$ and  $\alpha_{2}$ could thus be measured simultaneously,
eliminating the need for the above assumption for $\alpha_{1}$. 
This will reduce the error due to the choice of $\alpha_1$. 
While this is a nice extension of this measurement, it is not clear at
this point by how much it will numerically improve the 
significance of the measurement of $\alpha_{2}$.

In the above estimate of the errors, the binning in $p_T^W$ used for the
Run~I measurement was  kept unchanged. With larger statistics, one
would probably choose a finer binning in $p_T^W$, allowing for 
bins with larger mean  $p_T^W$. This would increase the sensitivity in 
the area where the deviation of the angular distribution due to QCD effects 
is most pronounced.

The Monte Carlo used in the current analysis, 
which was originally developed for the measurement of the 
$W$ mass~\cite{d0W} at D\O, treats hadronic jets as
point particles and the hadronic recoil is treated as a single jet.
This is clearly a simplification of the true processes involved and a
real next-to-leading order event generator would be useful. 

In addition to the experimental improvements discussed thus far, this 
measurement will be sensitive
to $W$ production models. These models have to be constrained by independent 
measurements.  

To summarize, in Run~II the measurement of the angular distribution of 
electrons from $W$ boson decays will be systematically limited.
While the recoil response and resolution will improve
with increased $Z$ statistics, the  estimate of the QCD background
fraction and shape becomes a limiting factor. 
It is difficult to estimate by how much  the other dominant error, the 
error due to the uncertainty of the electromagnetic scale, 
will be reduced in Run~II. Other improvements not quantified here
are expected from a finer binning in \pt\ and sign identification of 
electrons.

\section{Vector Boson Pair Production and Trilinear Gauge Boson Couplings --
Prospects for Run~II$^\S$\label{s:diboson}}
\footnotetext{Contributed by: U.~Baur, H.T.~Diehl and
D.~Rainwater}  

The Standard Model of electroweak interactions makes precise predictions
for the couplings between gauge bosons due to the non-abelian gauge 
symmetry of
$SU(2)_L\otimes U(1)_Y$. These self-interactions are described by the 
triple gauge boson (trilinear) $WW\gamma$, $WWZ$, $Z\gamma\gamma$, 
$ZZ\gamma$ and $ZZZ$ couplings and the quartic couplings.  
Vector boson pair production provides a sensitive ground for {\em direct
tests} 
of the trilinear couplings. Deviations of the couplings from the SM 
values would indicate the presence of new physical phenomena.

The purpose of this section is to present a brief overview of recent
theoretical advances in understanding di-boson production in hadronic
collisions, and to highlight Run~II opportunities for studying the 
physics of vector boson pair production.  Because of the 
large anticipated size of the data sample, $\int\!{\cal L}dt=2~{\rm fb}^{-1}$,
interesting processes 
and final states that were not studied in Run~I will become available.
These are discussed, as well as prospects available in extensions of the Run~I 
$W\gamma$, $WW$, and $WZ$ analyses to Run~II. This is meant to be 
an improvement over the forecasts of the TeV\_2000 Report~\cite{tev2000}, 
which was written in 1995 before we had the benefit of having performed the 
Run~Ib analyses. Indeed, some of the TeV\_2000 Report's prognostications for 
Run~II limits were achieved in Run~I. 

We begin with a brief summary of the trilinear gauge boson couplings and how
they are parameterized. Next, we give a 
short description of new theoretical developments. 
Following that, we summarize the anomalous coupling limits obtained in 
Run~Ia and Run~Ib, and compare the Run~Ib results with what we 
expected we would obtain, based on a simple extrapolation from Run~Ia. 
This exercise in hindsight provides both a calibration for, and a cross-check 
of, the extrapolation method.  The subsequent section provides 
expectations for anomalous coupling limits from the Run~II analyses based on 
extrapolation of the Run~I analyses to higher integrated luminosity.  
Next, we provide comments on some Run~II analyses, and, lastly, discuss
new channels and analyses which will become feasible in Run~II, in
particular the
prospects for measuring the $ZZV$ couplings via $ZZ$ production.

\subsection{Trilinear Couplings}
The $WWV \; (V = \gamma ~{\rm or}~ Z)$ vertices are described by a
general effective Lagrangian~\cite{Wcoupling,Baur88}
with two overall couplings, $g_{WW\gamma} = -e$ and
$g_{WWZ} = -e \cdot \cot \theta_{W}$, and
six dimensionless couplings
$g_{1}^{V}$, $\kappa_V$, and $\lambda_V$ $(V = \gamma$ or $Z)$,
after imposing {\it C}, {\it P}, and {\it CP} invariance. The 
$W^-_\alpha(q)\,W^+_\beta(\bar q)\,V_\mu(p)$ 
vertex function (where all momenta are outgoing, $p+q+\bar q=0$) in
presence of non-standard couplings is given by:
\begin{equation}
\begin{split}
{\Gamma^{\alpha\beta\mu}_{WWV}\over g_{WWV}} = 
& \,~\bar q^\alpha g^{\beta\mu}\left (g^V_1+\kappa_V+\lambda_V\,{q^2\over
M_W^2} \right ) \\
& -q^\beta
g^{\alpha\mu}\left(g^V_1+\kappa_V+\lambda_V\,{\bar q^2\over M_W^2}
\right ) \\
& -\left(\bar
q^\mu-q^\mu\right)~g^{\alpha\beta}\left(g_1^V+{\lambda_V\over
2}~{p^2\over M_W^2}\right)\\
& +\left(\bar
q^\mu-q^\mu\right){\lambda_V\over M_W^2}~p^\alpha p^\beta~.
\end{split}
\end{equation}
Here, $M_W$ is the $W$-boson mass.
Electromagnetic gauge invariance requires that $g_{1}^{\gamma} = 1$, which
we assume throughout this paper.
The SM Lagrangian is obtained by setting $g_1^{\gamma} = g_1^Z = 1$,
$\kappa_{V} = 1~(\Delta\kappa_{V} \equiv \kappa_V - 1 = 0)$ and
$\lambda_V = 0$.

A different set of parameters, motivated by $SU(2)\times U(1)$ gauge
invariance, had been used by the LEP collaborations~\cite{alpha} prior
to 1998. This set consists of three independent couplings
$\alpha_{B\phi}$, $\alpha_{W\phi}$ and $\alpha_W$:
$\alpha_{B\phi}\equiv \Delta\kappa_{\gamma} - \Delta g_1^Z \cos^{2}\theta_{W}$,
$\alpha_{W\phi}\equiv\Delta g_1^Z \cos^{2}\theta_{W}$ and
$\alpha_{W}\equiv\lambda_{\gamma}$.
The remaining $WWZ$ coupling parameters $\lambda_Z$ and $\Delta\kappa_Z$
are determined by the relations $\lambda_Z = \lambda_{\gamma}$ and
$\Delta\kappa_Z = -\Delta\kappa_{\gamma}\tan^{2}\theta_{W} + \Delta g_1^Z$.
The HISZ relations~\cite{HISZ}
which have been used by the D{\O} and CDF collaborations are
also based on this set with the additional constraint
$\alpha_{B\phi} = \alpha_{W\phi}$.

The di-boson production cross sections with non-SM couplings grow with
the parton center of mass energy 
$\sqrt{\hat s}$.  In order to avoid violation of $S$-matrix unitarity, the 
anomalous couplings $a=g_1^V,\,\Delta\kappa_V,\,\lambda_V$ 
are taken as momentum dependent form factors with a scale $\Lambda_{FF}$
\begin{equation}
a(\hat s)={a\over ( 1 + \hat{s}/\Lambda_{FF}^{2})^{n}}
\end{equation}
and $n=2$ (dipole form factor).

The $Z^\alpha(q_1)\,\gamma^\beta(q_2)\, V^\mu(P)$ $(V = \gamma ~{\rm
or}~ Z)$ vertices contributing to $Z\gamma$ production are described by 
a general vertex function~\cite{Wcoupling}
with eight dimensionless couplings
$h_{i}^{V} (i=1,\dots,4 ~; V=\gamma ~{\rm or}~ Z)$:
\begin{equation}
\begin{split}
\Gamma^{\alpha\beta\mu}_{Z\gamma V} = &{P^2-q_1^2\over M_Z^2}~
\Biggl\{h_1^V\left(q_2^\mu g^{\alpha\beta}-q_2^\alpha
g^{\mu\beta}\right) \\
& + {h_2^V\over M_Z^2}\,P^\alpha\left(\left(P\cdot q_2\right)
g^{\mu\beta}-q_2^\mu P^\beta\right) \\
& + h_3^V\epsilon^{\mu\alpha\beta\rho}q_{2\rho} \\
& + {h_4^V\over M_Z^2}\,P^\alpha\epsilon^{\mu\beta\rho\sigma}P_\rho
q_{2\sigma}\Biggr\}~. 
\end{split}
\label{eq:hiv}
\end{equation}

In the SM, all $h_{i}^{V}$'s are zero. The couplings $h_1^V$ and
$h_2^V$ violate $CP$; all couplings are $C$-odd.
The form factors for these couplings are
\begin{equation}
h_{i}^{V}(\hat{s}) = {h_{i0}^{V} \over ( 1 + \hat{s}/\Lambda_{FF}^{2})^{n}}~,
\label{dibeq:ff}
\end{equation}
where one usually assumes that $n = 3$ for $i = 1,3$ and $n = 4$ for 
$i = 2,4$~\cite{baurmc}.

In the SM, the $\ell^{+}\ell^{-}\gamma$ final state can be 
produced via radiative decays of the $Z$ boson or by production of a 
boson pair via $t$- or $u$-channel quark exchange. 
The former process is the dominant source of events with small 
opening angle between
the photon and charged lepton and for events with a low value of photon 
transverse energy, $E_T^{\gamma}$.  Events produced by the 
latter process have lepton-pair invariant mass, $m_{\ell\ell}$, close to $M_Z$ 
and three-body invariant mass, $m_{\ell\ell\gamma}$, larger than $M_Z$. 
Anomalous $ZZ\gamma$ or $Z\gamma\gamma$ couplings
would enhance the cross 
section for $Z\gamma$ production, particularly for high-$E_T$ photons, 
relative to the SM expectations. 

The most general form of the $Z^\alpha(q_1)\,Z^\beta(q_2)\, V^\mu(P)$
vertex function can be written in the form~\cite{Wcoupling}
\begin{equation}
\begin{split}
\Gamma^{\alpha\beta\mu}_{ZZV} = &{P^2-M_V^2\over
M_Z^2}~\biggl(if_4^V \left(P^\alpha g^{\mu\beta}+P^\beta
g^{\mu\alpha} \right) \\
& +if_5^V\epsilon^{\mu\alpha\beta\rho}\left(q_1-q_2\right)_\rho\biggr). 
\end{split}
\end{equation}
$CP$ invariance forbids $f_4^V$ and parity conservation requires
that $f_5^V$ vanishes. In the SM, $f_4^V=f_5^V=0$. 
$S$-matrix unitarity requires a form factor
behavior for $ZZV$ couplings similar to that of $h_1^V$ and
$h_3^V$~\cite{BR} (Eq.~(\ref{dibeq:ff}) with $n=3$). 

Although the $WWV$, $Z\gamma V$ and $ZZV$ couplings usually are assumed
to be real, they are in general complex quantities.

In theories which go beyond the SM, the $WWV$ couplings are expected to
be at most ${\cal O}(M_W^2/\Lambda^2)$ where $\Lambda\sim \Lambda_{FF}$
is the scale of new physics. $Z\gamma V$ and $ZZV$ couplings are at most 
${\cal O}(M_Z^4/\Lambda^4)$. 

\subsection{Recent Theoretical Developments}
\subsubsection{Parameterization of $\boldmath{Z\gamma V}$ Couplings}
In Ref.~\cite{renard} it was pointed out that the couplings $h_i^V$ have 
to be purely imaginary quantities in order to guarantee that an effective
Lagrangian which would lead to a vertex function of the form of
Eq.~(\ref{eq:hiv}) is hermitian. In contrast, the $WWV$ and $ZZV$
couplings are normalized such that real couplings automatically correspond to
a hermitian effective Lagrangian. Since one usually assumes real
couplings when placing limits on anomalous vector boson self-couplings,
it is useful to replace the couplings $h_i^V$ by 
\begin{equation}
{h'}_i^V=-ih_i^V
\end{equation}
for the Run~II analyses and beyond. 

\begin{figure}
\begin{center}
\mbox{\epsfxsize=8cm \epsffile{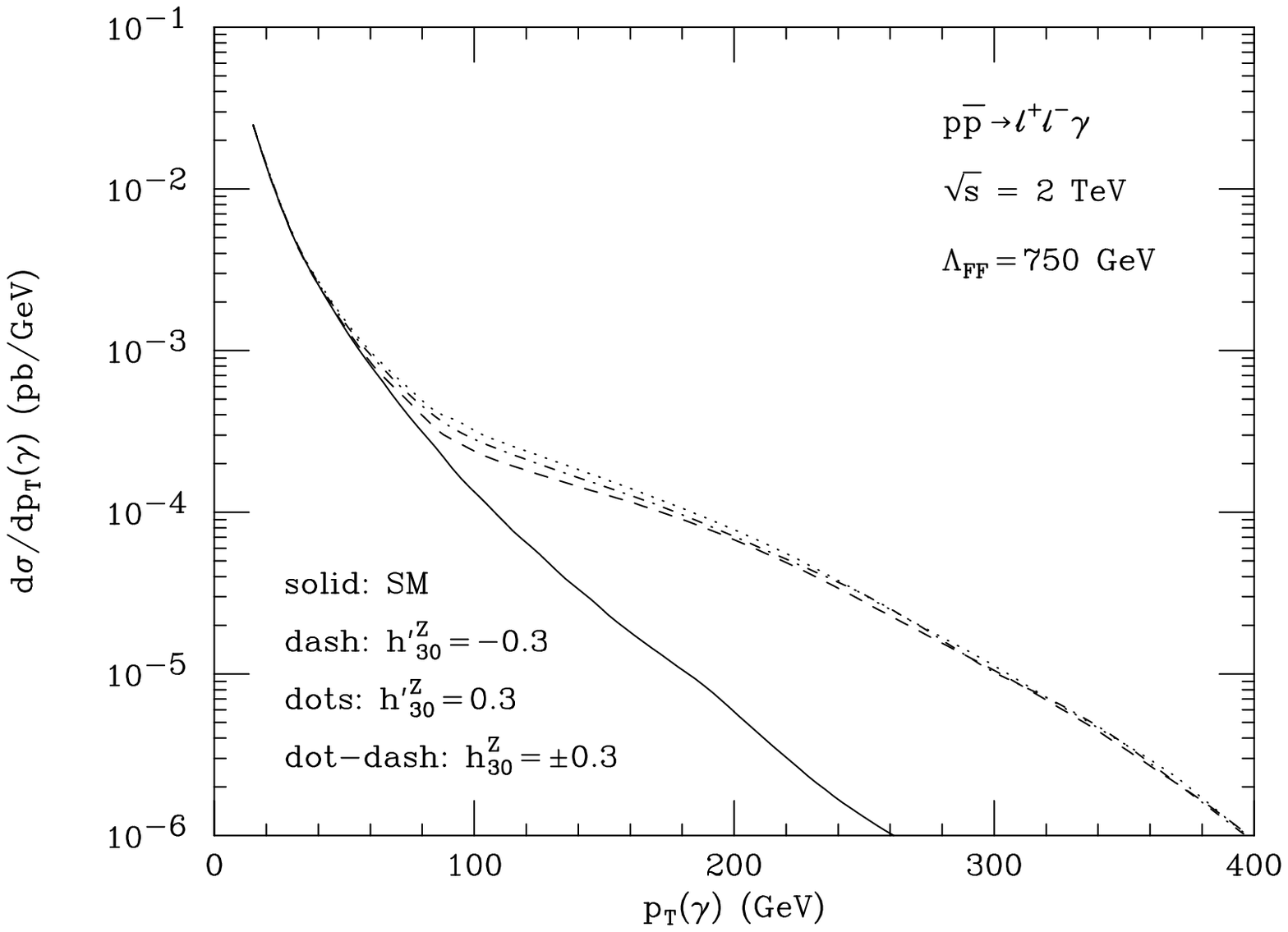}}
\end{center}
\vskip -8.mm
\caption{The photon transverse momentum distribution in $p\bar
p\to\ell^+\ell^-\gamma$ at the Tevatron in the SM and for anomalous
$ZZ\gamma$ couplings. The cuts imposed are described in the text.}
\label{dibfig:one}
\end{figure}

${h'}_i^V$'s and $h_i^V$'s of equal magnitude result in virtually the same 
differential cross sections at high energies. This is illustrated in
Fig.~\ref{dibfig:one} for the case $|{h'}_{30}^Z|=|h_{30}^Z|=0.3$. 
In order to simulate detector response, the following cuts have been
imposed in Fig.~\ref{dibfig:one}:
\begin{eqnarray}
p_T(\gamma)> 10~{\rm GeV,} &\qquad & |\eta(\gamma)|<2.5, \\
p_T(\ell)> 20~{\rm GeV,} &\qquad & |\eta(\ell)|<2.5, \\
m(\ell\ell)>75~{\rm GeV,}&\qquad & m(\ell\ell\gamma)>100~{\rm GeV,}
\end{eqnarray}
and
\begin{equation}
\Delta R(\ell\gamma)>0.7.
\end{equation}
The form factor scale has been chosen to be $\Lambda_{FF}=750$~GeV.

Unlike for real $h_{3,4}^V$ couplings, the interference terms between the SM
and the non-SM contributions do not vanish in the squared matrix element 
for real values of ${h'}_{3,4}^V$. Thus, for intermediate values of
$p_T(\gamma)$, the differential cross sections for values of
${h'}_{3,4}^V$ of equal magnitude but opposite sign slightly
differ. Since most of the sensitivity to anomalous couplings originates
from the high energy domain, the limits for ${h'}_i^V$ are expected to
be almost identical to those obtained for $h_i^V$. In the following we
therefore list limits only for $h_i^V$.

\subsubsection{NLO QCD Corrections to Vector Boson Pair Production}
\begin{figure}
\begin{center}
\mbox{\epsfxsize=8cm \epsffile{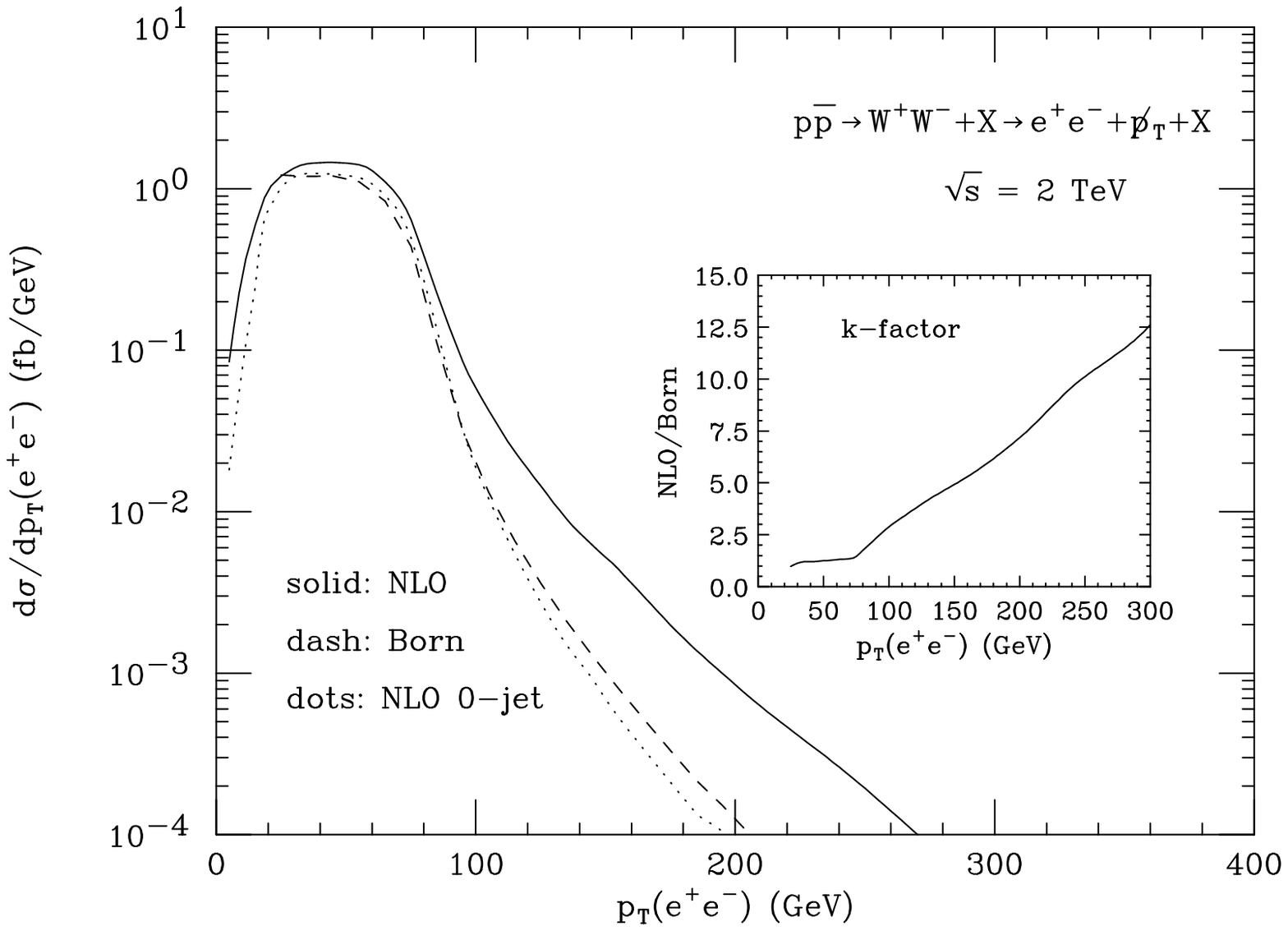}}
\end{center}
\vskip -8.mm
\caption{The $e^+e^-$ transverse momentum distribution in $p\bar p\to
W^+W^-+X\to e^+e^-p\llap/_T+X$ at the Tevatron in the SM. The following
cuts were imposed to simulate detector response: $p_T(e)>20$~GeV,
$|\eta(e)|<2.5$ and $p\llap/_T>20$~GeV. For the NLO 0-jet curve (dotted
line), jets with $p_T(j)>20$~GeV and $|\eta(j)|<3.5$ were vetoed.}
\label{dibfig:two}
\end{figure}

In the Run~I di-boson analyses, data were compared with leading order
production calculations to extract limits on the $WWV$ and $Z\gamma V$
couplings. The effect of higher order QCD corrections was simulated by 
multiplying the differential cross sections by a simple constant
$k$-factor
\begin{equation}
k=1+{8\pi\over 9}\alpha_s.
\end{equation}
NLO calculations have shown~\cite{older} that the ${\cal
O}(\alpha_s)$ QCD corrections in the SM depend logarithmically on $\hat
s$ and become large at high energies, due to
gluon-induced partonic subprocesses, which only enter at NLO. An example 
is shown in Fig.~\ref{dibfig:two}, where we display the transverse momentum 
distribution of the charged lepton pair in $p\bar p\to W^+W^-+X\to
e^+e^-p\llap/_T+X$. NLO corrections are seen to be very large (${\cal
O}(10)$) at high $p_T$, and dramatically alter the shape of the
distribution. Qualitatively this is precisely what one expects from
non-standard $WWV$ couplings.
Since the real emission diagrams are responsible for the increase of the 
QCD corrections with $p_T$, 
a jet veto drastically reduces the size of the QCD corrections (dotted
line). It should be noted that the NLO QCD corrections reduce the
$W^+W^-$ cross section when a jet veto is imposed.

With the more than 20-fold increase in statistics expected in Run~II, it is 
clear that the QCD corrections to di-boson production must properly be
taken into account when information on anomalous couplings is
extracted. Over the past seven years the NLO QCD corrections to 
$W\gamma$~\cite{BHO1}, $Z\gamma$~\cite{BHO2}, $WW$~\cite{BHO3} and $WZ$
production~\cite{BHO4} including non-standard $WWV$ and $Z\gamma V$
couplings have been calculated using the narrow width approximation and
ignoring spin correlations in the finite virtual corrections. Recently,
more complete calculations have become available which properly take
into account the 
previously ignored spin correlations. Ref.~\cite{CE} also includes 
single resonant diagrams and finite $W$ and 
$Z$ width effects in the calculation; however, no anomalous couplings are
taken into account. Refs.~\cite{DKS,FS} use the narrow width
approximation, but do include the option of non-standard $WWV$
couplings. 

The contribution of the finite virtual corrections to the NLO cross
section is smaller than about 10\% for all di-boson processes. The spin
correlations ignored in Refs.~\cite{BHO1,BHO2,BHO3,BHO4} therefore are
expected to have a rather small effect on the 
total cross section, as well as on most distributions. This expectation is
confirmed by an explicit comparison between the calculations of 
Refs.~\cite{DKS} (DKS) and~\cite{BHO3} (BHO) for $W^+W^-$ production
which is shown in Table~\ref{dibtab:one}. The two calculations are seen
to agree at the 1\% level. 
\begin{table}[tb]
\setlength{\tabcolsep}{1.5pc}
\caption{The $p\bar p\to W^+W^-+X\to e^+e^-p\llap/_T+X$ cross section
for $\sqrt{s}=2$~TeV with $p_T(e)>20$~GeV, $|\eta(e)|<2.5$, and 
$p\llap/_T>20$~GeV. Jets are required to have $p_T(j)>20$~GeV,
$|\eta(j)|<2.5$. Results are shown for the calculations of
Ref.~\cite{BHO3} (BHO) and Ref.~\cite{DKS} (DKS) with factorization
scale $Q^2=M_W^2$, and using the \hbox{CTEQ4M~\cite{cteq}} set
of parton distribution functions. \label{dibtab:one}
} 
\begin{tabular}{ccc}
\hline   
 & BHO & DKS \\ 
 & $\sigma$ (fb) & $\sigma$ (fb) \\ \hline
\multicolumn{3}{c}{Standard Model} \\ \hline
Born & 61.2 & 61.2 \\
NLO & 80.9 & 81.3 \\
NLO 0-jet & 65.6 & 65.3 \\ \hline \\[-3.mm]
\multicolumn{3}{c}{$\Delta g_1^Z=0.5$, $\lambda_Z=\lambda_\gamma=0.1$,
$\Delta\kappa_Z=\Delta\kappa_\gamma=0.3$} \\
\multicolumn{3}{c}{$\Lambda_{FF}=2$~TeV}\\ \hline
Born & 82.7 & 82.8 \\
NLO & 106.5 & 107.0 \\
NLO 0-jet & 84.2 & 83.7\\ \hline
\end{tabular}
\end{table}

A more detailed and careful comparison between the BHO and DKS
calculations for $W^+W^-$ and $W^\pm Z$ production at the LHC 
has been carried out in Ref.~\cite{lhcewk}. The $WZ$ cross sections of
the two calculations were found to agree within 1.5\%, whereas in the $WW$ 
case deviations of up to 3.8\% were observed at NLO. Subsequently, a
small error in the BHO $W^+W^-$ code was discovered~\cite{dobbs}. After
correction of this error, the BHO and DKS
calculations of $W^+W^-$ production at the LHC agree to better than
0.5\%~\cite{dobbs}. 

\subsection{Summary of Run~Ia and Run~Ib Analyses at CDF and D\O }
This subsection contains a summary of the published CDF and D\O \
analyses.  The intent is to provide an overview of the progress, a list
of published papers, and tables which provide the basis for the
comparison made in the next subsection. 

\subsubsection{$\boldmath{WW\gamma}$ and $\boldmath{WWZ}$ Couplings}

The D\O \ and CDF collaborations have performed several searches for
anomalous $WW\gamma$ and $WWZ$ couplings. Studies~\cite{wgcdf,D0WG01,D0WG02} 
of $p\bar{p}\rightarrow W\gamma +X$ have shown that the transverse energy 
spectrum of the photons agreed with that expected from SM production.
Searches~\cite{D0WW01,CDFWW,D0WW02} for an excess of 
$p\bar{p}\rightarrow WW+X$, where the $W$ bosons each decayed to 
$\ell \nu$ $(\ell=e$ or $\mu )$,  yielded events which matched the SM 
prediction. Further, the $p_T$ spectrum of the charged leptons 
agreed~\cite{D0WW02} with the prediction.
Studies~\cite{CDFWZ,D0WW03,D0WW04,D0WW05} of the processes 
$p\bar{p}\rightarrow WW + X$ and $p\bar{p}\rightarrow WZ + X$, where one 
$W$ boson decayed to a lepton or anti-lepton
and the corresponding anti-neutrino or neutrino and the other
vector boson decayed to a quark-antiquark pair manifested as jets, yielded no
excess of events and a $W$ boson transverse energy spectrum which matched the
expected background plus SM signal.  Lastly, D\O \ studied~\cite{D0WW05} 
the process $p\bar{p}\rightarrow WZ + X$ where the $Z$ boson decayed to $ee$ 
and the $W$ boson decayed to either $e\nu$ or $\mu \nu$.  Limits on
anomalous $WW\gamma$ and $WWZ$ couplings were derived from each of these
analyses.  Several~\cite{D0WG01,D0WW01,D0WW03} of these analyses were presented
in detail in Ref.~\cite{D01APRD}. The results of all of the D\O \ analyses were
combined~\cite{D0WW05}, using the maximum-likelihood 
method~\cite{D01APRD,D0WWWgWZ}, to form the Tevatron's most restrictive limits 
on anomalous $WW\gamma$ and $WWZ$ couplings.

Table~\ref{dibtab:tab1} shows the anomalous coupling limits achieved in
each of the analyses described above, the luminosity used, 
and the reference to the paper in which the result was published. 
It should be noted that many of the papers published limits under several
assumptions for the relations between the coupling parameters and with 
several values of the form factor $\Lambda_{FF}$. 
Only those limits from the
case $\lambda=\lambda_{\gamma} = \lambda_{Z}$ and $\Delta \kappa=\Delta
\kappa_{\gamma} = 
\Delta \kappa_Z $ are listed, except for $W\gamma$ and $WZ$ where 
only $WW\gamma$ and $WWZ$ couplings, respectively, are relevant. 
\begin{table*}[htb]
\setlength{\tabcolsep}{1.4pc}
\newlength{\digitwidth} \settowidth{\digitwidth}{\rm 0}
\catcode`?=\active \def?{\kern\digitwidth}
\caption{95\% confidence level $WW\gamma$ and $WWZ$ anomalous coupling
limits achieved in Run~I 
analyses by the D\O \ and CDF Collaborations. 
\label{dibtab:tab1}}
\begin{tabular*}{\textwidth}{@{}l@{\extracolsep{\fill}}cccccc}
\hline  
Analysis      & Ref.         & Run & Lum. (pb$^{-1}$)
                                        & $\Lambda_{FF}$
                                          & A.C. Limit (95\% CL)
                                                                      \\ \hline
                                                                         \hline
CDF $W\gamma \rightarrow$ 
              & \cite{wgcdf} & Ia  & 20 & 1.5 TeV
                                          & $-0.7\le \lambda \le 0.7$ \\
$e\nu\gamma$ and $\mu\nu\gamma$
              &              &     &    &  
                                          & $-2.2\le \dk \le 2.3$     \\ \hline
D\O \ $W\gamma \rightarrow$
              & \cite{D0WG01}& Ia  & 13.8  & 1.5 TeV 
                                          & $-0.6\le \lambda \le 0.6$ \\
$e\nu\gamma$ and $\mu\nu\gamma$
              &              &     &    & 
                                          & $-1.6\le \dk \le 1.8$     \\ \hline
D\O \ $W\gamma \rightarrow$
              & \cite{D0WG02}& Ia + Ib  
                                   & 92.8 
                                        & 1.5 TeV 
                                          & $-0.31\le \lambda \le 0.29$ 
                                                                      \\
$e\nu\gamma$ and $\mu\nu\gamma$
              &              &     &    & 
                                          & $-0.93\le \dk \le 0.94$   \\ \hline
D\O \ $WW \rightarrow$
              & \cite{D0WW01}& Ia  & 14 & 900 GeV 
                                          & $-2.1\le \lambda \le 2.1$ \\
Dilepton      &              &     &    & Equal Couplings
                                          & $-2.6\le \dk \le 2.8$     \\ \hline
CDF $WW \rightarrow$      
              & \cite{CDFWW} & Ia + Ib
                                   & 108& 1.0 TeV 
                                          & $-0.9\le \lambda \le 0.9$ \\
Dilepton      &              &     &    & Equal Couplings
                                          & $-1.0\le \dk \le 1.3$     \\ \hline
D\O \ $WW \rightarrow$    
              & \cite{D0WW02}& Ia + Ib
                                   & 97 & 1.5 TeV
                                          & $-0.53\le \lambda \le 0.56$ 
                                                                      \\
Dilepton      &              &     &    & Equal Coupling
                                          & $-0.62\le \dk \le 0.77$   \\ \hline
CDF $WW$ and $WZ\rightarrow$
              & \cite{CDFWZ} & Ia  & 19.6  
                                        & 1.0 TeV 
                                          & $-0.81\le \lambda \le 0.84$ 
                                                                      \\
leptons + jets&              &     &    & Equal Couplings
                                          & $-1.11\le \dk \le 1.27$   \\ \hline
D\O \ $WW$ and $WZ\rightarrow$
              & \cite{D0WW03}& Ia  & 13.7  
                                        & 1.5 TeV 
                                          & $-0.6\le \lambda \le 0.7$ \\
$e\nu$jj      &              &     &    & Equal Couplings
                                          & $-0.9\le \dk \le 1.1$     \\ \hline
D\O \ $WW$ and $WZ\rightarrow$
              & \cite{D0WW04}& Ia + Ib  
                                   & 96 & 1.5 TeV 
                                          & $-0.36\le \lambda \le 0.39$ 
                                                                      \\
$e\nu$jj      &              &     &    & Equal Couplings
                                          & $-0.47\le \dk \le 0.63$   \\ \hline
D\O \ $WW$ and $WZ\rightarrow$
              & \cite{D0WW04}& Ia + Ib  
                                   & 96 & 2.0 TeV 
                                          & $-0.33\le \lambda \le 0.36$ 
                                                                      \\
$e\nu$jj      &              &     &    & Equal Couplings
                                          & $-0.43\le \dk \le 0.59$   \\ \hline
D\O \ $WW$ and $WZ\rightarrow$
              & \cite{D0WW05}& Ib  & 81  
                                        & 2.0 TeV 
                                          & $-0.43\le \lambda \le 0.44$ 
                                                                      \\
$\mu\nu$jj    &              &     &    & Equal Couplings
                                          & $-0.60\le \dk \le 0.74$   \\ \hline
D\O \ $WZ\rightarrow$
              & \cite{D0WW05}& Ib  & 92
                                        & 1.0 TeV 
                                          & $-1.42\le \lambda_Z \le 1.42$ 
                                                                      \\
$ee\mu\nu$ and $eee\nu$
              &              &     &    & 
                                          & $-1.63\le \Delta g_1^Z \le 1.63$
                                                                      \\ \hline
D\O \ Combined& \cite{D0WW05}& Ia + Ib    
                                   & 96 & 2.0 TeV 
                                          & $-0.18\le \lambda \le 0.19$ 
                                                                      \\
              &              &     &    & Equal Couplings
                                          & $-0.25\le \dk \le 0.39$   \\ \hline
\end{tabular*}
\end{table*}

\subsubsection{$\boldmath{Z\gamma V}$ Couplings}
The D\O \ and CDF collaborations have also performed several searches for
anomalous $Z\gamma\gamma$ and $ZZ\gamma$ couplings. 
Studies~\cite{cdfzg1a,d0zg01,d0zg03} of the process $p\bar{p}\rightarrow 
Z\gamma +X \to\ell^+\ell^- \gamma + X$ have shown that the 
event yield and transverse energy spectrum of the photons agreed with that 
expected from SM $Z\gamma$ production, though it is noted that there were two 
$Z\gamma\rightarrow ee \gamma$ events with photons of $E_T$ greater 
than 70 GeV, expected in only 7.3\% of trial experiments. 

Studies of the process $Z\gamma\rightarrow \nu\bar{\nu} \gamma$ have the 
advantage of the higher branching fraction for decay to neutrinos than does 
the charged-lepton decay mode.  Furthermore, there is no final state 
radiation because the neutrinos are electrically neutral.  However, the 
signal-to-background ratio is rather lower than in the charged-lepton 
analysis.  D\O \ has published~\cite{d0zg02} the results of the Run~Ia 
$Z\gamma \rightarrow \nu\bar\nu\gamma$ analysis.  Again, the spectrum of the 
transverse energy of the photons, for $E_T^{\gamma} \ge 40$~GeV, agreed 
with the SM prediction.

D\O \ and CDF produced limits on anomalous $Z\gamma V$ ($V = Z, \gamma$) 
couplings using a fit to the $E_T^{\gamma}$ spectrum.  The 
D\O \ Run~Ia and Run~Ib results were combined in Ref.~\cite{d0zg03}.  
Table~\ref{dibtab:tab2} shows a compilation of all the Run~I CDF and D\O \
results.  The limits for $h_{30}^\gamma$ ($h_{40}^\gamma$) and
$h_{10}^V$ ($h_{20}^V$) are almost identical to those obtained for
$h_{30}^Z$ ($h_{40}^Z$) and are, therefore, not shown.
\begin{table*}[htb]
\setlength{\tabcolsep}{1.5pc}
\caption{95\% confidence level $ZZ\gamma$ and $Z\gamma\gamma$ anomalous
coupling limits 
achieved in the Run~I analyses by the D\O \ and CDF Collaborations. 
\label{dibtab:tab2}}
\begin{tabular*}{\textwidth}{@{}l@{\extracolsep{\fill}}cccccc}
\hline   
Analysis      & Ref.         & Run & Lum. (pb$^{-1}$) 
                                        & $\Lambda_{FF}$
                                          & A.C. Limit (95\% CL)
                                                                      \\ \hline
                                                        \hline \\[-3.mm]
CDF $Z\gamma \rightarrow$ 
              & \cite{cdfzg1a} 
                             & Ia  & 20 & 500 GeV
                                          & $-3.0\le h^Z_{30} \le 3.0$\\[1.mm]
$ee\gamma$ and $\mu\mu\gamma$
              &              &     &    & 
                                          & $-0.7\le h^Z_{40} \le 0.7$\\ 
\hline \\[-3.mm]
D\O \ $Z\gamma \rightarrow$ 
              & \cite{d0zg01}& Ia  & 14 & 500 GeV
                                          & $-1.8\le h^Z_{30} \le 1.8$\\[1.mm]
$ee\gamma$ and $\mu\mu\gamma$
              &              &     &    & 
                                          & $-0.5\le h^Z_{40} \le 0.5$\\ 
\hline \\[-3.mm]
D\O \ $Z\gamma \rightarrow$ 
              & \cite{d0zg03}& Ia  & 13 & 750 GeV
                                          & $-0.4\le h^Z_{30} \le 0.4$\\[1.mm]
$\nu\bar{\nu}\gamma$
              &              &     &    & 
                                          & $-0.06\le h^Z_{40} \le 0.06$\\ 
                                                                         
\hline \\[-3.mm]
D\O \ $Z\gamma \rightarrow$ 
              & \cite{d0zg02}& Ib  & 97 & 500 GeV
                                          & $-1.31\le h^Z_{30} \le
1.31$\\[1.mm] 
$ee\gamma$ and $\mu\mu\gamma$
              &              &     &    & 
                                          & $-0.26\le h^Z_{40} \le 0.26$\\ 
                                                                         
\hline \\[-3.mm]
D\O \ $Z\gamma \rightarrow$ 
              & \cite{d0zg02}& Ib  & 97 & 750 GeV
                                          & $-0.69\le h^Z_{30} \le
0.69$\\[1.mm] 
$ee\gamma$ and $\mu\mu\gamma$
              &              &     &    & 
                                          & $-0.08\le h^Z_{40} \le 0.08$\\ 
                                                                         
\hline \\[-3.mm]
D\O \ Combined& \cite{d0zg02}&Ia + Ib
                                   &    & 750 GeV
                                          & $-0.36\le h^Z_{30} \le
0.36$\\[1.mm]
              &              &     &    & & $-0.05\le h^Z_{40} \le
0.05$\\[1.mm]  
                                                                         \hline
\end{tabular*}
\end{table*}

\subsection{Hindsight: Extrapolating Run~Ia Results to Run~Ib}
It is interesting to see how well one can ``predict'' the Run~Ib limits
based on the Run~Ia results and a simple rule for scaling the limits based 
on the increase in the luminosity.

The $WW\gamma$ and $WWZ$ anomalous coupling limits should scale by 
\begin{equation}
\left(\int\!{\cal L}dt\right)^{1/4}.
\end{equation}
One square-root comes from the 
decrease in the statistical uncertainty of the cross section (as a function 
of $E_T$, for instance) and the other from the fact that the
differential cross section is a quadratic function of the anomalous
couplings. 

The $Z\gamma\gamma$ and $ZZ\gamma$ anomalous coupling limits would also 
scale by the fourth-root of the ratio of the luminosities, except that the
limits depend very strongly on the form-factor scale. 

\subsubsection{$\boldmath{W\gamma}$}
The integrated luminosity used in D\O's Run~Ia + Run~Ib $W\gamma$ analysis 
was 6.72 times larger than the Run~Ia sample alone. From that we expect 
the combined anomalous coupling limits to be $(1/6.72)^{1/4}=0.62$ as large
as the Run~Ia limits.  Scaling the Run~Ia results in
Table~\ref{dibtab:tab1}, 
we would expect limits $-1.0\le \Delta \kappa_{\gamma} \le 1.1$ and
$-0.38 \le \lambda_{\gamma} \le 0.38$. Instead, from
Table~\ref{dibtab:tab1}, 
the result was equivalent to a scaling of $\sim 0.52$. That does not seem
very different, but it is, for it corresponds to the equivalent of a factor 
of two more luminosity. The difference is attributed to an improvement 
in technique, the use of a three-body transverse mass criteria to remove 
events where the photon was radiated from a charged final state lepton. 
It's hard to predict improvement techniques because, if such improvements 
were a priori known, they would most likely have been applied. 

\subsubsection{$\boldmath{WW/WZ\rightarrow e\nu {\rm jj}}$}
We expect the Run~Ia + Run~Ib limits to scale by 0.61 from the ratio of the 
integrated luminosities. Consulting Table~\ref{dibtab:tab1}, we find
that this is essentially  
right on the nose for the $\Lambda_{FF} = 1.5$~TeV limits. 
The Run~Ia + Ib $\Lambda_{FF} = 2$~TeV limits, not available because of 
unitarity constraints in the Run~Ia sample, represent a slight 
($\sim 10\%$) improvement over the Run~Ia + Ib 
$\Lambda_{FF} = 1.5$~TeV results. 

\subsubsection{$\boldmath{WW\rightarrow {\rm dileptons}}$}
We expect the Run~Ia + Run~Ib limits to scale by 0.62 from the ratio
of the integrated luminosities. However, that is not what happened. An
important improvement in the technique, namely a 2-D fit to the lepton 
$E_T$ spectrum, plus the subsequent increase in the allowed form factor,
allowed the combined results to be almost a factor of four better. 
Here is a case where we have already challenged the Run~II limits predicted
by the TeV\_2000 report.

\subsubsection{$\boldmath{Z\gamma\rightarrow ee \gamma}$ and
$\boldmath{\mu\mu\gamma}$} 
Again, by the fourth-root rule, we expect the Run~Ib limits to scale by 0.62. 
For $\Lambda_{FF}=500$ GeV, we found that the $h^Z_{30}$ results scaled
by 0.72  
and the $h^Z_{40}$ by 0.52, averaging out to a scale factor of 0.62. 
But, because of the strong dependence on the form factor scale, the results 
at $\Lambda_{FF}=750$ GeV are 3.5 and 5 times better for $h^Z_{30}$ and
$h^Z_{40}$, respectively. 

\subsection{Expectations for Run~II Anomalous Coupling Limits} 
Having probed the usefulness and limitations of our scaling formula, we 
apply it to the Run~I analyses to determine the limits that might be attained 
with 2~fb$^{-1}$. Of course, any projections 
for anomalous coupling limits are merely sensitivity estimates.
Improvements in technique, such as multi-dimensional fits, or using
clever projection techniques (see Ref.~\cite{PP} for an example) may
yield more stringent limits.

For the $W\gamma$ and $WW/WZ$ analyses we will use an integrated luminosity 
scale factor $(2000/100)^{1/4}=2.1$. The slight improvement from the
$\approx 10\%$ increase in cross section available should the Tevatron
operate at center of mass energy 2000~GeV is ignored. Increasing the
form factor scale from 1.5~TeV to 2~TeV strengthens the limits by about
10\%. 
The $WZ\rightarrow {\rm trileptons}$ analysis will improve by about 
a factor of 6 because of the increased integrated luminosity and because 
of the improvement in limit-setting technique available by fitting the 
$E_T$ spectrum of the $Z$ bosons. Table~\ref{dibtab:tab3} contains the expected
results. 
%
\begin{table*}[htb]
\setlength{\tabcolsep}{1.5pc}
\caption{95\% confidence level $WW\gamma$ and $WWZ$ anomalous coupling
limits that might be achieved by D\O \ or CDF in Run~II.
\label{dibtab:tab3}}
\begin{tabular*}{\textwidth}{@{}l@{\extracolsep{\fill}}ccccc}
\hline   
Analysis      & Lum. (pb$^{-1}$)  & $\Lambda_{FF}$
                                        & A.C. Limit (95\% CL)
                                                                     \\ \hline
                                                                        \hline
$W\gamma \rightarrow$ 
              & 2000             & 1.5 TeV
                                          & $-0.14\le \lambda \le 0.14$ \\
$e\nu\gamma$ and $\mu\nu\gamma$
              &                  & Equal Couplings 
                                          & $-0.44\le \dk \le 0.44$  \\ \hline
$W\gamma \rightarrow$ 
              & 2000             & 2.0 TeV
                                          & $-0.12\le \lambda \le 0.12$ \\
$e\nu\gamma$ and $\mu\nu\gamma$
              &                  & Equal Couplings 
                                          & $-0.40\le \dk \le 0.40 $  \\ \hline
$WW$ and $WZ\rightarrow$
              & 2000             & 2.0 TeV 
                                          & $-0.16\le \lambda \le 0.17$ \\
$e\nu$jj      &                  & Equal Couplings
                                          & $-0.20\le \dk \le 0.28$  \\ \hline
$WZ\rightarrow {\rm trileptons}$
              & 2000             & 2.0 TeV& $-0.2\le \lambda_Z \le 0.2$ \\
              &                  &        & $-0.3\le \Delta g_1^Z \le 0.3$
                                                                      \\ \hline
      Combined& 2000             & 2.0 TeV& $-0.086\le \lambda \le 0.090$ \\
 (per experiment)         
              &                  & Equal Couplings
                                          & $-0.12\le \dk \le 0.19$   \\
\hline \\
\end{tabular*}
\end{table*}
In order to put these bounds into perspective, we list the most recent
LEP2 (95\% CL) limits from a 3-parameter fit~\cite{mor2000}, 
assuming $\Delta\kappa_Z=\Delta g_1^Z-\Delta\kappa_\gamma\tan^2\theta_W$
and $\lambda_Z=\lambda_\gamma$:
\begin{eqnarray}
-0.073 < & \Delta g_1^Z & < 0.075, \\
-0.12 < & \Delta\kappa_\gamma & < 0.16, \\
-0.15 < & \lambda_\gamma & < 0.01.
\end{eqnarray}
It should be noted that form factor effects are {\sl not} included in the
bounds obtained at LEP2. Taking into account the form factor behavior of 
anomalous couplings weakens the limits obtained. For a dipole form
factor with $\Lambda_{FF}=2$~TeV, this is a 2\% effect.

For anomalous $Z\gamma\gamma$ and $ZZ\gamma$ couplings we forecast limits 
which are very similar to those given in the Tev\_2000 Report. 
Those predictions
are based on 1~fb$^{-1}$ integrated luminosity and
$\Lambda_{FF}=1500$~GeV and are listed in 
Table~\ref{dibtab:tab4}. The bounds obtained for $h_{10}^V$ ($h_{20}^V$)
almost coincide with those found for $h_{30}^V$ ($h_{40}^V$).
\begin{table*}[hbt]
\setlength{\tabcolsep}{1.5pc}
\caption{95\% confidence level $Z\gamma V$ anomalous coupling limits 
that might be achieved by D\O \ or CDF experiments in
Run~II.\label{dibtab:tab4} }
\begin{tabular*}{\textwidth}{@{}l@{\extracolsep{\fill}}cccc}
\hline   
Analysis  & Lum. (pb$^{-1}$)  & $\Lambda_{FF}$ & A.C. Limit (95\% CL) \\ 
\hline \hline \\[-3.mm]          
$Z\gamma \rightarrow$ 
          &1000              & 1.5~TeV   & $-0.105\le h^V_{30}\le
0.105$\\[1.mm]  
$ee\gamma$
          &                  &            & $-0.0064\le
h^V_{40} \le 0.0064$ 
                                                                    \\[1.mm]
\hline \\[-3.mm]
$Z\gamma \rightarrow$ 
          &1000              & 1.5~TeV
                                          & $-0.038\le h^V_{30}
\le 0.038$\\[1.mm]
$\nu\nu\gamma$
          &                  &            & $-0.0027\le
h^V_{40} \le 0.0027$\\ [1.mm]
                                                                       \hline
\end{tabular*}
\end{table*}
For comparison, the most recent 95\% CL limits on $h_i^V$ from LEP2 are:
\begin{alignat}{4}
-0.17 < & h_1^\gamma & < 0.08 & \qquad &  -0.26 < & h_1^Z & < 0.09,
\qquad & \\
-0.11 < & h_2^\gamma & < 0.10 & \qquad & -0.11 < & h_2^Z & <
0.16,\qquad  &  \\
-0.027 < & h_3^\gamma & < 0.041 & \qquad &  -0.29 < & h_3^Z & <
0.21, \qquad & \\
%
-0.026 < & h_4^\gamma & < 0.022 & \qquad & -0.12 < & h_4^Z & <
0.20. \qquad &
\end{alignat}
Only one coupling at a time is varied here.
Correcting for form factor effects, the limits for $h_{1,3}$ ($h_{2,4}$)
weaken by about 5\% (7\%) for $\Lambda_{FF}=1500$~GeV. 

In Run~II, the Tevatron will thus be able to improve the existing bounds 
on anomalous $Z\gamma V$ couplings mostly for $h_{2,4}^V$. 
If an integrated luminosity of 10~fb$^{-1}$ can be achieved in
Run~II, the limits listed in Tables~\ref{dibtab:tab3}
and~\ref{dibtab:tab4} would improve by approximately a factor~1.5.

A few additional comments are in order at this point:

\subsubsection{$\boldmath{WW/WZ\rightarrow e\nu {\rm jj}}$}
Note that the expected Run~II anomalous coupling limit has nearly been 
ruled out by the Run~I combined analysis measurement. Nevertheless, 
if the Run~II combined measurement is to scale based on the increase in the 
luminosity, all of the analyses must be carried out again.

\subsubsection{$\boldmath{Z\gamma\rightarrow ee \gamma}$ and
$\boldmath{\mu\mu\gamma}$} 
Scaling the Run~I yield, totalling 29 $ee\gamma + \mu\mu\gamma$ candidates 
at D\O, by the increase in luminosity, one expects about 600 
$ee\gamma + \mu\mu \gamma$ events per experiment. The QCD background
and final state radiation background will be reduced compared to Run~I 
through the application of a di-lepton invariant mass criteria, reducing the
samples to $\sim 250$ events.  
The Run~II data will settle once and for all, whether there is a bump in the 
$Z\gamma$ invariant mass spectrum, as is not very strongly suggested by the 
Run~I data. That is, unless a new one crops up. 

\subsubsection{$\boldmath{Z\gamma\rightarrow \nu\bar\nu \gamma}$}
This is a powerful tool for studying $Z\gamma\gamma$ and 
$ZZ\gamma$ couplings because of the large (20\%) branching fraction for 
$Z\to\nu\bar{\nu}$.  However, because the $Z$ boson is undetectable,
there aren't  
any other kinematic handles.  There are common backgrounds which produce 
the same signature as the signal: a photon recoiling against missing
transverse energy
(neutrinos). In order to reduce the backgrounds, a higher $E_T^\gamma$ 
cut is used than in the other $Z\gamma$ analyses (40 GeV instead of 7~-- 
10 GeV). Understanding the normalization of the background
from cosmic ray muons that happened to deposit energy in the calorimeter 
in such a way as to mimic a photon was the main difficulty in this analysis 
in Run~I.   This background should be more tractable in  Run~II using the 
new central and forward preshower detectors and a technique similar to that 
described in D\O's Run~Ia publications~\cite{D01APRD,d0zg02}. 

\subsection{New Directions in Di-boson Production for Run~II}

Besides improving limits from final states analyzed in Run~I, a number of
new channels will become accessible in Run~II, either due to the
increased data 
sample, or because of detector improvements. In addition, it will be
possible to search for the so-called ``radiation zero'' in $W\gamma$
production. In this subsection, we first briefly describe the search for the
radiation zero and the prospects for using di-boson final states
involving $b$-quarks. This is followed by a somewhat more detailed
analysis of $ZZ$ production, the main new di-boson channel which will
become accessible in Run~II.

\subsubsection{Radiation Zero in $\boldmath{W\gamma}$ Production}
$W\gamma$ production is of special interest because of the ``radiation zero''
in the helicity amplitudes~\cite{BM}. The Tev\_2000 Report describes the
situation very 
eloquently and completely. The SM helicity amplitudes of the process
$q_1\bar{q}_2\rightarrow W^{\pm}\gamma$ vanish for 
\begin{equation}
\cos{\theta^*}=\frac{Q_1+Q_2}{Q_1-Q_2}=\pm\frac{1}{3}
\end{equation}
where $\theta^*$ is the scattering 
angle of the $W$ boson with respect to the quark $(q_1)$ direction in the 
$W\gamma$ rest frame and $Q_1$ and $Q_2$ are the quark and anti-quark 
electric charges normalized by the proton electric charge. Anomalous
couplings destroy the radiation zero as do higher-order QCD corrections,
backgrounds, finite $W$-width effects, and events where the photon is
radiated from the charged lepton instead of the $W$. 

The trick in reconstructing $\theta^*$ is in determining the parton 
center-of-mass frame because there are two solutions for the 
$z$-component of the neutrino momentum. CDF discussed~\cite{dougb} 
a possible solution in selecting the minimum of the $p_z(\nu)$ solutions 
for $W^-\gamma$ and the maximum of the $p_z(\nu)$ solutions for $W^+\gamma$. 
This is correct 73\% of the time because of the high $W$ polarization at 
Tevatron production energies. CDF saw a hint~\cite{dougb} of the 
radiation zero in Run~Ib but the signal was not definitive. 

The twofold ambiguity in reconstructing 
$\cos{\theta^*}$ can be avoided by studying rapidity correlations such as 
$\Delta y(\gamma,\ell)=y(\gamma)-y(\ell)$, which manifests the radiation
zero at  $\Delta y(\gamma,\ell)\approx -0.3$~\cite{BEL}. 

\begin{figure}
\begin{center}
\epsfclipon
\mbox{\epsfxsize=8cm \epsffile[66 415 300 649]{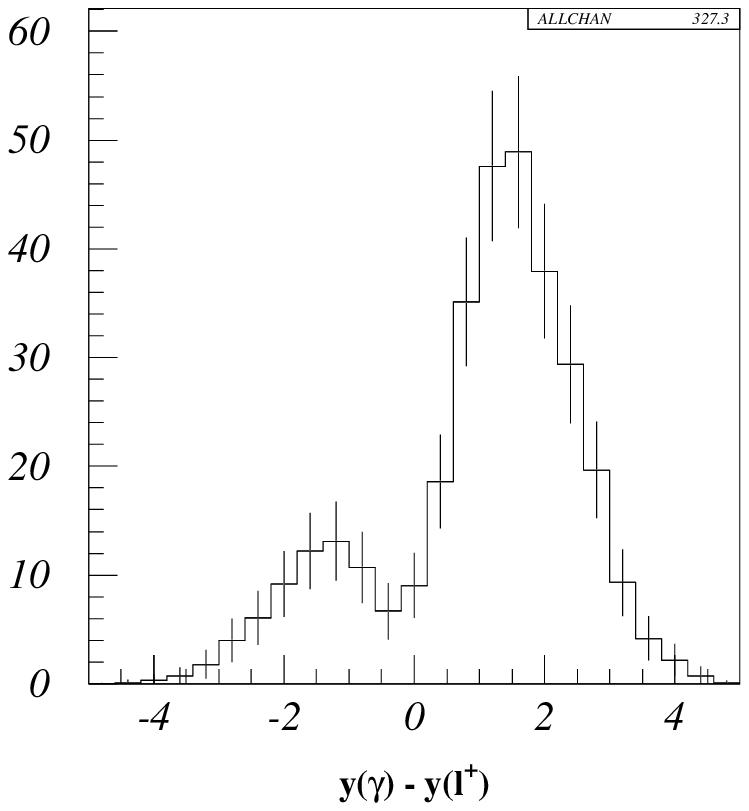}}
\epsfclipoff
\end{center}
\vskip -8.mm
\caption{Simulation of the SM $\Delta y(\gamma,e)$ distribution in
$p\bar p\to W^+\gamma\to e^+p\llap/_T\gamma$ in the D\O\ detector (from
Ref.~\cite{tev2000}). In addition to the standard $p\llap/_T$, electron
and photon $p_T$ and rapidity cuts, a $\Delta R(\gamma,\ell)>0.7$ cut, and 
a cluster transverse mass cut of $m_T(\ell\gamma;p\llap/_T)>90$~GeV are
imposed (to reduce the $W\to e\nu\gamma$ background). The statistical
error bars for an integrated luminosity of 1~fb$^{-1}$ are also shown.}
\label{dibfig:three}
\end{figure}

The Tevatron is operating at the ideal energy for observing the 
radiation zero because the zero is not smeared out by NLO processes 
expected from $W\gamma$ production at higher energy
accelerators. Fig.~\ref{dibfig:three} shows the $\Delta y(\gamma,\ell)$ 
distribution, together with the statistical errors for 1~fb$^{-1}$.
The radiation zero will be observed in Run~II if it is there. 

\subsubsection{$\boldmath{WZ\rightarrow \ell \nu {\rm {b\bar{b}}}}$ }
This channel has not been studied in Run~I. It 
will be examined very closely in Run~II because 
it is a background in the search for associated Higgs boson production
($W+H^0$ where $W\rightarrow \ell \nu$ and $H^0\rightarrow {\rm b\bar{b}}$).
The SM cross section for $WZ$ production, including NLO QCD corrections, 
is about 3.7~pb. The 
branching fraction for $Z\rightarrow {\rm b\bar{b}}$ is $\sim 15\%$.
This is 2.5~times as much as that of $Z\rightarrow \mu\bar{\mu}$ and 
$Z\rightarrow  e\bar{e}$ combined.
So we can expect about 250 $e\nu {\rm b \bar{b}}+\mu \nu {\rm b \bar{b}}$
events per experiment, not counting acceptance, lepton ID, and b-tagging 
efficiencies, which can be expected to amount to $\sim 0.20$ for such 
a final state. A cut on $b\bar{b}$ invariant mass will reduce the 
$W+$jets background. 

This is ripe for an anomalous coupling analysis.  
To produce $WWZ$ anomalous coupling limits, one can fit the 
$E_T$ spectrum of the $W$ boson and of the final-state lepton, such 
as was done in the Run~I $WW/WZ\rightarrow \ell\nu +$jets analyses.

\subsubsection{$\boldmath{Z\gamma\rightarrow {\rm b\bar{b}} \gamma}$}
Scaling the Run~I yield, totalling 29 $ee\gamma + \mu\mu\gamma$ candidates 
at D\O, by the increase in luminosity and a factor for the larger 
branching ratio of $Z\rightarrow {\rm b\bar{b}}$ to $Z\rightarrow ee (\mu\mu)$,
one might expect about 1000 $Z\gamma\rightarrow {\rm b\bar{b}} \gamma$ events. 
Background from $\gamma$jj and three jet events where a jet mimics a photon 
are larger than the signal and may constrain this to a limit-setting 
analysis (see also Sec.~\ref{s:other}).

\subsubsection{$\boldmath{ZZ}$ Production}
For Run~I, $ZZ$ production has not been analyzed. The total
cross section for $p\bar p\to ZZ$ at $\sqrt{s}=2$~TeV, including NLO QCD 
corrections is approximately 1.5~pb. For an integrated 
luminosity of 2~fb$^{-1}$ one thus expects a few
$ZZ\to\ell_1^+\ell_1^-\ell_2^+\ell_2^-$ ($\ell_1,\,\ell_2=e,\,\mu$)
events, if realistic lepton $p_T$ and pseudo-rapidity
cuts are imposed. Larger event rates are expected for
$ZZ\to\ell^+\ell^-\bar\nu\nu$, $ZZ\to\ell^+\ell^-$jj and
$ZZ\to\bar\nu\nu$jj. These channels, however, suffer from non-trivial
background contributions. In this subsection we briefly discuss the
signals of anomalous $ZZV$ couplings in the four 
channels, and derive sensitivity bounds on $f_4^V$ and
$f_5^V$ which one expects to achieve with 2~fb$^{-1}$ (Run~IIa) and
10~fb$^{-1}$ (Run~IIb). More details will be given
elsewhere~\cite{BR}. 

The results reported here are based on a tree level calculation of $ZZ$
production in the double pole approximation. Timelike photon exchange
and the decays of the $Z$
bosons, including full decay correlations and finite $Z$ width effects,
are taken into account in the calculation. To simulate detector
response, we impose the 
following transverse momentum, pseudo-rapidity and separation cuts:
\begin{eqnarray}
p_T(\ell)>15~{\rm GeV}, & \qquad & |\eta(\ell)|<2.5, \label{eq:lep_cuts} 
\\
p_T(j)>20~{\rm GeV}, & \qquad & |\eta(j)|<2.5, \\
\Delta R(\ell j)>0.6, & \qquad & \Delta R(jj)>0.6.
\end{eqnarray}
In the $ZZ\to\bar\nu\nu$jj case, Eq.~(\ref{eq:lep_cuts}) is replaced by
a charged lepton veto
\begin{equation}
p_T(\ell)<10~{\rm GeV} \qquad {\rm for} \qquad |\eta(\ell)|<2.5. 
\label{eq:lep_veto}
\end{equation}
In addition to the cuts imposed on the leptons and jets, we require 
\begin{eqnarray}
p\llap/_T > 20~{\rm GeV} & {\rm for} & ZZ\to\ell^+\ell^-p\llap/_T, \\
p\llap/_T < 20~{\rm GeV} & {\rm for} & ZZ\to\ell^+\ell^- {\rm jj}, \\
p\llap/_T > 60~{\rm GeV} & {\rm for} & ZZ\to p\llap/_T{\rm jj}
\end{eqnarray}
and 
\begin{eqnarray}
76~{\rm GeV} & < m(\ell\ell) & < 106~{\rm GeV}, \\
76~{\rm GeV} & < m(jj) & < 106~{\rm GeV}.
\end{eqnarray}
Finally, in the $ZZ\to\ell^+\ell^-\bar\nu\nu$ case, we require that the
angle in the transverse plane between a charged lepton and the missing
transverse momentum is between $20^\circ$ and $160^\circ$ if the missing 
$p_T$ is $p\llap/_T<50$~GeV. This suppresses backgrounds from $b\bar b$
production and $Z\to\tau^+\tau^-$ decays to a negligible level.

Uncertainties in the energy measurements are taken into account in the
numerical simulations by Gaussian smearing of the particle momenta
according to the resolutions of the CDF~II detector. For the form factor 
we use the form of Eq.~(\ref{dibeq:ff}) with $n=3$ and 
$\Lambda_{FF}=750$~GeV. We
use the CTEQ4L parton distribution functions with $Q^2=M_Z^2$. Unless
stated otherwise, only one $ZZV$ coupling at a time is chosen different from
its zero SM value. For simplicity, we only consider real $ZZV$
couplings. \\

\paragraph{$\boldmath{ZZ\to 4~leptons}$}

\begin{figure*}
\begin{center}
\mbox{\epsfxsize=14cm 
\epsffile{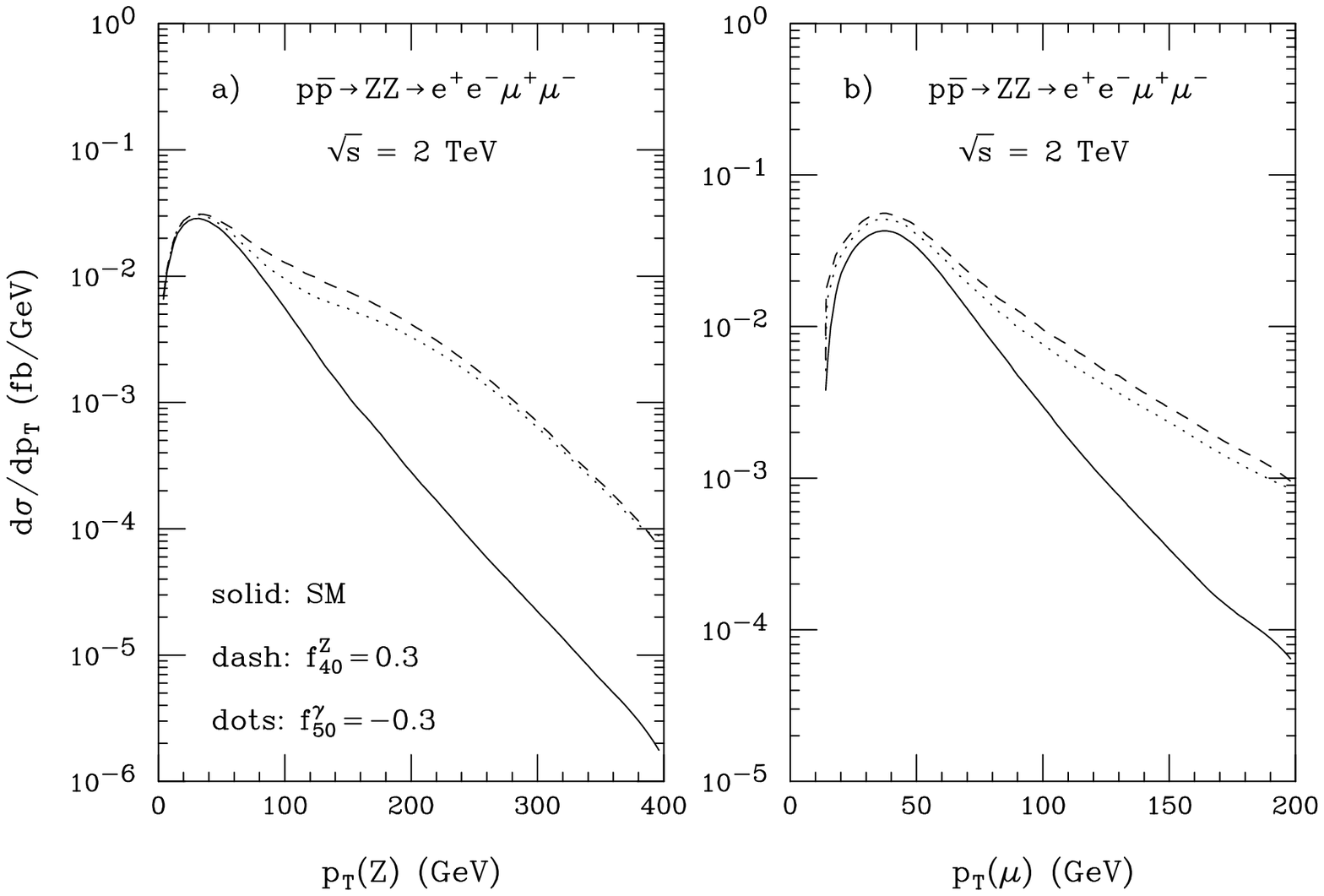}}
\end{center}
\vskip -8.mm
\caption{The $p_T(Z)$ and $p_T(\mu)$ distributions in $p\bar p\to ZZ\to
e^+e^-\mu^+\mu^-$ in the SM and in the presence of non-standard $ZZV$
couplings. }
\label{dibfig:four}
\end{figure*}

Similar to the $WWV$ and $Z\gamma V$ couplings, the effects of anomalous 
$ZZV$ couplings are enhanced at large energies. A typical signal of
nonstandard $ZZZ$ and $ZZ\gamma$ couplings thus will be a broad increase 
in the $ZZ$ invariant mass distribution, the $Z$ transverse momentum
distribution and the $p_T$ distribution of the $Z$ decay leptons. This
is illustrated in Fig.~\ref{dibfig:four} for the $p_T(Z)$ and the
$p_T(\mu)$ distributions in $p\bar p\to ZZ\to e^+e^-\mu^+\mu^-$. Results 
are shown for the SM, $f_{40}^Z=0.3$ and $f_{50}^\gamma=-0.3$. 
%
Terms proportional to $f_4^V$ and $f_5^V$ in the matrix elements have 
identical high energy behavior. Differences in the differential cross
sections at high energies between $ZZZ$ and $ZZ\gamma$
couplings are thus controlled by the $Zf\bar f$ and $\gamma f\bar{f}$
couplings, and by the parton distribution functions. At the Tevatron
these result in
differential cross sections which differ by only a few percent for
$\hat s\gg M_Z^2$ if $|f_i^Z|=|f_i^\gamma|$ ($i=4,5$). 
Slightly larger differences are observed at intermediate
energies and transverse momenta. Since $f_4^V$ violate $CP$
conservation, terms in the helicity 
amplitudes proportional to those couplings do not interfere with the SM
terms as long as $f_4^V$ is real. Cross sections thus are independent of 
the sign of $f_4^V$.

\begin{figure*}
\begin{center}
\mbox{\epsfxsize=14cm 
\epsffile{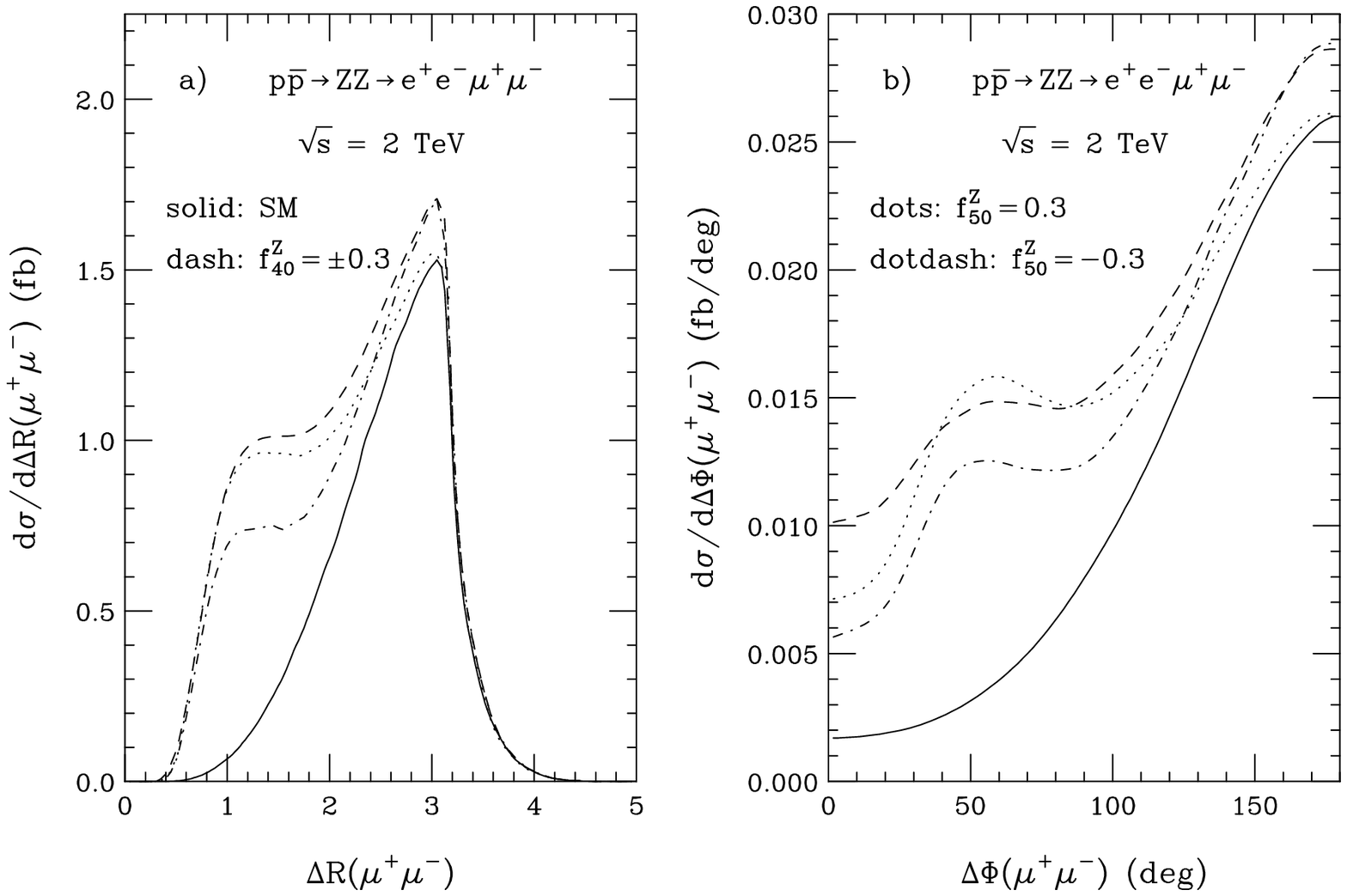}}
\end{center}
\vskip -8.mm
\caption{The $\Delta R(\mu^+\mu^-)$ and
$\Delta\Phi(\mu^+\mu^-)$ distributions in $p\bar p\to ZZ\to
e^+e^-\mu^+\mu^-$ in the SM and in the presence of non-standard $ZZZ$
couplings. }
\label{dibfig:five}
\end{figure*}

To distinguish $f_4^V$ and $f_5^V$, and to determine the sign of $f_5^V$,
the $\Delta 
R(\ell_i^+\ell_i^-)$ and $\Delta\Phi(\ell_i^+\ell_i^-)$ ($i=1,2$) 
distributions may be helpful, if deviations from the SM predictions
should be found in the $p_T$ or the $m_{ZZ}$ differential cross
sections. Fig.~\ref{dibfig:five} shows the $\Delta R(\mu^+\mu^-)$ and
$\Delta\Phi(\mu^+\mu^-)$ distributions for $p\bar p\to ZZ\to
e^+e^-\mu^+\mu^-$ in the SM and for non-standard $ZZZ$
couplings. The shape of the distributions for non-zero $f_4^V$,
$f_5^V>0$ and $f_5^V<0$ are quite different. 
Similar results are obtained for the corresponding
distributions of the $e^+e^-$ pair, and for the $ZZ\gamma$ couplings 
$f_{4,5}^\gamma$. 

Anomalous couplings mostly affect the cross section at large $Z$-boson 
transverse momentum. Due to the Lorentz boost, the relative opening
angle between the leptons originating from the $Z$ decay decreases with
increasing $p_T$. The deviations due to non-standard $ZZV$ couplings in the 
$\Delta R(\ell_i^+\ell_i^-)$ and $\Delta\Phi(\ell_i^+\ell_i^-)$
distributions thus are therefore concentrated at rather small values. The SM 
$\Delta R(\ell_i^+\ell_i^-)$ and $\Delta\Phi(\ell_i^+\ell_i^-)$
differential cross sections are dominated by the threshold region, 
$\sqrt{\hat s} \approx 2m_Z$, where the $Z$ boson 
momenta are small and the decay leptons tend to be back-to-back, {\it
i.e.} the distributions are strongly peaked at $\Delta R\approx 3$ and
$\Delta\Phi=180^\circ$. 

Using the $\Delta R(\ell_i^+\ell_i^-)$ and $\Delta\Phi(\ell_i^+\ell_i^-)$
distributions, it may be possible to distinguish $f_4^V$ and $f_5^V$
and to determine the sign of $f_5^V$, provided 
a sufficient number of events are observed.\\

\paragraph{$\boldmath{ZZ\to\ell^+\ell^-\bar\nu\nu}$}

In contrast to the $ZZ\to 4$~leptons mode which is almost background
free, there are several potentially important background processes if
one of the two $Z$ bosons decays into neutrinos. The advantage of the 
$ZZ\to\ell^+\ell^-\bar\nu\nu$ channel is its larger branching
fraction. Summing over the three neutrino species, the number of
$ZZ\to\ell^+\ell^-\bar\nu\nu$ signal events is about a factor~6 larger
than the number of $ZZ\to 4$~leptons events. 

\begin{figure*}
\begin{center}
\mbox{\epsfxsize=13.9cm 
\epsffile{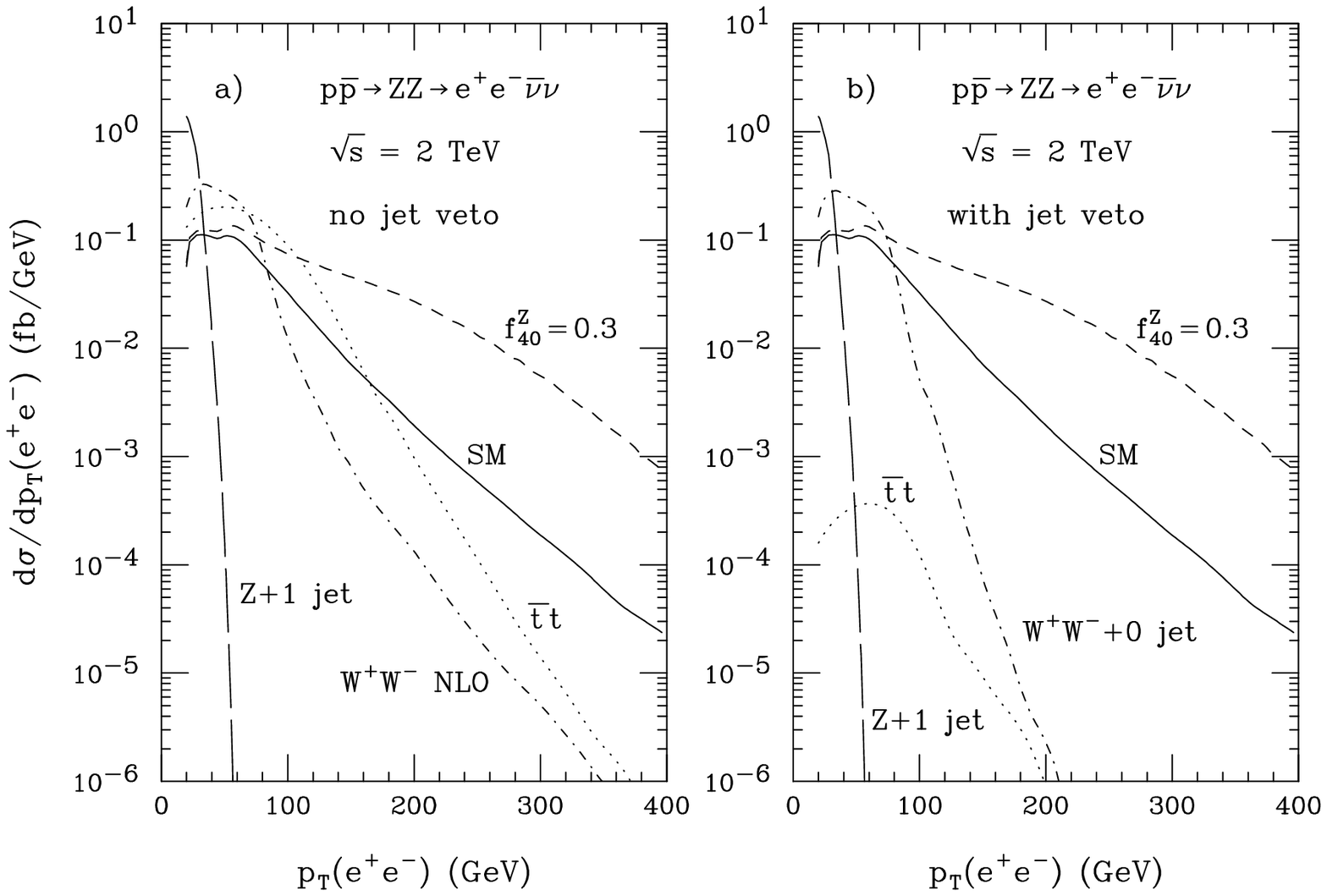}}
\end{center}
\vskip -8.mm
\caption{Transverse momentum distribution of the $e^+e^-$ pair in $p\bar
p\to ZZ\to e^+e^-\bar\nu\nu$ at the Tevatron, together with the
differential cross sections from several background processes a) without 
and b) with a jet veto applied.}
\label{dibfig:six}
\end{figure*}

The most important background processes contributing to the
$ZZ\to\ell^+\ell^-\bar\nu\nu$ channel are $t\bar t\to W^+W^-b\bar b$,
standard electroweak $W^+W^-+X$ production with
$W^+W^-\to\ell^+\nu\ell^-\bar\nu$, and $Z(\to\ell^+\ell^-)+1$~jet
production with the jet rapidity outside the range covered by the
detector and thus faking missing $p_T$. Our results for signal and
backgrounds are summarized in Fig.~\ref{dibfig:six} for the $ZZ\to
e^+e^-\bar\nu\nu$ case. 

The two most important backgrounds are $t\bar t$ and $W^+W^-+X$
production. If no additional cuts are imposed to suppress the $t\bar t$
background, its differential cross section is larger than the SM signal
for $e^+e^-$ transverse momenta as large as 200~GeV, and may thus reduce 
the sensitivity to anomalous $ZZV$ couplings (see
Fig.~\ref{dibfig:six}a). Requiring that no jets with $p_T(j)>20$~GeV and 
$|\eta(j)|<3.5$ are present almost completely eliminates the $t\bar t$
background. It also reduces the $W^+W^-+X$ background at large
transverse momenta. As shown in Fig.~\ref{dibfig:two}, NLO QCD
corrections strongly affect the $p_T(e^+e^-)$ distribution in $W^+W^-\to 
e^+\nu e^-\bar\nu$. The enhancement at large $p_T$ is mostly due to real 
emission diagrams, leading to events which contain a hard jet.

To calculate the $Z+1$~jet background, we have assumed that jets with a
rapidity $|\eta(j)|>3.5$ are misidentified as $p\llap/_T$. With this
rather conservative assumption, the $Z+1$~jet background is much larger
than the $ZZ$ signal at small transverse momenta. Due to 
kinematical constraints, however, it drops rapidly with $p_T$. Since
non-standard $ZZV$ couplings lead to large deviations from the SM only
at high transverse momentum, essentially no sensitivity is lost by
requiring $p_T(\ell^+\ell^-)>40$~GeV when testing for $ZZV$
couplings. \\


\paragraph{$\boldmath{ZZ\to\ell^+\ell^-jj}$ and
$\boldmath{ZZ\to\bar\nu\nu jj}$} 

\begin{figure*}
\begin{center}
\mbox{\epsfxsize=13.9cm 
\epsffile{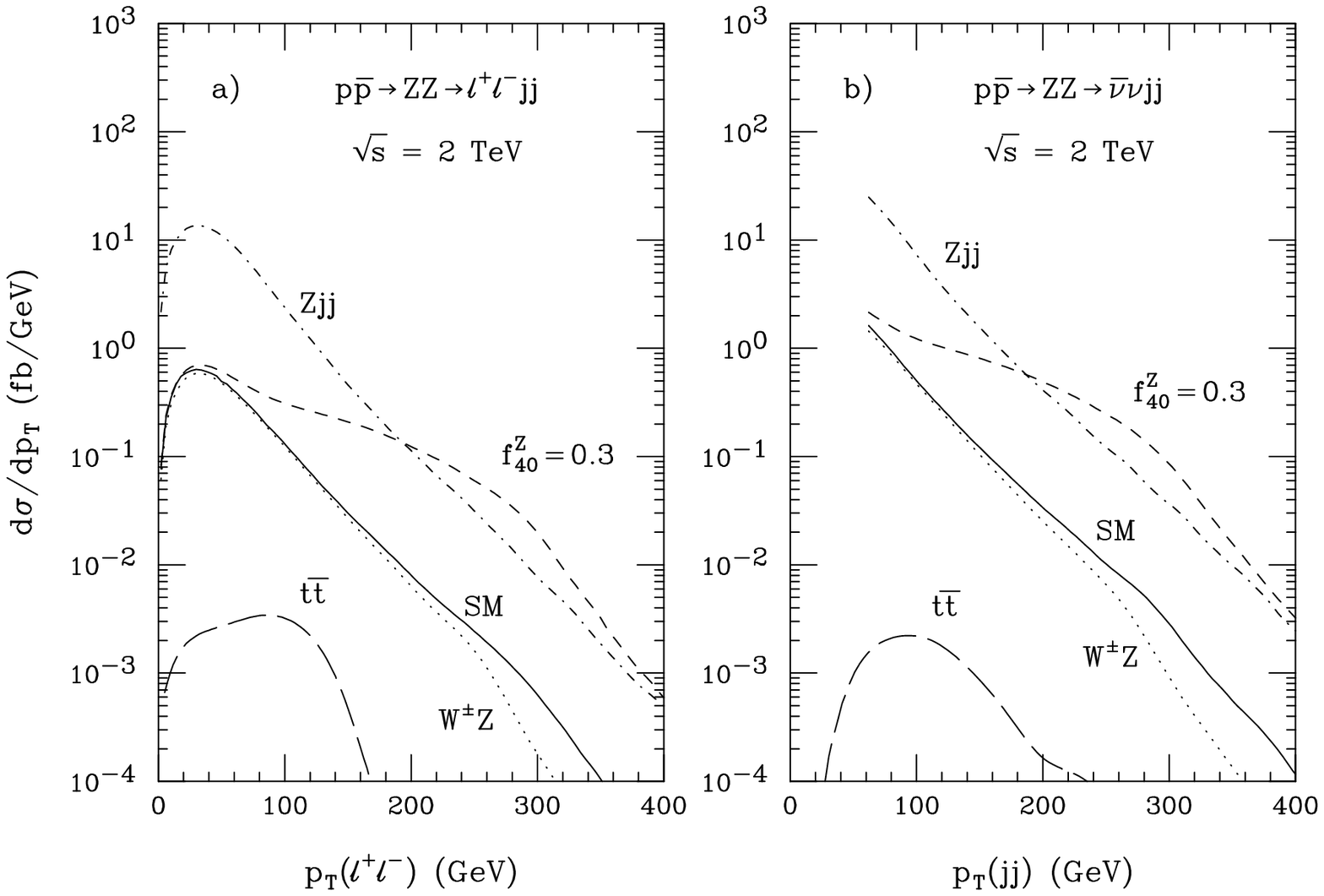}}
\end{center}
\vskip -8.mm
\caption{Transverse momentum distribution of a) the $\ell^+\ell^-$ pair
in $p\bar p\to ZZ\to\ell^+\ell^-jj$, and b) of the jet pair in
$ZZ\to\bar\nu\nu jj$ at the Tevatron, together with the
differential cross sections from $t\bar t$, $W^\pm Z$ and $Z+2$~jet
production.} 
\label{dibfig:seven}
\end{figure*}

The $ZZ\to\ell^+\ell^-$jj and $ZZ\to\bar\nu\nu$jj channels have 
larger branching ratios than the $ZZ\to 4$~leptons and the
$ZZ\to\ell^+\ell^-\bar\nu\nu$ channels, but also much higher backgrounds. The
main background sources are QCD $Z+2$~jet production and $W^\pm 
Z$ production with the $W$ decaying into two jets. The
$p_T(\ell^+\ell^-)$ distribution for $ZZ\to\ell^+\ell^-$jj is shown in
Fig.~\ref{dibfig:seven}a. Fig.~\ref{dibfig:seven}b shows the $p_T({\rm jj})$
distribution for $ZZ\to\bar\nu\nu$jj. In each case we display the SM
cross section together with the two main backgrounds, $Z$jj and $W^\pm
Z$ production. We also show the $ZZ$ cross section for $f_{40}^Z=0.3$.

The $p\llap/_T>60$~GeV cut imposed in the $ZZ\to p\llap/_T$jj case
helps to suppress the $b\bar b$ and $Z\to\tau^+\tau^-$ backgrounds.
The ``kink'' in the $WZ$ and $ZZ$ differential cross sections at
$p_T\approx 250$~GeV is due to the $\Delta R(jj)>0.6$ cut which
becomes effective only at sufficiently high transverse momenta. 
The $W^\pm Z$ differential cross section is very similar to that of the
SM signal over most of the $p_T$ range considered. The
$p\llap/_T<20$~GeV cut imposed in the $ZZ\to\ell^+\ell^-$jj channel 
effectively
eliminates the $t\bar t\to\ell^+\nu\ell^-\bar\nu$jj background.
The charged lepton veto (see Eq.~(\ref{eq:lep_veto})) required in the
$ZZ\to p\llap/_T$jj case rejects backgrounds from $t\bar t$ production, 
$W\to\ell\nu$, and $Z\to\ell^+\ell^-$ decays. 
The $Z$jj background is uniformly about a factor~10 larger than the SM
$ZZ$ signal. It will therefore be very difficult to observe $ZZ$
production in the semi-hadronic channels, if the SM prediction is
correct. However, for sufficiently large anomalous $ZZV$ couplings, the $ZZ$
cross section exceeds the background at large transverse
momenta. $ZZ\to\ell^+\ell^-$jj and $ZZ\to\bar\nu\nu$jj therefore may
still be useful in obtaining limits on the $ZZV$ couplings, similar to
the semi-hadronic $WW$ and $WZ$ channels used by CDF and D\O\ in
Run~I to extract limits on the $WWV$ couplings.\\

\paragraph{Sensitivity Bounds}

\begin{table*}[thb]
\setlength{\tabcolsep}{1.5pc}
\caption{Sensitivities achievable at $95\%$ CL for the
anomalous $ZZV$ couplings in $p\bar p\to ZZ\to 4$~leptons, $p\bar p\to
ZZ\to\ell^+\ell^-\bar\nu\nu$, $p\bar p\to ZZ\to\ell^+\ell^-$jj, and
$p\bar p\to ZZ\to\bar\nu\nu$jj 
at the Tevatron a) for an integrated luminosity of
2~fb$^{-1}$, and b) for an integrated luminosity of
10~fb$^{-1}$.\label{dibtab:two} }
\begin{tabular*}{\textwidth}{@{}l@{\extracolsep{\fill}}ccccc}
\hline   
\multicolumn{5}{c}{} \\ [-2.mm]
\multicolumn{5}{c}{a) $\int\!{\cal L}dt=2$~fb$^{-1}$} \\ [1.mm]
coupling & $ZZ\to 4$~leptons & $ZZ\to \ell^+\ell^-\bar\nu\nu$ & $ZZ\to
\ell^+\ell^-$jj & $ZZ\to \bar\nu\nu$jj \\ \hline 
$f_{40}^Z$ & -- & $ \begin{matrix} +0.169  \\ -0.169 \end{matrix}$ &
$ \begin{matrix} +0.219 \\ -0.218\end{matrix}$ & $\begin{matrix} +0.156 \\ 
-0.155 \end{matrix}$ \\[4.mm]
$f_{40}^\gamma$ & -- & $ \begin{matrix} +0.175  \\ -0.174 \end{matrix}$ &
$ \begin{matrix} +0.222 \\ -0.221\end{matrix}$ & $\begin{matrix} +0.157 \\ 
-0.157 \end{matrix}$ \\[4.mm]
$f_{50}^Z$ & -- & $ \begin{matrix} +0.171  \\ -0.204 \end{matrix}$ &
$ \begin{matrix} +0.220 \\ -0.244\end{matrix}$ & $\begin{matrix} +0.157 \\ 
-0.179 \end{matrix}$ \\[4.mm]
$f_{50}^\gamma$ & -- & $ \begin{matrix} +0.184  \\ -0.202 \end{matrix}$ &
$ \begin{matrix} +0.229 \\ -0.241\end{matrix}$ & $\begin{matrix} +0.166 \\ 
-0.174 \end{matrix}$ \\[4.mm]
\hline
\multicolumn{5}{c}{} \\ [-2.mm]
\multicolumn{5}{c}{b) $\int\!{\cal L}dt=10$~fb$^{-1}$} \\ [1.mm]
coupling & $ZZ\to 4$~leptons & $ZZ\to \ell^+\ell^-\bar\nu\nu$ & $ZZ\to
\ell^+\ell^-$jj & $ZZ\to \bar\nu\nu$jj \\ \hline 
$f_{40}^Z$ & $ \begin{matrix} +0.180  \\ -0.179 \end{matrix}$  & $
\begin{matrix} +0.097  \\ -0.097 \end{matrix}$ & 
$ \begin{matrix} +0.146 \\ -0.145\end{matrix}$ & $\begin{matrix} +0.104 \\ 
-0.103 \end{matrix}$ \\[4.mm]
$f_{40}^\gamma$ & $ \begin{matrix} +0.185  \\ -0.185 \end{matrix}$  & $
\begin{matrix} +0.100  \\ -0.099 \end{matrix}$ & 
$ \begin{matrix} +0.148 \\ -0.147\end{matrix}$ & $\begin{matrix} +0.104 \\ 
-0.104 \end{matrix}$ \\[4.mm]
$f_{50}^Z$ & $ \begin{matrix} +0.178  \\ -0.216 \end{matrix}$  & $
\begin{matrix} +0.092  \\ -0.120 \end{matrix}$ & 
$ \begin{matrix} +0.144 \\ -0.167\end{matrix}$ & $\begin{matrix} +0.102 \\ 
-0.124 \end{matrix}$ \\[4.mm]
$f_{50}^\gamma$ & $ \begin{matrix} +0.192  \\ -0.213 \end{matrix}$  & $
\begin{matrix} +0.103  \\ -0.115 \end{matrix}$ & 
$ \begin{matrix} +0.151 \\ -0.163\end{matrix}$ & $\begin{matrix} +0.109 \\ 
-0.118 \end{matrix}$ \\[1.mm]
\hline
\end{tabular*}
\end{table*}
In order to derive sensitivity limits for anomalous $ZZV$ couplings
which one can hope to achieve in Run~II, we use the $p_T(\ell^+\ell^-)$
distribution for $ZZ\to 4$~leptons, $ZZ\to\ell^+\ell^-\bar\nu\nu$ and
$ZZ\to\ell^+\ell^-$jj. For the $ZZ\to\bar\nu\nu$jj channel we use the
$p_T({\rm jj})$ distribution. Other distributions, such as the $ZZ$ invariant
mass distribution (useful only for $ZZ\to 4$~leptons), or the maximum or 
minimum transverse momenta of the charged leptons or jets, yield similar
results. In deriving our sensitivity limits, we combine channels with
electrons and muons in the final state.

We calculate 95\% confidence level (CL) limits performing a $\chi^2$
test. The statistical significance is calculated by splitting the $p_T$
distribution into a number of bins, each with more than five events
typically. In each bin the Poisson statistics is approximated by a
Gaussian distribution. In order to derive realistic limits, we allow for 
a normalization uncertainty of 30\% of the SM cross section. Backgrounds 
in the $ZZ\to\ell^+\ell^-\bar\nu\nu$, $ZZ\to\ell^+\ell^-$jj and
$ZZ\to\bar\nu\nu$jj channels are included in our calculation. In the
$ZZ\to\ell^+\ell^-\bar\nu\nu$ case we assume that a jet veto has been
imposed to reduce the $t\bar t$ background and require
$p_T(\ell^+\ell^-)>40$~GeV to eliminate the $Z+1$~jet background. 
As before, we use a form
factor of the form of Eq.~(\ref{dibeq:ff}) with $n=3$ and
$\Lambda_{FF}=750$~GeV. Non-negligible interference effects are 
found between $f_4^Z$ and $f_4^\gamma$, and between $f_5^Z$ and
$f_5^\gamma$. As a result, different anomalous contributions to the
helicity amplitudes may cancel partially, resulting in weaker bounds
than if only one coupling at a time is allowed to deviate from its SM
value. 

In Table~\ref{dibtab:two} we display sensitivity limits for the Tevatron
and integrated luminosities of 2~fb$^{-1}$ and 10~fb$^{-1}$, taking into
account the correlations between $f_4^Z$ and $f_4^\gamma$, and between
$f_5^Z$ and $f_5^\gamma$. 
No limits for the $ZZ\to 4$~leptons case with 2~fb$^{-1}$ are given. The 
limited number of events in this case does not allow for an analysis of
the $p_T(\ell^+\ell^-)$ distribution using the method chosen here. The
bounds obtained from $ZZ\to\ell^+\ell^-\bar\nu\nu$ and
$ZZ\to\bar\nu\nu$jj are quite similar. The cross section for 
$ZZ\to\bar\nu\nu$jj is
about a factor~10 larger than that for $ZZ\to\ell^+\ell^-\bar\nu\nu$,
however, the large background from $Z$jj production considerably limits the
sensitivity to $ZZV$ couplings for $ZZ\to\bar\nu\nu$jj. The limits from 
the $ZZ\to\ell^+\ell^-$jj and $ZZ\to 4$~leptons channels are about a 
factor~1.5 and~2 weaker than those from $ZZ\to\ell^+\ell^-\bar\nu\nu$
and $ZZ\to\bar\nu\nu$jj.

We have not made any attempt to combine the limits from different
channels. From Table~\ref{dibtab:two} it is clear that this would result in a
significant improvement of the bounds. 

The sensitivity limits which can be achieved at the Tevatron in Run~II
should be compared with the bounds from recent measurements at
LEP2~\cite{mor2000}:
\begin{eqnarray}
|f_4^Z|< 0.49 & \qquad & |f_4^\gamma| < 0.82 \\
|f_5^Z| < 1.1 & \qquad & |f_5^\gamma|  < 1.1
\end{eqnarray}
Only one coupling at a time is varied here.
(The LEP2 limits do not contain any form factor effects. For the form
and scale chosen here, form factor effects weaken the limits by about
20\%.) 
In Run~II, CDF and D\O\ will be able to improve these bounds by at least 
a factor~4 to~8.

\subsubsection{Measuring Form Factors}
The limits on anomalous $WWV$, $Z\gamma V$ and $ZZV$ couplings all
depend on the power, $n$, and the scale, $\Lambda_{FF}$, of the form
factor. These parameters are {\it a priori} unknown. 
In Ref.~\cite{renard} it was pointed out that in final states without
missing transverse momentum one can in principle determine the form factor 
by measuring the $\sqrt{\hat s}$ distribution. For $W\gamma$ production, 
the longitudinal momentum of the neutrino can only be reconstructed with 
a twofold ambiguity. Selecting the minimum of the two reconstructed
values of $\hat s$, a similar measurement can be performed in the
$W\gamma$ case~\cite{FS}. Alternatively, the photon $p_T$ distribution
can be used.

In Ref.~\cite{lhcewk} a detailed study of the method was performed for
$W(\to e\nu,\,\mu\nu)\gamma$ production at the LHC. Assuming
$\lambda_0^\gamma=0.025$, $n=2$, $\Lambda_{FF}=2$~TeV and an integrated
luminosity of 300~fb$^{-1}$, $\lambda_0^\gamma$ and $\Lambda_{FF}$ were
reconstructed using a binned maximum likelihood fit to the $p_T(\gamma)$ 
distribution. The power of the form factor, $n=2$, was not varied and
no detector simulation was included in the study. The
reconstructed coupling and form factor scale were determined to
$\lambda_0^\gamma=0.0295\pm 0.0022$ and $\Lambda_{FF}=1.67\pm 0.22$~TeV,
{\it ie.} they can  
be measured with a relative precision of about 10 -- 15\%. 
The central values of the reconstructed parameters differ by about 20\%
from the input parameters. Including
detector response, and treating the form factor power $n$ as an
additional free parameter is expected to increase the relative error.

The study demonstrates that, due to the restricted number of events in
each bin, the method will not produce competitive limits. However, if
non-zero anomalous couplings are observed, the method may be useful in
determining the shape of the form factor which provides indirect
information on the dynamics of the underlying new physics. 

At the Tevatron, the limits on the $WWV$ couplings exhibit only a modest 
dependence on $n$ and $\Lambda_{FF}$. Direct measurement of the form
factor will thus be very difficult for these couplings. The situation is 
more promising for $Z\gamma V$ and $ZZV$ couplings, where the
sensitivity bounds depend more strongly on the form factor parameters.

\section*{Acknowledgements}

We acknowledge contributions from the other members of the working
group including P.~Aurenche, C.~Bal\'{a}zs, L.~Dixon, S.~Eno, G.~Gomez,
N.~Kidonakis, P.~Nason, J.~Qiu, and G.~Sterman.  We also thank
T.~Ferbel, M.~Fontannaz, G.~Ginther, B.~Kniehl, A.~D.~Martin,
R.~Roberts, P.~Slattery, J.~Stirling, R.~Thorne and W.K.~Tung. We also wish to
thank A.~Beretvas, C.~Bigongiari, J.~Campbell, O.~Lobban, and 
W.~Wester for their help. Special thanks go to M.~Mangano
for some enlightening discussions.

\end{document}